%% file: main.tex
\documentclass{lmcs}

\usepackage{enumerate}
\usepackage{hyperref}
\usepackage{pstricks}
\usepackage{pst-node}
\usepackage{algorithm}
\usepackage[noend]{algorithmic}
\usepackage{bm}
\usepackage{tikz}
\usepackage{graphicx}
\usepackage{booktabs}
\usepackage{array}
\usepackage{pifont}
\usepackage{stmaryrd}
\usepackage{paralist}
\usepackage{rotating}

\SetSymbolFont{stmry}{bold}{U}{stmry}{m}{n}

\newcommand\lmcshevea[2]{#1}
\newcommand\lmcscorrection[2]{#1}

\input{macros}

\input{basic-defs}

\newif\ifdraft\draftfalse

\input{drafting-macros}

\renewenvironment{itemize}{
    \setdefaultleftmargin{1em}{1em}{}{}{}{}
    \begin{olditemize}
}{
    \end{olditemize}
}

\renewenvironment{enumerate}{
    \setdefaultleftmargin{1em}{1em}{}{}{}{}
    \begin{oldenumerate}
}{
    \end{oldenumerate}
}

\begin{document}

\title[\cshore: Saturation for CPDS]
      {\cshore: Higher-Order Verification via Collapsible Pushdown System Saturation}

\author[C. Broadbent et al.]{Christopher Broadbent}	
\address{Institut f\"ur Informatik (I7), Technische Universit\"at M\"unchen}	
\email{broadben@in.tum.de}  

\author[]{Arnaud Carayol}
\address{Laboratoire d'informatique de l'Institut Gaspard Monge, Universit\'e Paris-Est, and CNRS}
\email{Arnaud.Carayol@univ-mlv.fr}

\author[]{Matthew Hague}
\address{Department of Computer Science, Royal Holloway, University of London}
\email{matthew.hague@rhul.ac.uk}
\thanks{}

\author[]{Olivier Serre}
\address{IRIF, Universit\'e Paris Diderot - Paris 7, and CNRS}
\email{olivier.serre@cnrs.fr}

\titlecomment{This article gives a full account of the \cshore
tool~\cite{cshore} whose algorithms and implementation have been
published in ICALP 2012~\cite{BCHS12} and ICFP 2013~\cite{BCHS13}.}



\begin{abstract}
    \noindent
    \input{abstract}

\end{abstract}

\maketitle

\input{contents}

\bibliographystyle{plain}
\bibliography{dblprefs}

\end{document}

%% file: macros.tex
\usepackage{amsfonts}
\usepackage{amsmath}
\usepackage{amssymb}
\usepackage{xspace}
\usepackage{stmaryrd}
\usepackage{pifont}

\newcommand\neworrenewcommand[1]{%
    \let#1\relax
    \newcommand#1%
}

\providecommand\lmcshevea[2]{#1}

\newcommand\cshore{C-SHORe\xspace}
\newcommand\trecs{TRecS\xspace}

\newcommand\gtrecst{GTRecS2\xspace}
\newcommand\gtrecsot{GTRecS(2)\xspace}
\newcommand\travmc{TravMC\xspace}
\newcommand\mochi{MoCHi\xspace}
\newcommand\ehmtt{EHMTT Verifier\xspace}
\newcommand\horsat{HorSat\xspace}
\newcommand\preface{Preface\xspace}
\newcommand\horsattwo{HorSat2\xspace}

\lmcshevea{
    \providecommand\qed{\hfill\square}
}{
    \providecommand\qed{\square}
}

\newcommand{\sembr}[1]{\llbracket #1 \rrbracket}
\newcommand{\semdesc}[2]{\llbracket #2 \rrbracket_{#1}}

\newcommand\etal{~\textit{et al.}\xspace}

\newcommand\wlogen{w.l.o.g.\xspace}

\newcommand\setcomp[2]{\left\{{#1}\ \left|\ {#2}\right.\right\}}
\newcommand\set[1]{\left\{{#1}\right\}}
\newcommand\tuple[1]{\brac{#1}}
\newcommand\brac[1]{\left({#1}\right)}
\newcommand\apply[2]{{#1}\mathord{\brac{#2}}}
\newcommand\idxi{i}
\newcommand\idxj{j}
\newcommand\idxz{z}

\newcommand\numof{\ell}
\lmcshevea{
    \neworrenewcommand\ordinal{\alpha}
}{
    \renewcommand\ordinal{\alpha}
}

\newcommand\exptower[2]{2 \uparrow_{#1} \brac{#2}}
\newcommand\cardinality[1]{\left|{#1}\right|}

\newcommand\naturals{\mathbb{N}}
\newcommand\sizeof[1]{\left|{#1}\right|}
\newcommand\sequence[2]{\left({#1}\right)_{#2}}
\newcommand\maxfun{\mathrm{max}}

\newcommand\tick{\ding{51}\xspace}
\newcommand\cross{\ding{55}\xspace}

\newcommand\stacksalpha[2]{\mhchanged{\apply{Stacks_{#1}}{#2}}}
\newcommand\prestar[2]{\apply{Pre^*_{#1}}{#2}}

\newcommand\poststarraw{Post^*_{\mathcal{C}}}
\newcommand\postdepth[1]{{Post^{#1}_{\mathcal{C}}}}
\newcommand{\errorconfigs}{\boldsymbol{\mathcal{E}}}

\newcommand\pre[3]{\apply{Pre^{#1}_{#2}}{#3}}
\newcommand\cpds{\mathcal{C}}
\newcommand\controls{\mathcal{P}}
\newcommand\controlset{P}
\newcommand\cpdsrules{\mathcal{R}}
\newcommand\cpdsord{n}
\newcommand\opord{k}
\newcommand\cops[1]{\mathcal{O}_{#1}}

\newcommand\cha{a}
\newcommand\chb{b}
\newcommand\chc{c}
\newcommand\chd{d}
\newcommand\che{e}
\newcommand\cpush[2]{push^{#2}_{#1}}
\newcommand\push[1]{push_{#1}}

\newcommand\pop[1]{pop_{#1}}
\newcommand\collapse[1]{collapse_{#1}}
\newcommand\rew[1]{rew_{#1}}

\newcommand\cpdsruler{r}
\newcommand\cpdsrule[4]{\tuple{{#1},{#2},{#3},{#4}}}

\newcommand\cpdsaltrule[2]{\tuple{{#1},{#2}}}
\newcommand\control{p}
\newcommand\genop{o}

\newcommand\configc{c}
\newcommand\config[2]{\langle {#1}, {#2} \rangle}
\newcommand\bigconfig[2]{\left\langle {#1}, {#2} \right\rangle}
\newcommand\configset{C}
\newcommand\stackw{w}
\newcommand\stacku{u}
\newcommand\stackv{v}
\newcommand\stackr{w}
\newcommand\ctop[1]{top_{#1}}
\newcommand\cbottom[2]{bot^{#2}_{#1}}
\newcommand\cpdstran{\longrightarrow}

\newcommand\alphabet\Sigma

\newcommand\ccompose[3]{{#1} :_{#2} {#3}}

\newcommand\stack[1]{\left[{#1}\right]}

\newcommand\annot[2]{{#1}^{#2}}

\newcommand\cpdsalttran[2]{{#1} \rightarrow {#2}}

\newcommand\oalphabet\Gamma
\newcommand\ocha\gamma
\newcommand\cpdatran[1]{\xrightarrow{#1}}

\newcommand{\trivialise}[1]{\mathrm{Triv}(#1)}

\newcommand\substacks[1]{\apply{\mathrm{Subs}}{#1}}

\newcommand{\approxgraph}{\mathcal{G}}
\newcommand{\backrules}[1]{\mathrm{BackRules}(#1)}
\newcommand{\backrulesG}[1]{\mathrm{BackRulesG}(#1)}
\newcommand{\heads}[1]{\mathrm{Heads}(#1)}
\newcommand{\stackdes}{\mathrm{SDesc}}
\newcommand{\stackdesk}[1]{\mathrm{SDesc}_{#1}}

\newcommand\sastates{\mathbb{Q}}
\newcommand\sastateset{Q}
\newcommand\sadelta{\Delta}
\newcommand\safinals{\mathcal{F}}
\newcommand\sastate{q}

\newcommand\saauta{A}

\newcommand\sopen[1]{[_{#1}}
\newcommand\sclose[1]{]_{#1}}
\newcommand\sbrac[2]{[{#1}]_{#2}}

\newcommand\lang{\mathcal{L}}
\newcommand\langof[1]{\apply{\lang}{#1}}
\newcommand\slang[2]{\apply{\lang_{#1}}{#2}}
\newcommand\satran[1]{\xrightarrow{#1}}
\newcommand\satrancol[2]{\xrightarrow[{#2}]{#1}}
\newcommand\satranfull[4]{{#1} \xrightarrow[{#3}]{#2} \brac{{#4}}}
\newcommand\satranfullk[3]{{#1} \xrightarrow{#2} \brac{{#3}}}

\newcommand\branch{{col}}

\newcommand\sat{t}
\newcommand\satset{T}
\newcommand\satfull{\tau}
\newcommand\tjust{J}
\newcommand\unjust{0}

\newcommand\saprojection[1]{\apply{\mathrm{Proj}}{#1}}
\newcommand\saruncontained[2]{\apply{\mathrm{Contained}}{#1, #2}}
\newcommand\tunexpand[1]{\apply{\mathrm{Top}_1}{#1}}
\newcommand\srcorig{e}

\newcommand\satstep{\Pi}

\newcommand\countdown{\sastateset^C}
\newcommand\qerror{{\control_{\mathit{err}}}}
\newcommand\transdone{\Delta_{done}}
\newcommand\transtodo{\Delta_{new}}

\newcommand\sources{\mathcal{U}_{src}}
\newcommand\targs{\mathcal{U}_{targ}}

\newcommand\createtrip{\mathrm{CreateTripWire}\xspace}
\newcommand\updatetrip{\mathrm{UpdateTripWires}\xspace}
\newcommand\extractshort{\mathrm{Extract}\xspace}
\newcommand\proctarg{\mathrm{ProcTargAgainstTran}\xspace}
\newcommand\procsrccomptarg{\mathrm{ProcSourceCompleteTarg}\xspace}

\newcommand\updatebyrules{\mathrm{UpdateRules}\xspace}
\newcommand\addworklist{\mathrm{AddToWorklist}\xspace}

\newcommand\addtarget{\mathrm{AddTarget}\xspace}

\newcommand\sastatelabel{\mathit{lbl}}

\newcommand\targmarker{\diamond}

\newcommand\rsrew{\hookrightarrow}

\newcommand{\popg}[2]{\pop{#1}^{#2}}
\newcommand{\collapseg}[2]{\collapse{#1}^{#2}}
\newcommand{\alphaset}{S} 

\newcommand{\gentr}{t}

\newcommand{\getcountereg}{\mathrm{GetWitness}}
\newcommand\nodelab[1]{\left[{#1}\right]}
\newcommand\nodelabel{\alpha}
\newcommand\treeap[2]{\apply{#1}{#2}}

\newcommand\treedom{D}
\newcommand\treelab{\lambda}
\newcommand\treenode{\eta}
\newcommand\treelabset{\Gamma}

\newcommand{\runindexrel}{\hookrightarrow}



%% file: basic-defs.tex
\newtheorem{example}[thm]{Example}

\let\oldparagraph\paragraph
\renewcommand\paragraph[1]{\oldparagraph{\textit{#1}}}

%% file: drafting-macros.tex
\usepackage{pstricks}
\usepackage{manfnt}

\ifdraft
\else

\newcommand\todo[1]{}
\newcommand\os[1]{}
\newcommand\ac[1]{}
\newcommand\mh[1]{}
\newcommand\cb[1]{}

\newcommand\mhchanged[1]{#1}

\newcommand\reviewP[2]{}
\newcommand\reviewW[2]{}
\newcommand\reviewF[2]{}
\newcommand\answer[1]{}
\newcommand{\remove}[1]{}
\fi

%% file: abstract.tex
Higher-order recursion schemes (HORS) have received much attention as a
useful abstraction of higher-order functional programs with a number of new verification
techniques employing HORS model-checking as their centrepiece.  We give an account of the \cshore tool, which contributed to the ongoing quest for a truly scalable model-checker for HORS by
offering a different, automata theoretic perspective.
\cshore implements the first
practical model-checking algorithm that acts on a generalisation of pushdown
automata equi-expressive with HORS called \emph{collapsible pushdown systems}
(CPDS). At its core is a backwards saturation algorithm for CPDS.
Additionally, it is able to use information
gathered from an approximate forward reachability analysis to guide its backward
search. Moreover, it uses an algorithm that prunes the CPDS prior to
model-checking and a method for extracting counter-examples in negative
instances.  We provide an up-to-date comparison of \cshore with the state-of-the-art verification tools for
HORS.
The tool and additional material are available from
\url{http://cshore.cs.rhul.ac.uk}.

%% file: contents.tex
\input{introduction}

\input{modelling}

\input{preliminaries}

\input{saturation}

\input{corrcomp}

\input{forward}

\input{fastalg}

\input{experiments}

\input{conclusion}

\input{acknowledgments}

%% file: introduction.tex
\section{Introduction}
\label{sec:introduction}

Functional languages such as Haskell, OCaML and Scala strongly encourage the use
of higher-order functions. This represents a challenge for software
verification, which usually does not model recursion accurately, or models only
first-order calls (e.g.  SLAM~\cite{BR02} and Moped~\cite{S02}).  However, there
has recently been much interest in a model called \emph{higher-order recursion
schemes (HORS)} (see e.g.~\cite{O06}), which offers a way of abstracting
functional programs in a manner that precisely models higher-order control-flow.

The execution trees of HORS enjoy decidable $\mu$-calculus theories~\cite{O06}.
Even `reachability' properties (subsumed by the $\mu$-calculus) are very useful
in practice. As a simple example, the safety of incomplete pattern matching
clauses could be checked by asking whether the program can `reach a state'
where a pattern match failure occurs.  More complex `reachability' properties
can be expressed using a finite automaton and could, for example, specify that
the program respects a certain discipline when accessing a particular resource
(see \cite{K09}).  Despite even reachability being $(n-1)$-EXPTIME complete,
recent research has revealed that useful properties of HORS can be checked in
practice.

Kobayashi's \trecs~\cite{K09b} tool, which checks properties expressible by a
deterministic trivial B\"uchi automaton (all states  accepting), was the first
to achieve this. It works by determining whether a HORS is typable in an
intersection-type system characterising the property to be checked~\cite{K09}.
In a bid to improve scalability, a number of other algorithms have subsequently
been designed and implemented such as Kobayashi \etal's
\gtrecsot~\cite{K11b,gtrecs2} and Neatherway \etal's \travmc~\cite{NRO12} tools,
all based on intersection type inference.
A recent overview of HORS model-checking was given by Ong~\cite{O15}.

This work is the basis of various techniques for verifying functional programs.
In particular, Kobayashi\etal have developed \mochi~\cite{KSU11} that checks safety properties
of (OCaML) programs, and \ehmtt~\cite{Unno10APLAS} for tree processing
programs.  Both use a recursion schemes model-checker as a central
component.  Similarly, Ong and Ramsay~\cite{OR11}
analyse programs with pattern matching employing recursion schemes as an
abstraction.

Achieving scalability while accurately tracking higher-order control-flow is a
challenging problem.  This article offers an automata-theoretic perspective on
this challenge, providing a fresh set of tools that contrast with previous
intersection-type approaches.

\emph{Collapsible pushdown systems (CPDS)}~\cite{HMOS08,HMOS17} are an alternative
representation of the class of execution trees that can be generated by
recursion schemes (with linear-time mutual-translations between the two
formalisms~\cite{HMOS08,HMOS17,CS12}).  While pushdown systems augment a finite-state
machine with a stack and provide an ideal model for first-order
programs~\cite{JM77}, collapsible pushdown systems model higher-order programs
by extending the stack of a pushdown system to a nested ``stack-of-stacks''
structure. The nested stack structure enables one to represent closures. Indeed
the reader might find it helpful to view a CPDS as being a Krivine's Abstract
Machine in a guise making it amenable to the generalisation of techniques for
pushdown model-checking. Salvati and Walukiewicz have studied in detail the
connection with the Krivine abstract machine~\cite{SW12,SW16}.

For ordinary (`order-$1$') pushdown systems, a model-checking approach called
\emph{saturation} has been successfully implemented by tools such as
Moped~\cite{S02} and PDSolver~\cite{HO10b}. Given a regular set of
configurations of the pushdown system (represented by a finite automaton
$\saauta$ acting on stacks), saturation can solve the `backward reachability
problem' by computing another finite automaton recognising a set of
configurations from which a configuration in $\mathcal{\lang}(\saauta)$ can be
reached.  This is a fixed-point computation that gradually adds transitions to
$\saauta$ until it is `saturated'. If $\saauta$ recognises a set of error
configurations, one can determine whether the pushdown system is `safe' by
checking if its initial configuration is recognised by the automaton computed
by saturation.

The first contribution of this article was first presented in ICALP 2012.
We extend the saturation method to a backward reachability analysis of
collapsible pushdown systems~\cite{BCHS12}. This runs in PTIME when the number
of control states is bounded. Crucially, this condition is satisfied when
translating from recursion schemes of bounded arity with properties represented
by automata of bounded size~\cite{HMOS08,HMOS17}. Whilst the HORS/intersection-type
based tool \gtrecsot{} also enjoys this  fixed-parameter tractability, it times
out on  many benchmarks that our tool solves quickly.  We remark also
Ramsay\etal introduced a third fixed-parameter tractable algorithm in 2014
underlying their \preface tool~\cite{RNO14}.

In this work, we revisited the foundations of higher-order
verification tools and introduced \cshore~\cite{cshore} --- the first model-checking tool for
the (direct) analysis of collapsible pushdown systems.
This work was presented in ICFP 2013~\cite{BCHS13}.
To achieve an efficient implementation, some substantial
modifications and additions were made to the algorithm, leading to several novel
practical and theoretical contributions:
\begin{enumerate}
\label{enum:contributions}
\item
    An approximate \emph{forward} reachability algorithm providing data
    \begin{enumerate}
    \item
        \dots allowing the CPDS to be pruned so that
        saturation receives a smaller input.

    \item
        \dots employed by a modified saturation algorithm to guide its
        \emph{backward} search.
    \end{enumerate}
    This is essential for termination on most of our benchmarks.

\item
    A method for extracting witnesses to reachability.

\item
    A complete rework of the saturation algorithm to speed up
    fixed-point computation.
\item
    Experimental results comparing our approach with other tools.
\end{enumerate}
We remark that the tools mentioned above propagate information forwards WRT the
evaluation of the model. In contrast, the raw saturation algorithm works
backwards, but we also show how forward and backward propagation can be
combined.

Here we give a full account of the \cshore tool.
This covers the saturation algorithm presented at ICALP 2012 as well as efficient algorithms implemented by \cshore in ICFP 2013.
To prove soundness, we diverge from the ICALP 2012 proof, and instead base our proof on the witness generation algorithm presented in ICFP 2013.
In particular, we present novel generalisations of witness generation, the forwards analysis, and the efficient fixed-point calculation to \emph{alternating} CPDSs.
These were only given for non-alternating CPDSs in ICFP 2013.
The tool is available at \url{http://cshore.cs.rhul.ac.uk}.

Since \cshore was released, two new tools were released.
Broadbent\etal introduced \horsat, which is an application of the saturation technique and initial forward analysis directly to intersection type analysis of HORS~\cite{BK13}.
Recently \horsattwo improved the forwards analysis and made other algorithmic improvements~\cite{horsat2}.
Secondly, in POPL 2014, Ramsay\etal introduced \preface~\cite{RNO14}.
This is a type-based abstraction-refinement algorithm that attempts to simultaneously prove and disprove the property of interest.
Both \horsattwo and \preface perform significantly better than previous tools.

Even though both \preface and \horsattwo both outperform \cshore, we consider the CPDS approach to offer a different perspective by providing a link between successful pushdown model-checking tools and higher-order model-checking.
Moreover, CPDS have been instrumental in proving a number of results about higher-order languages.
Hence, it is very natural to consider the implementation of a model-checker using these automata techniques and the challenges and opportunities therein.
This article provides an account of a significant effort to extend pushdown model-checking to the higher-order case, and therefore we hope it will be instructive to readers interested in building verification tools for higher-order programming languages.

Section~\ref{sec:functionalIntro} is an informal introduction to HORS and CPDS.
In Section~\ref{sec:preliminaries} we describe CPDS and how to represent sets of their configurations.
The basic saturation algorithm introduced in ICALP 2012 is presented in Section~\ref{sec:saturation-alg} and proven correct in Section~\ref{sec:correctness}.
Section~\ref{sec:counter-egs} gives our generalised witness generation algorithm (that also implies soundness of saturation).
We describe two optimisations to the saturation algorithm used by \cshore:
an initial forwards analysis in Section~\ref{section:optimisations} and an efficient fixed point computation in Section~\ref{sec:fastalgorithm}.
Experimental results are in Section~\ref{sec:experiments}.

%% file: modelling.tex
\section{Modelling Higher-Order Programs}
\label{sec:functionalIntro}

In this section we give an informal introduction to the process of modelling
higher-order programs for verification.  In particular, we show how a simple
example program can be modelled using a higher-order recursion scheme, and then
we show how this scheme is evaluated using a collapsible pushdown system.  For a
more systematic approach to modelling higher-order programs with recursion
schemes, we refer the reader to work by Kobayashi\etal~\cite{KSU11}.  This
section is for background only, and can be safely skipped.

For this section, consider the toy example below.
\begin{verbatim}
  Main = MakeReport Nil
  MakeReport x = if * (Commit x)
                 else (AddData x MakeReport)
  AddData y f = if * (f Error) else (f Cons(_, y))
\end{verbatim}
In this example, \verb+*+ represents a non-deterministic choice (that may, for
example, be a result of some input by the user). Execution begins at
\verb+Main+ which aims to make a report which is a list. It sends an empty
report to \verb+MakeReport+.  Either \verb+MakeReport+ finishes and commits the
report somehow, or it adds an item to the head of the list using
\verb+AddData+, which takes the report so far, and a continuation.
\verb+AddData+ either detects a problem with the new data (maybe it is
inconsistent with the rest of the report) and flags an error by passing
\verb+Error+ to the continuation, or extends the report with some item.  In
this case, since there is no error handling in \verb+MakeReport+, an
\verb+Error+ may be committed.

\subsection{Higher-Order Recursion Schemes}

\begin{figure*}
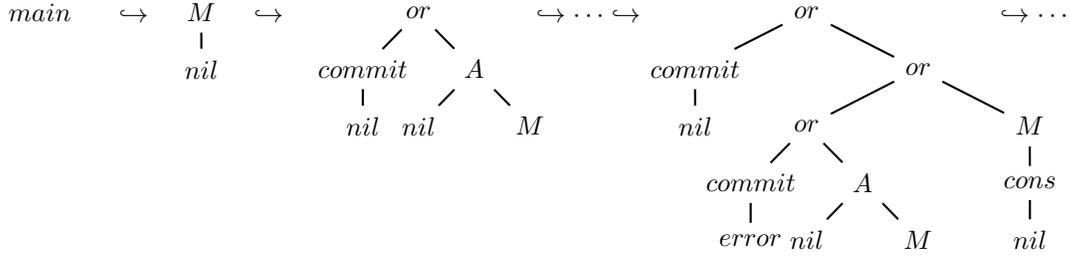

    \small
    \begin{center}
    \psset{xunit=4.9ex,yunit=5ex,nodesep=1ex}%
    \begin{tabular}{cccccccc}
        \pspicture(1,5)(1,9)
            \rput(1,9){\rnode{R0N0}{$main$}}%
        \endpspicture
    & $\qquad$\raisebox{2.9cm}{$\rsrew$}$\quad$ &
        \psset{nodesep=1ex}%
        \pspicture(1,5)(1,9)
            \rput(1,9){\rnode{R0N0}{$M$}}%
            \rput(1,8){\rnode{R1N0}{$nil$}}%

            \ncline{R0N0}{R1N0}
        \endpspicture
    & \quad\raisebox{2.9cm}{$\rsrew$}$\qquad$ &
        \psset{nodesep=1ex}%
        \pspicture(0,5)(3.25,9)
            \rput(1,9){\rnode{R0N0}{$or$}}%
            \rput(0,8){\rnode{R1N0}{$commit$}}%
            \rput(2,8){\rnode{R1N1}{$A$}}%
            \rput(0,7){\rnode{R2N0}{$nil$}}%
            \rput(1,7){\rnode{R2N1}{$nil$}}%
            \rput(3,7){\rnode{R2N2}{$M$}}%

            \ncline{R0N0}{R1N0}%
            \ncline{R0N0}{R1N1}%
            \ncline{R1N0}{R2N0}%
            \ncline{R1N1}{R2N1}%
            \ncline{R1N1}{R2N2}%
        \endpspicture
    & \hspace{-3ex}\raisebox{2.9cm}{$\rsrew \cdots \rsrew$} &
        \psset{nodesep=1ex}%
        \pspicture(-.5,5)(5,9)
            \rput(2,9){\rnode{R0N0}{$or$}}%
            \rput(0,8){\rnode{R1N0}{$commit$}}%
            \rput(4,8){\rnode{R1N1}{$or$}}%
            \rput(0,7){\rnode{R2N0}{$nil$}}%
            \rput(2,7){\rnode{R2N1}{$or$}}%
            \rput(6,7){\rnode{R2N2}{$M$}}%
            \rput(1,6){\rnode{R3N0}{$commit$}}%
            \rput(3,6){\rnode{R3.2}{$A$}}%
            \rput(6,6){\rnode{R3N2}{$cons$}}%
            \rput(1,5){\rnode{R4N0}{$error$}}%
            \rput(2,5){\rnode{R4N1}{$nil$}}%
            \rput(4,5){\rnode{R4N2}{$M$}}%
            \rput(6,5){\rnode{R4N3}{$nil$}}%
            \ncline{R0N0}{R1N0}%
            \ncline{R0N0}{R1N1}%
            \ncline{R1N0}{R2N0}%
            \ncline{R1N1}{R2N1}%
            \ncline{R1N1}{R2N2}%
            \ncline{R2N1}{R3N0}%
            \ncline{R2N1}{R3.2}%
            \ncline{R2N2}{R3N2}%
            \ncline{R3N0}{R4N0}%
            \ncline{R3.2}{R4N1}%
            \ncline{R3.2}{R4N2}%
            \ncline{R3N2}{R4N3}%
        \endpspicture
    & \raisebox{2.9cm}{$\rsrew \cdots$}
    \end{tabular}
    \end{center}
    \caption{\label{fig:executing-toy}The behaviour of a toy recursion scheme.}
\end{figure*}

We introduce, informally, higher-order recursion schemes.  These are rewrite
systems that generate the computation tree of a functional program.  A rewrite
rule takes the form
\[
    N\ \phi\ x\ \rsrew\ t
\]
where $N$ is a (simply) typed non-terminal with (possibly higher-order)
arguments $\phi$ and $x$.  A term $N\ t_\phi\ t_x$ rewrites to $t$ with
$t_\phi$ substituted for $\phi$ and $t_x$ substituted for $x$.  Note that
recursion schemes require $t$ to be of ground type.  We illustrate recursion
schemes and their use in analysis using the toy example from above.
We can directly model our example with the scheme
\[
    \begin{array}{rcl}
        main & \rsrew & M\ nil  \\
        M\ x & \rsrew & or\ (commit\ x)\ (A\ x\ M) \\
        A\ y\ \phi & \rsrew & or\ (\phi\ error)\ (\phi\ (cons\ y))
    \end{array}
\]
where $M$ is the non-terminal associated with the \verb+MakeReport+ function, and $A$ is
the non-terminal associated with the \verb+AddData+ function; $nil$, $or$,
$commit$, $error$ and $cons$ are terminal symbols of arity 0, 2, 1, 0 and 1 respectively (e.g. in the second rule, $or$ takes the two arguments $(commit\ x)$ and $(A\ x\ M)$).  The scheme above begins with the non-terminal $main$ and, through a
sequence of rewrite steps, generates a tree representation of the evolution of
the program.  Figure~\ref{fig:executing-toy}, described below, shows such a
sequence.

Beginning with the non-terminal $main$, we apply the first rewrite rule to
obtain the tree representing the term $(M\ nil)$.  We then apply the second
rewrite rule, instantiating $x$ with $nil$ to obtain the next tree in the
sequence.  This continues \textit{ad infinitum} to produce a possibly infinite
tree labelled only by terminals.

We aim to show the correctness of the program.  I.e.~the program never
tries to $commit$ an $error$.  The rightmost tree in
Figure~\ref{fig:executing-toy}, has a branch labelled $or, or, or,
commit, error$.  Note, $commit$ is being called
with an $error$ report.  In general we define the regular language
$\lang_{err} = or^\ast commit\ or^\ast error$.  If the tree generated by the
HORS contains a branch labelled by a word appearing in
$\lang_{err}$, then we have identified an error in the program.

\subsection{Collapsible Pushdown Automata}

Previous research into the verification of HORS has used \emph{intersection
types} (e.g.~\cite{K11,NRO12}).  Here we investigate a radically different
approach exploiting the connection between HORS and an automata model called
\emph{collapsible pushdown automata} (CPDA).  These two formalisms are, in
fact, equivalent.

\begin{thm}[Equi-expressivity~\cite{HMOS08,HMOS17}]
    For each order-$n$ recursion scheme, there is an order-$n$ collapsible
    pushdown automaton generating the same tree, and vice-versa.
    Furthermore, the translation from recursion schemes to collapsible pushdown automata is linear, and the opposite translation is polynomial.
    \qed
\end{thm}

We describe at a high level the structure of a CPDA and how they can be used to
evaluate recursion schemes.  In our case, this means outputting a sequence of
non-terminals representing each path in the tree.  More formal definitions are
given in Section~\ref{sec:preliminaries}.  At any moment, a CPDA is in a
\emph{configuration} $\config{\control}{\stackw}$, where $\control$ is a
control state taken from a finite set $\controls$, and $\stackw$ is a
higher-order collapsible stack.  In the following we will focus on the stack.
Control states are only needed to ensure that sequences of stack operations occur in
the correct order and are thus elided for clarity.

\begin{figure*}
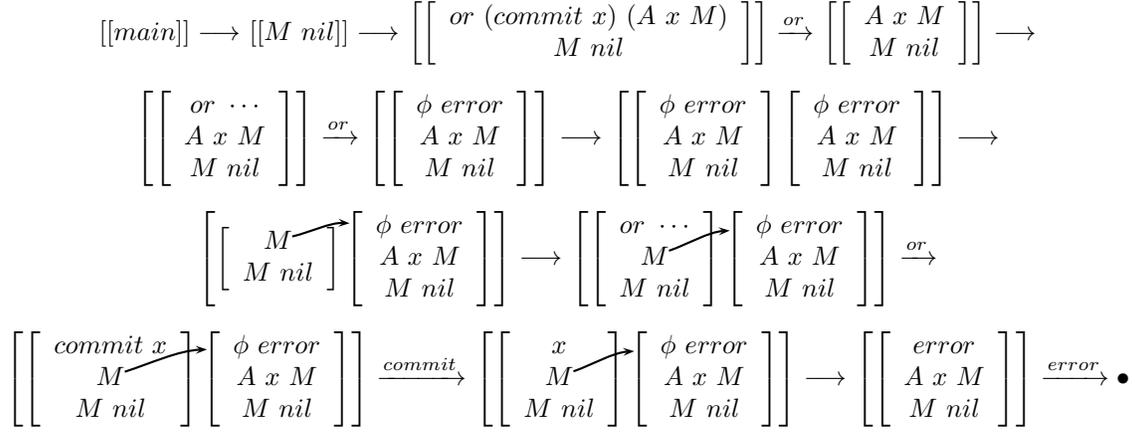

\small
    \begin{center}
    \begin{psmatrix}[rowsep=2ex,colsep=1ex]
        $\stack{\stack{main}} \cpdstran$
        $\stack{\stack{M\ nil}} \cpdstran$
        $\stack{\stack{\begin{array}{c}
            or\ (commit\ x)\ (A\ x\ M) \\
            M\ nil
        \end{array}}} \cpdatran{or}$
        $\stack{\stack{\begin{array}{c}
            A\ x\ M \\
            M\ nil
        \end{array}}} \cpdstran$ \\
        $\stack{\stack{\begin{array}{c}
            or\ \cdots \\
            A\ x\ M \\
            M\ nil
        \end{array}}} \cpdatran{or}$
        $\stack{\stack{\begin{array}{c}
            \phi\ error \\
            A\ x\ M \\
            M\ nil
        \end{array}}} \cpdstran$
        $\stack{
            \stack{\begin{array}{c}
                \rnode{PHI1}{\phi\ error} \\
                A\ x\ M \\
                M\ nil
            \end{array}}
            \stack{\begin{array}{c}
                \rnode{PHI1}{\phi\ error} \\
                A\ x\ M \\
                M\ nil
            \end{array}}
        } \cpdstran$ \\
        $\stack{
            \stack{\begin{array}{c}
                \rnode{P1}{M} \\
                M\ nil
            \end{array}}
            \stack{\begin{array}{c}
                \rnode{PHI1}{\phi\ error} \\
                A\ x\ M \\
                M\ nil
            \end{array}}
        } \cpdstran$
        $\stack{
            \stack{\begin{array}{c}
                or\ \cdots \\
                \rnode{P2}{M} \\
                M\ nil
            \end{array}}
            \stack{\begin{array}{c}
                \rnode{PHI2}{\phi\ error} \\
                A\ x\ M \\
                M\ nil
            \end{array}}
        } \cpdatran{or}$ \\
        $\stack{
            \stack{\begin{array}{c}
                commit\ x \\
                \rnode{P3}{M} \\
                M\ nil
            \end{array}}
            \stack{\begin{array}{c}
                \rnode{PHI3}{\phi\ error} \\
                A\ x\ M \\
                M\ nil
            \end{array}}
        } \cpdatran{commit}$
        $\stack{
            \stack{\begin{array}{c}
                x \\
                \rnode{P4}{M} \\
                M\ nil
            \end{array}}
            \stack{\begin{array}{c}
                \rnode{PHI4}{\phi\ error} \\
                A\ x\ M \\
                M\ nil
            \end{array}}
        } \cpdstran$
        $\stack{
            \stack{\begin{array}{c}
                error \\
                A\ x\ M \\
                M\ nil
            \end{array}}
        } \cpdatran{error} \bullet$
    \end{psmatrix}
    \psset{nodesepB=2.3ex}
    \ncarc{->}{P1}{PHI1}
    \ncarc{->}{P2}{PHI2}
    \ncarc{->}{P3}{PHI3}
    \ncarc{->}{P4}{PHI4}
    \end{center}
    \caption{\label{fig:executing-stack}A stack evaluating the toy example.}
\end{figure*}

In our toy example, we have an order-$2$ HORS and hence
an order-$2$ stack.  An order-$1$ stack is a stack of characters
from a
finite alphabet $\alphabet$.  An order-$2$ stack is a stack of order-$1$ stacks.
Thus $\stack{\stack{main}}$ denotes the order-$2$ stack
containing only the order-$1$ stack $\stack{main}$; $\stack{main}$ is an order-$1$
stack containing only the character $main$.  In general $\alphabet$ will contain
all subterms appearing in the original statement of our toy example recursion
scheme.  The evolution of the CPDA stack is given in
Figure~\ref{fig:executing-stack} and explained below.

The first step is to rewrite $main$ using $main \rsrew M\ nil$.  Since $(M\
nil)$ is a subterm of our recursion scheme, we have $(M\ nil) \in \alphabet$
and we rewrite the stack $\stack{\stack{main}}$ to $\stack{\stack{M\ nil}}$.
Next, we call $M$.  As usual, a function call necessitates a new stack frame.
In particular, we push the body of $M$ (that is $(or\ (commit\ x)\ (A\ x\ M))$)
onto the stack, giving the third stack in Figure~\ref{fig:executing-stack}.
Note, we do not instantiate the variable $x$, hence we use only the subterms
appearing in the recursion scheme.

Recall that we want to obtain a CPDA that outputs a sequence of terminals
representing each path in the tree.  To evaluate $or\ (\cdots)\
(\cdots)$ we output the terminal $or$ and then (non-deterministically)
choose a branch of the tree to follow.  Let us choose $(A\ x\ M)$.  Hence, the
CPDA outputs $or$ and rewrites the top term to $(A\ x\ M)$.  Next
we call $A$, pushing its body to the stack, then
pick out the $(\phi\ error)$ branch of the $or$ terminal.  This takes us to the
beginning of the second row of Figure~\ref{fig:executing-stack}.

To proceed, we evaluate $(\phi\ error)$.  To do
this, we have to know the value of $\phi$.  We can obtain this information by
inspecting the stack and seeing that the second argument of the call of $A$ is
$M$.  However, since we can only see the top of a stack, we would have to remove
the character $(\phi\ error)$ to determine that $\phi = M$, thus
losing our place in the computation.

However, an order-$2$ stack is able
--- via a $\push{2}$ operation --- to create a copy of its topmost order-$1$
stack.  After this copy (note that the top of the stack is written
on the left) we delve into the copy of the stack to find the value of
$\phi$.  Simultaneously we create a \emph{collapse link}, pictured as an
arrow from $M$ to the stack with the term $(\phi\ error)$ on top.  This collapse link points from
$M$ to the context in which $M$ will be evaluated.  In particular, if we need to
know the value of $x$ in the body of $M$, we need to know that $M$ was
called with the $error$ argument, within the term $(\phi\ error)$; the collapse
link points to this information (i.e.~encodes a
closure in the stack).  We can access this information via a \emph{collapse}
operation.  These are the two main features of a higher-order collapsible stack,
described formally in the next section.

To continue, we push the body of $M$ on to the stack, output the
$or$ symbol and choose the $(commit\ x)$ branch.  Since $commit$ is a
terminal, we output it and evaluate $x$.  To compute
$x$, we look into the stack and follow the collapse link from $M$ to
the stack with $(\phi\ error)$ on top.  We do not create a copy of the
stack here because $x$ is an order-$0$ variable and thus represents a
self-contained execution.  Since $x$ has value $error$, we output it and terminate.  This
completes the execution corresponding to the error branch identified in
Figure~\ref{fig:executing-toy}.

\subsection{Collapsible Pushdown Systems}

The CPDA output $or, or, or, commit, error$ in the execution above.  This is an
error sequence in $\lang_{err}$ and should be flagged.  In general, we take the
finite automaton $\saauta$ representing the regular language $\lang_{err}$ and
form a synchronised product with the CPDA.  This results in a CPDA that does
not output any symbols, but instead keeps in its control state the progression
of $\saauta$.  Thus we are interested in whether the CPDA is able to reach an
accepting state of $\saauta$, not the language it generates.  We call a CPDA
without output symbols a \emph{collapsible pushdown system} (CPDS), and the
question of whether a CPDS can reach a given state is the reachability problem.
This is the subject of the remainder of the paper.

%% file: preliminaries.tex
\section{Preliminaries}
\label{sec:preliminaries}

\lmcscorrection{
    \subsection{Collapsible Pushdown Systems}

    We give the definition of higher-order collapsible stacks and their operations,
    before giving the definition of collapsible pushdown systems.

    \subsubsection{Higher-Order Collapsible Stacks}
}{
    We give the definition of higher-order collapsible stacks before describing
    stack automata.

    \subsection{Higher-Order Collapsible Stacks}
}

Higher-order collapsible stacks are a nested ``stack-of-stacks'' structure
over a stack alphabet $\alphabet$.  Each stack character
contains a pointer --- called a ``link'' --- to a position lower down in the
stack.  The stack operations, defined below, create copies of
sub-stacks.  The link is intuitively a pointer to the context in
which the stack character was first created.
These links will be defined as tuples, the meaning of which is expanded
upon after the following definition.
Let the natural numbers $\naturals$ be
$\set{0, 1, 2, \ldots}$.
We will write stacks with the top of the stack appearing on the left.

\begin{defi}[Order-$\cpdsord$ Collapsible Stacks]
    An \emph{order-$\opord$ link} is a tuple
    $\tuple{\opord, \idxi}$
    where $\opord \geq 1$ and $\idxi$ are natural numbers.
    If
    $\opord \in \set{1, \ldots, \cpdsord}$
    we say the link is \emph{up-to} order-$\cpdsord$.
    Given a finite set of stack characters $\alphabet$, an \emph{order-$0$
    stack} with an up-to order-$\cpdsord$ link is
    $\annot{\cha}{\tuple{\opord, \idxi}}$
    where $\cha \in \alphabet$ and $\tuple{\opord, \idxi}$ is an up-to order-$\cpdsord$ link.
    An \emph{order-$\opord$ stack} with up-to order-$\cpdsord$ links is a sequence
    $\stackw = \sbrac{\stackw_1 \ldots \stackw_\numof}{\opord}$
    such that each $\stackw_\idxj$ is an order-$(\opord-1)$ stack with up-to order-$\cpdsord$ links.
    Moreover, for each $\stackw_\idxj$ and each order-$\opord$ link
    $\tuple{\opord, \idxi}$
    appearing on a character in $\stackw_\idxj$, we have
    $0 \leq \idxi \leq \numof - \idxj$.
    Let $\stacksalpha{\cpdsord}{\alphabet}$ denote the set of order-$\cpdsord$ stacks over $\alphabet$ with up-to order-$\cpdsord$ links.
\end{defi}

In the sequel we will refer to order-$\cpdsord$ stacks with up-to order-$\cpdsord$ links simply as order-$\cpdsord$ stacks.
We will use order-$\opord$ stack to mean an order-$\opord$ stack with up-to order-$\cpdsord$ links, where $\cpdsord$ is clear from the context.
We define the interpretation of the collapse links formally below.
Intuitively, the collapse links point to a position lower down in the stack.
In a link
$\tuple{\opord, \idxi}$
the first component $\opord$ indicates that the link points to a location inside the order-$\opord$ stack where the link is contained.
The second component $\idxi$ gives the distance from the bottom of the stack of the targeted position.
For example, a link $\tuple{2,0}$ in an order-$2$ stack
$\sbrac{\stackw_1\ldots\stackw_\numof}{2}$
would point to the bottom of the order-$2$ stack.
That is, after $\stackw_\numof$.
Hence, we can represent collapse links informally with arrows as shown below.

\begin{example}
    An example order-$3$ stack is
    $\sbrac{
        \sbrac{\sbrac{\cha^{\tuple{3,1}} \chb^{\tuple{1,0}}}{1}}{2}
        \sbrac{
            \sbrac{\chc^{\tuple{2,1}}}{1}
            \sbrac{\chd^{\tuple{1,1}} \che^{\tuple{1,0}}}{1}
        }{2}
    }{3}$
    where the topmost character is $\cha$.
    This could be written
    {%
        \vspace{2ex}
        \[
            \sbrac{\quad
                \sbrac{\quad\sbrac{\quad\rnode{a}{\cha}\quad\rnode{b}{\chb}\ \ \pnode{d2}\ \ }{1}\quad}{2}
                \ \ \pnode{d1}\ \ %
                \sbrac{\quad%
                    \sbrac{\quad\rnode{c}{\chc}\quad}{1}%
                    \ \ \pnode{d3}\ \ %
                    \sbrac{\quad\rnode{d}{\chd}\ \ \pnode{d4}\ \ \rnode{e}{\che}\ \ \pnode{d5}\ \ }{1}\quad%
                }{2}\quad%
            }{3} \ .
        \]
        \psset{arcangleA=100,arcangleB=80,nodesep=1pt}%
        \ncarc[ncurv=.7]{->}{a}{d1}%
        \ncarc[ncurv=1.5]{->}{b}{d2}%
        \ncarc[ncurv=1.2]{->}{c}{d3}%
        \ncarc[ncurv=1.5]{->}{d}{d4}%
        \ncarc[ncurv=1.5]{->}{e}{d5}%
    }
\end{example}

The collapse operation, defined below, will remove all parts of the stack above the destination of the topmost collapse link.
Collapse on the stack in the example above gives
$\sbrac{
    \sbrac{
        \sbrac{\chc^{\tuple{2,1}}}{1}
        \sbrac{\chd^{\tuple{1,1}} \che^{\tuple{1,0}}}{1}
    }{2}
}{3}$.
Note, we will often omit the collapse link annotations for readability.
In particular, we will often write
$\apply{\ctop{1}}{\stackw} = \cha$
instead of
$\apply{\ctop{1}}{\stackw} = \annot{\cha}{\tuple{\opord, \idxi}}$
when we are not interested in the link.

Given an order-$\cpdsord$ stack
$\sbrac{\stackw_1\ldots\stackw_\numof}{\cpdsord}$, we define
\[
    \begin{array}{rcll}
        \apply{\ctop{\cpdsord}}{\sbrac{\stackw_1 \ldots
        \stackw_\numof}{\cpdsord}} &=& \stackw_1 & \text{when $\numof > 0$} \\ 
        \apply{\ctop{\opord}}{\sbrac{\stackw_1 \ldots \stackw_\numof}{\cpdsord}}
        &=& \apply{\ctop{\opord}}{\stackw_1} & \text{when $\opord < \cpdsord$ 
        and $\numof > 0$}
    \end{array}
\]
noting that $\apply{\ctop{\opord}}{\stackw}$ is undefined if
$\apply{\ctop{\opord'}}{\stackw}$ is empty for any $\opord' > \opord$.
For technical reasons, we also define
$\apply{\ctop{\cpdsord+1}}{\stackw} = \sbrac{\stackw}{\cpdsord+1}$
when $\stackw$ is an order-$\cpdsord$ stack.
We remove the top portion of a $\ctop{\opord}$ stack using, where $\idxi > 0$,
\[
    \begin{array}{rcll}%
        \apply{\cbottom{\cpdsord}{\idxi}}{\sbrac{\stackw_1 \ldots %
        \stackw_\numof}{\cpdsord}} &=& \sbrac{\stackw_{\numof - \idxi + 1} %
        \ldots \stackw_\numof}{\cpdsord} & \text{when $\idxi \leq \numof$} \\ %
        \apply{\cbottom{\opord}{\idxi}}{\sbrac{\stackw_1 \ldots %
        \stackw_\numof}{\cpdsord}} &=& %
        \sbrac{\apply{\cbottom{\opord}{\idxi}}{\stackw_1} \stackw_2 \ldots %
        \stackw_\numof}{\cpdsord} & \text{when $\opord < \cpdsord$ and $\numof > %
        0$} \ .%
    \end{array}%
\]

For
$\apply{\ctop{1}}{\stackw} = \annot{\cha}{\tuple{\opord, \idxi}}$,
the destination of the link is
$\apply{\cbottom{\opord}{\idxi}}{\stackw}$.

When $\stacku$ is an order-$(\opord-1)$ stack and $\stackv=\sbrac{\stackv_1\ldots\stackv_\numof}{\cpdsord}$ is an $\cpdsord$-stack with $\opord\leq \cpdsord$, we define $\ccompose{\stacku}{\opord}{\stackv}$ as the stack obtained by adding $\stacku$ on top of the topmost $\opord$-stack of $\stackv$. Formally, we let
\[
    \begin{array}{rcll}
        \ccompose{\stacku}{\opord}{\stackv} &=& \sbrac{\stacku\stackv_1\ldots\stackv_\numof}{\cpdsord} & \text{when $\opord= \cpdsord$} \\ 
        \ccompose{\stacku}{\opord}{\stackv} &=& \sbrac{(\ccompose{\stacku}{\opord}{\stackv_1})\stackv_2\ldots\stackv_\numof}{\cpdsord} & \text{when $\opord< \cpdsord$ and $\numof > 0$} \\ 
    \end{array}
\]

\lmcscorrection{
    \subsubsection{Operations on Order-$\cpdsord$ Collapsible Stacks}

    The following operations may be performed on an order-$\cpdsord$ collapsible
    stack.
    \[
        \begin{array}{rcl}%
            \cops{\cpdsord} &=& \set{\pop{1},\ldots,\pop{\cpdsord}} \cup %
                                \set{\push{2},\ldots,\push{\cpdsord}}\ \cup \\ %
                            & & \set{\collapse{2},\ldots,\collapse{\cpdsord}} \cup %
                                \setcomp{\cpush{\cha}{2},\ldots,\cpush{\cha}{\cpdsord},\rew{\cha}}{\cha \in \alphabet}%
        \end{array}%
    \]
    We say $\genop \in \cops{\cpdsord}$ is of order-$\opord$ when $\opord$ is
    minimal such that $\genop \in \cops{\opord}$.  E.g., $\push{\opord}$ is
    of order $\opord$.

    The $\collapse{\opord}$ operation is non-standard in the sense of
    Hague\etal~\cite{HMOS08,HMOS17} and has the semantics of a normal collapse, with the
    additional constraint that the top character has an order-$\opord$ link.  The
    standard version of collapse can be simulated with a non-deterministic choice on
    the order of the stack link.  In the other direction, we can store in the stack
    alphabet the order of the collapse link attached to each character on the stack.
    Note, we do not allow order-$1$ links to be created or used.
    In effect, these links are ``null''.

    We define each stack operation in turn for an order-$\cpdsord$ stack $\stackw$.
    Collapse links are created by the $\cpush{\cha}{\opord}$ operations, which add a
    character to the top of a given stack $\stackw$ with a link pointing to
    $\apply{\pop{\opord}}{\stackw}$.
    \begin{enumerate}
        \item We set $\apply{\pop{\opord}}{\stackw} = \stackv$ when $\stackw$
              decomposes into $\ccompose{\stacku}{\opord}{\stackv}$.

        \item We set $\apply{\push{\opord}}{\stackw} =
              \ccompose{\stacku}{\opord}{\ccompose{\stacku}{\opord}{\stackv}}$ when
              $\stackw = \ccompose{\stacku}{\opord}{\stackv}$.

        \item We set $\apply{\collapse{\opord}}{\stackw} =
              \apply{\cbottom{\opord}{\idxi}}{\stackw}$ when
              $\apply{\ctop{1}}{\stackw} = \cha^{\tuple{\opord,\idxi}}$ for some
              $\idxi$.

        \item We set $\apply{\cpush{\chb}{\opord}}{\stackw} =
              \ccompose{\chb^{\tuple{\opord,\numof-1}}}{1}{\stackw}$ where
              $\apply{\ctop{\opord+1}}{\stackw} =
              \sbrac{\stackw_1\ldots\stackw_\numof}{\opord+1}$.

        \item We set $\apply{\rew{\chb}}{\stackw} =
              \ccompose{\chb^{\tuple{\opord,\idxi}}}{1}{\stackv}$ where $\stackw =
              \ccompose{\cha^{\tuple{\opord,\idxi}}}{1}{\stackv}$.

    \end{enumerate}
    Note that, for a $\push{\opord}$ operation, links outside of $\stacku =
    \apply{\ctop{\opord}}{\stackw}$ point to the same destination in both copies of
    $\stacku$, while links pointing within $\stacku$ point within the respective
    copies of $\stacku$.  For full introduction, we refer the reader to
    Hague\etal~\cite{HMOS08,HMOS17}.  In Section~\ref{sec:examples} we give several example
    stacks and show how the stack operations affect them.

    \subsubsection{Collapsible Pushdown Systems}

    We define alternating collapsible pushdown systems.

    \begin{defi}[Collapsible Pushdown Systems]
        An alternating order-$\cpdsord$ \emph{collapsible pushdown system (collapsible PDS)} is
        a tuple $\cpds = \tuple{\controls, \alphabet, \cpdsrules}$ where $\controls$
        is a finite set of control states, $\alphabet$ is a finite stack alphabet,
        and $\cpdsrules \subseteq \brac{\controls \times \alphabet \times
        \cops{\cpdsord} \times \controls} \cup \brac{\controls \times 2^\controls}$
        is a set of rules.
    \end{defi}

    We write \emph{configurations} of a collapsible PDS as a pair $\config{\control}{\stackw}$
    where $\control \in \controls$ and $\stackw \in \stacksalpha{\cpdsord}{\alphabet}$.  We write
    $\config{\control}{\stackw} \cpdstran \config{\control'}{\stackw'}$ to denote a
    transition from a rule $\cpdsrule{\control}{\cha}{\genop}{\control'}$ with
    $\apply{\ctop{1}}{\stackw} = \cha$ and $\stackw' = \apply{\genop}{\stackw}$.
    Furthermore, we have a transition $\config{\control}{\stackw} \cpdstran
    \setcomp{\config{\control'}{\stackw}}{\control' \in \controlset}$ whenever we
    have a rule $\cpdsalttran{\control}{\controlset}$.  A non-alternating collapsible PDS has
    no rules of this second form.  We write $\configset$ to denote a set of
    configurations.

    We will be interested in the configurations that may reach a particular target set.
    That is, given a set $\configset_f$ of target configurations, we define the set
    $\prestar{\cpds}{\configset_f}$
    of configurations which can eventually reach $\configset_f$.
    A configuration can reach $\configset_f$ if it is contained in $\configset_f$ or there is a transition to a configuration that can reach $\configset_f$.
    In the case of alternating transitions to a set of configurations $\configset$, we require all configurations in $\configset$ to be able to reach $\configset_f$.
    This is formally defined as
    $\prestar{\cpds}{\configset_f}
     =
    \bigcup\limits_{\ordinal < \omega}
        \pre{\ordinal}{\cpds}{\saauta_0}$
    where
    \[
        \begin{array}{rcl}%
            \pre{0}{\cpds}{\configset_f} &=& \configset_f%
            \\%
            \pre{\ordinal+1}{\cpds}{\configset_f} &=&%
            \setcomp{\config{\control}{\stackw}}{
                \begin{array}{l}
                    \exists
                        \config{\control}{\stackw}
                        \cpdstran
                        \config{\control'}{\stackw'}
                        \in \pre{\ordinal}{\cpds}{\configset_f}
                    \ \lor \\ 
                    \exists
                        \config{\control}{\stackw}
                        \cpdstran
                        \configset
                        \subseteq
                        \pre{\ordinal}{\cpds}{\configset_f}
                \end{array}}
            \cup
            \pre{\ordinal}{\cpds}{\configset_f} \ .
         \end{array}
    \]
}{
}

\subsection{Regularity of Collapsible Stacks}
\label{sec:stackaut}

\lmcscorrection{
    We will present an algorithm that operates on sets of configurations.
}{
    We are interested in regular representations of sets of collapsible
    pushdown stacks.
}
For this we use order-$\cpdsord$ stack automata, thus defining a notion of
regular sets of stacks.  These have a nested structure based on a similar
automata model by Bouajjani and Meyer~\cite{BM04}.  The handling of collapse
links is similar to automata introduced by Broadbent\etal~\cite{BCOS10}, except
we read stacks top-down rather than bottom-up.
Note, the second condition in the definition below is a uniqueness condition that will be technically convenient throughout this article.
It can be shown that it does not restrict the expressive power of the automata.

\begin{defi}[Order-$\cpdsord$ Stack Automata]
    An \emph{order-$\cpdsord$ stack automaton}
    \[
        \saauta = \tuple{
                      \sastates_\cpdsord,\ldots,\sastates_1,
                      \alphabet,
                      \sadelta_\cpdsord,\ldots,\sadelta_1,
                      \safinals_\cpdsord,\ldots,\safinals_1
                  }
    \]
    is a tuple where
        $\alphabet$ is a finite stack alphabet,
        $\sastates_\cpdsord, \ldots, \sastates_1$ are finite disjoint statesets,
    and
    \begin{enumerate}
        \item
            for all $\opord \in \set{2,\ldots,\cpdsord}$, we have that
            $\sadelta_\opord \subseteq
                \sastates_\opord \times \sastates_{\opord-1} \times 2^{\sastates_\opord}$
            is a transition relation, and
            $\safinals_\opord \subseteq \sastates_\opord$ is a set of accepting states,

        \item
            for all
            $\opord \in \set{2,\ldots,\cpdsord}$
            and
            $\sastate_{\opord-1} \in \sastates_{\opord-1}$,
            if
            $\tuple{\sastate_\opord, \sastate_{\opord-1}, \sastateset_\opord} \in \sadelta_\opord$
            and
            $\tuple{\sastate'_\opord, \sastate_{\opord-1}, \sastateset'_\opord} \in \sadelta_\opord$
            then
            $\sastate_\opord = \sastate'_\opord$
            and
            $\sastateset_\opord = \sastateset'_\opord$, and

         \item
            $\sadelta_1 \subseteq
                \bigcup\limits_{2 \leq \opord \leq \cpdsord}\brac{
                    \sastates_1 \times
                    \alphabet \times
                    2^{\sastates_\opord} \times
                    2^{\sastates_1}
                }$
            is a transition relation, and
            $\safinals_1 \subseteq \sastates_1$
            a set of accepting states.
    \end{enumerate}
\end{defi}

Stack automata are alternating automata that read the stack in a nested fashion.
Order-$\opord$ stacks are recognised from states in $\sastates_\opord$.  A
transition $\tuple{\sastate, \sastate', \sastateset} \in \sadelta_\opord$ from
$\sastate$ to $\sastateset$ for some $\opord > 1$ can be fired when the
topmost order-$(\opord-1)$ stack is accepted from $\sastate' \in \sastates_{(\opord-1)}$.
The remainder of the stack must be accepted from all states in $\sastateset$.
At order-$1$, a transition $\tuple{\sastate, \cha, \sastateset_\branch,
\sastateset}$ is a standard alternating $\cha$-transition with the additional
requirement that the stack pointed to by the collapse link of $\cha$ is accepted
from all states in $\sastateset_\branch$.  A stack is accepted if a subset of
$\safinals_\opord$ is reached at the end of each order-$\opord$ stack.  In
Section~\ref{sec:formal-sa-run}, we formally define the runs of a stack automaton.  We write
$\stackw \in \slang{\sastate}{\saauta}$ whenever $\stackw$ is accepted from a
state $\sastate$. For ease of presentation, we write $\sastate \satran{\sastate'}\sastateset\in \sadelta_\opord$ instead of $\tuple{\sastate, \sastate', \sastateset} \in \sadelta_\opord$ and $\sastate \satrancol{\cha}{\sastateset_\branch} \sastateset
\in \sadelta_1$ instead of $\tuple{\sastate, \cha, \sastateset_\branch,
\sastateset}\in \sadelta_1$.
Note that a transition to the empty set is distinct from having no transition.

We give two informal examples of runs below.
The first is more schematic, while the second is concrete.
Further examples can be found in Section~\ref{sec:examples}.

\begin{figure}
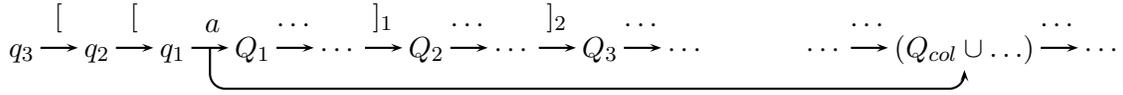

    \begin{center}
        \vspace{3ex}
        \begin{psmatrix}[nodealign=true,colsep=2ex,rowsep=1.25ex]
            \Rnode{N1}{$\sastate_3$} &             & \Rnode{N2}{$\sastate_2$}         &  &
            \Rnode{N3}{$\sastate_1$} & \pnode{N34} & \Rnode{N4}{$\sastateset_1$}      &  &
            \Rnode{N5}{$\cdots$}     &             & \Rnode{N6}{$\sastateset_2$}      &  &
            \Rnode{N7}{$\cdots$}     &             & \Rnode{N8}{$\sastateset_3$}      &  &
            \Rnode{N12}{$\cdots$}    &             &                                  &  &
            \Rnode{N9}{$\cdots$}     &             & \Rnode{N10}{$\brac{\sastateset_\branch \cup \ldots}$} &  &
            \Rnode{N11}{$\cdots$} \\

            \psset{nodesep=.5ex,angle=-90,linearc=.2}
            \ncline{->}{N1}{N2}^{$\sopen{}$}
            \ncline{->}{N2}{N3}^{$\sopen{}$}
            \ncline{->}{N3}{N4}^{$\cha$}
            \ncbar[arm=1.5ex,nodesepA=0]{->}{N34}{N10}
            \ncline{->}{N4}{N5}^{$\cdots$}
            \ncline{->}{N5}{N6}^{$\sclose{1}$}
            \ncline{->}{N6}{N7}^{$\cdots$}
            \ncline{->}{N7}{N8}^{$\sclose{2}$}
            \ncline{->}{N8}{N12}^{$\cdots$}
            \ncline{->}{N9}{N10}\Aput{$\cdots$}
            \ncline{->}{N10}{N11}\Aput{$\cdots$}
        \end{psmatrix}
    \end{center}
    \caption{\label{fig:schematic-run}An informal depiction of a partial stack automaton run.}
\end{figure}

\begin{example} \label{eg:schematic}
    A (partial) run is informally pictured in Figure~\ref{fig:schematic-run}, reading an order-$3$ stack using
    $\sastate_3 \satran{\sastate_2} \sastateset_3 \in \sadelta_3, \sastate_2
    \satran{\sastate_1} \sastateset_2 \in \sadelta_2$ and $\sastate_1
    \satrancol{\cha}{\sastateset_\branch} \sastateset_1 \in \sadelta_1$.
    Note, the transition
    $\sastate_3 \satran{\sastate_2} \sastateset_3$
    reads the topmost order-$2$ stack, with the remainder of the stack being read from $\sastateset_3$.
    The node labelled $\sastateset_\branch$ begins a run on the stack pointed to by the
    collapse link of $\cha$.  Note that the label of this node may contain other
    elements apart from $\sastateset_\branch$.  These additional elements come from
    the part of the run coming from the previous node (and other collapse links).
\end{example}

\begin{figure}
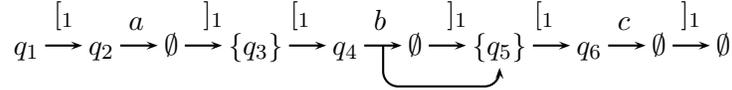

\begin{center}
    \vspace{3ex}
    \begin{psmatrix}[nodealign=true,colsep=2ex,rowsep=1.25ex]
        \Rnode{N10}{$\sastate_1$} &&
        \Rnode{N9}{$\sastate_2$} &&
        \Rnode{N8}{$\emptyset$} &&
        \Rnode{N7}{$\set{\sastate_3}$} &&
        \Rnode{N6}{$\sastate_4$} &\pnode{N65}&
        \Rnode{N5}{$\emptyset$} &&
        \Rnode{N4}{$\set{\sastate_5}$} &&
        \Rnode{N3}{$\sastate_6$} &&
        \Rnode{N2}{$\emptyset$} &&
        \Rnode{N1}{$\emptyset$} \\

        \psset{nodesep=.5ex,angle=-90,linearc=.2}
        \ncline{->}{N10}{N9}^{$\sopen{1}$}
        \ncline{->}{N9}{N8}^{$\cha$}
        \ncline{->}{N8}{N7}^{$\sclose{1}$}
        \ncline{->}{N7}{N6}^{$\sopen{1}$}
        \ncline{->}{N6}{N5}^{$\chb$}
        \ncbar[arm=1.5ex,nodesepA=0]{->}{N65}{N4}
        \ncline{->}{N5}{N4}^{$\sclose{1}$}
        \ncline{->}{N4}{N3}^{$\sopen{1}$}
        \ncline{->}{N3}{N2}^{$\chc$}
        \ncline{->}{N2}{N1}^{$\sclose{1}$}
    \end{psmatrix}
    \caption{\label{fig:example-run-for-stack}A complete example stack automaton run.}
\end{center}
\end{figure}

\begin{example}\label{eg:concrete}
    Figure~\ref{fig:example-run-for-stack} shows a stack automaton run over
    $\sbrac{\sbrac{\cha}{1}\sbrac{\annot{\chb}{\tuple{2,1}}}{1}\sbrac{\chc}{1}}{2}$
    using our informal graphical depiction.
    Note, we only show a collapse link on $\chb$, with the others omitted for readability.
    This run uses the transitions (in order from left-to-right)
    $\sat_1 = \brac{\sastate_1 \satran{\sastate_2} \set{\sastate_3}}$,
    $\sat'_1 = \brac{\sastate_2 \satrancol{\cha}{\emptyset} \emptyset}$,
    $\sat_2 = \brac{\sastate_3 \satran{\sastate_4} \set{\sastate_5}}$,
    $\sat'_2 = \brac{\sastate_4 \satrancol{\chb}{\set{\sastate_5}} \emptyset}$,
    $\sat_3 = \brac{\sastate_5 \satran{\sastate_6} \emptyset}$,
    and
    $\sat'_3 = \brac{\sastate_6 \satrancol{\chc}{\emptyset} \emptyset}$.
    Observe that the
    $\set{\sastate_5}$
    is the result of
    $\set{\sastate_5} \cup \set{\sastate_5}$.
    This is because both $\sat_2$ and $\sat'_2$ target $\set{\sastate_5}$, the latter via the collapse link.
\end{example}

\subsubsection{Notation and Conventions}

Before we define runs formally, we introduce the notions of a substack and initial states, and show how a run can be represented only by considering $\sadelta_1$ by defining the $\opord$-expansion.
Fix a stack automaton
\[
    \saauta = \tuple{
                  \sastates_\cpdsord,\ldots,\sastates_1,
                  \alphabet,
                  \sadelta_\cpdsord,\ldots,\sadelta_1,
                  \safinals_\cpdsord,\ldots,\safinals_1
              } \ .
\]

\paragraph{Substacks}

We define the set of substacks of a stack.
Intuitively, the set of substacks is the set of all suffixes of the stack.

\begin{defi}[$\substacks{\stackw}$]
    Given an order-$\cpdsord$ stack $\stackw$, we denote by
    $\substacks{\stackw}$
    the smallest set of stacks such that
    $\stackw \in \substacks{\stackw}$
    and if
    $\ccompose{\stacku}{\opord}{\stackv} \in \substacks{\stackw}$
    for some
    $\opord \in \set{1, \ldots, \cpdsord}$
    then
    $\stackv \in \substacks{\stackw}$.
\end{defi}

\paragraph{Initial States}

We say a state is \emph{initial} if it is a state
$\sastate_{\opord} \in \sastateset_{\opord}$ for $\opord < \cpdsord$ such that
there exists a transition $\sastate_{\opord+1} \satran{\sastate_{\opord}}
\sastateset_{\opord+1}$ in $\sadelta_{\opord+1}$.

\paragraph{Expanding Transitions}

Given a transition
$\sastate_1 \satrancol{\cha}{\sastateset_\branch} \sastateset_1$
we can define the \emph{$\opord$-expansion} of the transition.
This expansion may not always exist, and it is a consequence of the fact that a state may label at most one transition.
Moreover, it shows how a single transition in $\sadelta_1$ can represent uniquely a number of transitions.
The $\opord$-expansion of
$\sastate_1 \satrancol{\cha}{\sastateset_\branch} \sastateset_1$
is
\[
    \satranfull{\sastate_\opord}
               {\cha}
               {\sastateset_\branch}
               {\sastateset_1, \ldots, \sastateset_\opord}
\]
and it exists if $\sastate_1$ is initial and the $\opord$-expansion of $\sastate_1$ is
 $\satranfullk{\sastate_\opord}
              {\sastate_1}
              {\sastateset_2, \ldots, \sastateset_\opord}$.
The $\opord$-expansion of $\sastate_{\opord'}$ for $\opord' < \opord$ is
\[
    \satranfullk{\sastate_\opord}
                {\sastate_{\opord'}}
                {\sastateset_{\opord'+1}, \ldots, \sastateset_\opord}
\]
and it exists if $\sastate_{\opord'}$ is initial,
$\sastate_{\opord'+1}
 \satran{\sastate_{\opord'}}
 \sastateset_{\opord'+1} \in \sadelta_{\opord'+1}$
is the (unique) transition labelled by $\sastate_{\opord'}$, and, moreover, when $\opord' + 1 < \opord$, we have that
$\satranfullk{\sastate_\opord}
             {\sastate_{\opord'+1}}
             {\sastateset_{\opord'+2}, \ldots, \sastateset_\opord}$
is the $\opord$-expansion of $\sastate_{\opord'+1}$ (and it exists).

\begin{example}
    The $3$-expansion of
    $\sastate_1
     \satrancol{\cha}{\sastateset_\branch}
     \sastateset_1 \in \sadelta_1$
    from Example~\ref{eg:schematic} is
    $\satranfull{\sastate_3}
                {\cha}
                {\sastateset_\branch}
                {\sastateset_1, \sastateset_2, \sastateset_3}$.
    The $2$-expansion of
    $\sastate_2 \satrancol{\cha}{\emptyset} \emptyset$
    in Example~\ref{eg:concrete} is
    $\satranfull{\sastate_1}
                {\cha}
                {\emptyset}
                {\emptyset, \set{\sastate_3}}$.
\end{example}

For technical convenience we allow for all
$\sastate \in \sastates_\cpdsord$
the $\cpdsord$-expansion
\[
    \satranfullk{\sastate}{\sastate}{} \ .
\]
This allows us to refer to the $\cpdsord$-expansion of an order-$\cpdsord$ state using the same notation as an order-$\opord$ state with $\opord < \cpdsord$.
That is using,
$\satranfullk{\sastate_\cpdsord}
             {\sastate_{\opord}}
             {\sastateset_{\opord+1}, \ldots, \sastateset_\cpdsord}$
where $\opord$ may equal $\cpdsord$.

We also define the $\opord$-expansion of a set of order-$1$ transitions.
In fact, we define a refined notion that is
$(\opord, \sastateset)$-expansion for
$\sastateset \subseteq \sadelta_\opord$,
as it will become useful in our proofs.
Intuitively, it is the $\opord$-expansion limited to transitions whose expansions begin with states in $\sastateset$.
Given
$\satset \subseteq \sadelta_1$.
for each
$\sat \in \satset$
suppose we have the $\opord$-expansion
$\satranfull{\sastate^\sat_\opord}
            {\cha}
            {\sastateset^\sat_\branch}
            {\sastateset^\sat_1, \ldots, \sastateset^\sat_\opord}$.
Take
$\satset' = \setcomp{\sat}
                    {\sat \in \satset \land
                     \sastate^\sat_\opord \in \sastateset}$.
The $(\opord, \sastateset)$-expansion of $\satset$ is
\[
    \satranfull{\sastateset}
               {\cha}
               {\sastateset_\branch}
               {\sastateset_1, \ldots, \sastateset_\opord}
\]
where
$\sastateset_\branch =
 \bigcup\limits_{\sat \in \satset'}
     \sastateset^\sat_\branch$
and for each
$\opord' \in \set{1, \ldots, \opord}$
we have
$\sastateset_{\opord'} =
 \bigcup\limits_{\sat \in \satset'}
    \sastateset^\sat_{\opord'}$.
The expansion is \emph{strict} whenever $\satset = \satset'$.

We will use symbols $\satfull, \satfull'$ \&c. to refer to expansions.
Finally, we also define a convenience function for extracting the (unique) order-$1$ transition from an expansion.
Let
$\satfull = \satranfull{\sastate_\opord}
                       {\cha}
                       {\sastateset_\branch}
                       {\sastateset_1, \ldots, \sastateset_\opord}$
then
\[
    \tunexpand{\satfull} = \sat
\]
where $\sat$ is the (unique) transition in $\sadelta_1$ with $\opord$-expansion $\satfull$.

\subsubsection{Formal Definition of a Run}
\label{sec:formal-sa-run}

A stack automaton is essentially a stack- and collapse-aware
alternating automaton, where collapse links are treated as special cases of the
alternation.
Again, let
\[
    \saauta = \tuple{
                  \sastates_\cpdsord,\ldots,\sastates_1,
                  \alphabet,
                  \sadelta_\cpdsord,\ldots,\sadelta_1,
                  \safinals_\cpdsord,\ldots,\safinals_1
              } \ .
\]

We represent a run over a stack as another stack of (sets of) transitions of
$\saauta$.
This representation simplifies the kinds of run manipulations we will need later in the article.
Formally, then, a run over an order-$\opord$ {stack} {$\stackw$} is an order-$\opord$ stack $\stackv$ over the alphabet
$\alphabet \times 2^{\sadelta_1}$
such that when projecting on the $\Sigma$-component we {retrieve} the stack {$\stackw$}.
More formally, we require
$\saprojection{\stackv} = \stackw$
where
\[
    \begin{array}{rcl}
        \saprojection{\annot{\tuple{\cha, \satset}}{\tuple{\opord, \idxi}}}
        &=&
        \annot{\cha}{\tuple{\opord, \idxi}}, \text{\ and}
        \\
        \saprojection{\ccompose{\stackw}{\opord}{\stackw'}}
        &=&
        \ccompose{\saprojection{\stackw}}{\opord}{\saprojection{\stackw'}} \ .
    \end{array}
\]

In addition, the run stack must satisfy certain conditions that we will explain below.
First we will give a formal description of the run in Example~\ref{eg:concrete} shown in Figure~\ref{fig:example-run-for-stack}.

\begin{example}
    Recall from Example~\ref{eg:concrete} that
        $\sat_1 = \brac{\sastate_1 \satran{\sastate_2} \set{\sastate_3}}$,
        $\sat'_1 = \brac{\sastate_2 \satrancol{\cha}{\emptyset} \emptyset}$,
        $\sat_2 = \brac{\sastate_3 \satran{\sastate_4} \set{\sastate_5}}$,
        $\sat'_2 = \brac{\sastate_4 \satrancol{\chb}{\set{\sastate_5}} \emptyset}$,
        $\sat_3 = \brac{\sastate_5 \satran{\sastate_6} \emptyset}$,
        and
        $\sat'_3 = \brac{\sastate_6 \satrancol{\chc}{\emptyset} \emptyset}$.

    Intuitively, a run over a stack is obtained by adding to each character in the stack the set of order-$1$ transitions that read it.
    We can represent this run as the stack
    \[
        \sbrac{\sbrac{\tuple{\cha, \set{\sat'_1}}}{1}
               \sbrac{\tuple{\chb, \set{\sat'_2}}}{1}
               \sbrac{\tuple{\chc, \set{\sat'_3}}}{1}}{2} \ .
    \]
    Note, the $2$-expansions of $\sat'_1$, $\sat'_2$, and $\sat'_3$ uniquely determine $\sat_1$, $\sat_2$, and $\sat_3$ respectively.
    Thus, we do not need to explicitly store these transitions in our stack representation of runs.
\end{example}

The definition of an accepting run requires two consistency notions, detailed below.

Let $\stackr$ be an order-$\opord$ run of $\saauta$. For a set $\sastateset \subseteq \sastates_{\opord}$ of
order-$\opord$ states, we say that \emph{$\stackr$ is $Q$-valid} if the following holds.
If the run $\stackr$ is an empty stack, then, $Q$ must be a subset of $\safinals_{\opord}$.
I.e.\ the end of a run is accepting.
Assume now that $\stackr$ is not empty.
If $\opord=1$ and $\stackr=\ccompose{(\cha,T)}{1}{\stackr'}$, then for all $\sastate \in \sastateset$,
there must exist a transition in $T$ of the form
$\sastate \satrancol{\cha}{\sastateset_\branch}\sastateset'$
such that
$\stackr'$ is $\sastateset'$-valid.
If $\opord>1$ and $\stackr=\ccompose{\stacku}{\opord}{\stackr'}$ then for all $\sastate \in \sastateset$,
there must exist a transition $\sastate \satran{\sastate_{\sastateset'}} {\sastateset'} \in \sadelta_{\opord}$ such that $\stacku$ is  $\{\sastate_{\sastateset'}\}$-valid and $\stackr'$ is $\sastateset'$-valid.

Note that $Q$-validity does not check the constraint imposed by the $\sastateset_{\branch}$ {component} appearing in order-1 transitions. This is done by
\emph{link-validity} which is only meaningful on order-$\cpdsord$ runs:
An order-$\cpdsord$ run $\stackr$ is \emph{link-valid} if for every substack
$\stackr' \in \substacks{\stackr}$ of the form $\stackr'= (a,T)^{(\opord,i)} : \stackr''$ and for every transition ${\sastate} \satrancol{\cha}{\sastateset_\branch} \sastateset$ in $T$ we have
$\ctop{\opord + 1}({\collapse{k}(\stackr')})$ is $\sastateset_{\branch}$-valid.

For $\sastate \in \sastates_{\cpdsord}$, an order-$\cpdsord$ run $\stackr$ is \emph{$\sastate$-accepting} if it is both $\{\sastate\}$-valid and link-valid.
In addition, we require that if $w$ is non-empty, and hence of the form
$\ccompose{(a,T)}{1}{\stackr'}$, then $T$ is a singleton $\{t\}$
and we refer to $t$ as the head transition of the run.

\begin{defi}[$\slang{\sastate}{\saauta}$ and $\slang{\sastateset}{\saauta}$]
    Given a stack automaton $\saauta$ we define
    $\slang{\sastate}{\saauta}$
    to be the set of all order-$\cpdsord$ stacks $\stackw$ which have a $\sastate$-accepting run of $\saauta$.
    Moreover, we define
    $\slang{\sastateset}{\saauta}$
    to be the set of order-$\cpdsord$ stacks $\stackw$ such that for all
    $\sastate \in \sastateset$
    we have
    $\stackw \in \slang{\sastate}{\saauta}$.
    Note, if
    $\sastateset = \emptyset$
    then all stacks are accepted
\end{defi}

\begin{example}
    In our example run,
    $\sbrac{\sbrac{\tuple{\cha, \set{\sat'_1}}}{1}
            \sbrac{\tuple{\chb, \set{\sat'_2}}}{1}
            \sbrac{\tuple{\chc, \set{\sat'_3}}}{1}}{2}$,
    the head transition is $\sat'_1$.
    It is immediate that the empty stacks
    $\sbrac{}{2}$ and $\sbrac{}{1}$
    are $\emptyset$-valid.
    Thus
    $\sbrac{\tuple{\chc, \set{\sat'_3}}}{1}$
    is $\set{\sastate_6}$-valid as witnessed by $\sat'_3$.
    From this we can deduce
    $\sbrac{\sbrac{\tuple{\chc, \set{\sat'_3}}}{1}}{2}$
    is $\set{\sastate_5}$-valid via $\sat_3$.
    Continuing in this way we can determine
    $\sbrac{\sbrac{\tuple{\cha, \set{\sat'_1}}}{1}
            \sbrac{\tuple{\chb, \set{\sat'_2}}}{1}
            \sbrac{\tuple{\chc, \set{\sat'_3}}}{1}}{2}$
    is $\set{\sastate_1}$-valid.
    We can also deduce link-validity since only $\sat'_2$ has a non-empty condition on the link.
    This condition is $\set{\sastate_5}$ and this link targets
    $\sbrac{\sbrac{\tuple{\chc, \set{\sat'_3}}}{1}}{2}$
    which we already know is $\set{\sastate_5}$-valid.
\end{example}

\lmcscorrection{
    \subsubsection{Properties of Stack Automata}

    We show here that stack automata membership is polynomial time.
    Several further results can also be shown~\cite{BCHS12}: the sets of stacks accepted
    by these automata form an effective Boolean algebra (note that complementation
    causes a blow-up in the size of the automaton); and they accept the same family
    of collapsible stacks as the automata used by Broadbent\etal~\cite{BCOS10}.  We
    omit these here for space reasons.

    We also report that our PSPACE emptiness algorithm for stack automata~\cite{BCHS12} is not correct\footnote{%
        We thank an anonymous reviewer for pointing this out.
    }.
    Indeed, we will show in a forthcoming article that, given a stack automaton $\saauta$, deciding whether there exists a collapsible pushdown stack accepted by $\saauta$, is NEXPTIME-complete.
    We again omit this proof for space reasons.
    In the sequel, we will primarily be interested in membership rather than emptiness.

    \begin{prop}[Stack Automata Membership]
        Membership of order-$\cpdsord$ stack automata can be tested in polynomial time
        in the size of the input stack and stack automaton.
    \end{prop}
    \proof
        Take a stack $\stackw$ and let
        \[
        \saauta = \tuple{
                      \sastates_\cpdsord,\ldots,\sastates_1,
                      \alphabet,
                      \sadelta_\cpdsord,\ldots,\sadelta_1,
                      \safinals_\cpdsord, \ldots,\safinals_1
                  } \ .
        \]
        The membership algorithm iterates from the bottom (end) of the stack to the top (beginning).
        We start at the bottom of the order-$\cpdsord$ stack with
        $\sbrac{}{\cpdsord}$
        and observe that this stack is trivially $\safinals_\cpdsord$-valid and link-valid.

        Now suppose, in previous iterations, we have constructed a link-valid run and decomposition
        $\stackv =
         \ccompose{\stackv_\opord}
                  {(\opord+1)}
                  {\ccompose{\cdots}{\cpdsord}{\stackv_\cpdsord}}$
        such that for each
        $\opord' \in \set{\opord, \ldots, \cpdsord}$
        we have
        $\stackv_{\opord'}$
        is
        $\sastateset_{\opord'}$-valid
        for some
        $\sastateset_{\opord'}$.
        Note, in all cases, we consider the largest sets for which validity holds.
        There are now several cases.
        \begin{itemize}
        \item
            When $\opord = 1$ we have
            $\stackv =
             \ccompose{\stackv_1}{2}{
                 \ccompose{\cdots}{\cpdsord}{
                     \stackv_\cpdsord
                 }
             }$
            valid for
            $\sastateset_1, \ldots, \sastateset_\cpdsord$
            respectively and there are two cases.
            \begin{itemize}
            \item
                Suppose, for some
                $\annot{\cha}{\tuple{\opord', \idxi}}$
                we have
                $\ccompose{\annot{\cha}{\tuple{\opord', \idxi}}}
                          {1}
                          {\saprojection{\stackv}}
                 \in
                 \substacks{\stackw}$.
                Moreover, suppose in a previous iteration we showed
                $\apply{\collapse{\opord'}}
                       {\ccompose{\annot{\cha}{\tuple{\opord', \idxi}}}
                                 {1}
                                 {\saprojection{\stackv}}}$
                is $\sastateset_\branch$-valid.
                Let
                $\satset =
                 \setcomp{\sastate
                          \satrancol{\cha}{\sastateset'_\branch}
                          \sastateset'_1 \in \sadelta_1}
                         {\sastateset'_1 \subseteq \sastateset_1
                          \land
                          \sastateset'_\branch \subseteq \sastateset_\branch}$.
                We construct
                $\stackv' =
                 \ccompose{\stackv'_1}
                          {2}
                          {\ccompose{\stackv_2}
                                    {3}
                                    {\ccompose{\cdots}{\cpdsord}{\stackv_\cpdsord}}}$
                where
                $\stackv'_1 =
                 \brac{\ccompose{\annot{\tuple{\cha, \satset}}
                                       {\tuple{\opord', \idxi}}}
                                {1}
                                {\stackv_1}}$
                and observe $\stackv'_1$ is $\sastateset'_1$-valid, where
                $\sastateset'_1 =
                 \setcomp{\sastate}
                         {\sastate
                          \satrancol{\cha}{\sastateset'}
                          \sastateset \in \satset}$.
                We then continue to the next iteration with
                $\ccompose{\stackv'_1}
                          {2}
                          {\ccompose{\stackv_2}
                                    {3}
                                    {\ccompose{\cdots}{\cpdsord}{\stackv_\cpdsord}}}$.
                By construction, $\stackv'$ is also link-valid.

            \item
                When the previous case does not apply, then we can adjust the decomposition as follows.
                We observe that
                $\ccompose{\stackv_1}{2}{\stackv_2}$
                is $\sastateset'_2$-valid for
                $\sastateset'_2 =
                 \setcomp{\sastate}
                         {\sastate
                            \satran{\sastate'}
                            \sastateset \in \sadelta_2
                          \land
                          \sastate' \in \sastateset_1
                          \land
                          \sastateset \subseteq \sastateset_2}$.
                We thus set
                $\stackv'_2 = \ccompose{\stackv_1}{2}{\stackv_2}$
                and continue the iteration with the decomposition
                $\ccompose{\stackv'_2}
                          {3}
                          {\ccompose{\stackv_3}
                                    {4}
                                    {\ccompose{\cdots}
                                              {\cpdsord}
                                              {\stackv_\cpdsord}}}$.
            \end{itemize}

        \item
            When $\opord > 1$ there are again two cases.
            \begin{itemize}
            \item
                If
                $\ccompose{\sbrac{}{\opord-1}}
                          {\opord}
                          {\saprojection{\stackv}}
                 \in \substacks{\stackw}$
                then we observe
                $\sbrac{}{\opord-1}$
                is trivially $\safinals_{\opord-1}$-valid and continue the iteration with the decomposition
                $\ccompose{\sbrac{}{\opord-1}}
                          {\opord}
                          {\ccompose{\stackv_\opord}
                                    {(\opord+1)}
                                    {\ccompose{\cdots}
                                              {\cpdsord}
                                              {\stackv_\cpdsord}}}$,
                which immediately remains link-valid.
            \item
                When the previous case does not apply and
                $\opord < \cpdsord$,
                then we can adjust the decomposition as follows.
                We observe that
                $\ccompose{\stackv_\opord}{(\opord+1)}{\stackv_{\opord+1}}$
                is $\sastateset'_{\opord+1}$-valid for
                $\sastateset'_{\opord+1} =
                 \setcomp{\sastate}
                         {\sastate
                            \satran{\sastate'}
                            \sastateset \in \sadelta_{\opord+1}
                          \land
                          \sastate' \in \sastateset_\opord
                          \land
                          \sastateset \subseteq \sastateset_{\opord+1}}$.
                We thus set
                $\stackv'_{\opord+1} =
                 \ccompose{\stackv_\opord}
                          {(\opord+1)}
                          {\stackv_{\opord+1}}$
                and continue the iteration with the decomposition
                $\ccompose{\stackv'_{\opord+1}}
                          {(\opord+2)}
                          {\ccompose{\stackv_{\opord+2}}
                                    {(\opord+3)}
                                    {\ccompose{\cdots}
                                              {\cpdsord}
                                              {\stackv_\cpdsord}}}$.

            \item
                If neither of the previous cases apply, then
                $\opord = \cpdsord$
                and we have constructed a run over $\stackw$.
                Let
                $\apply{\ctop{1}}{\stackv} = \tuple{\cha, \satset}$.
                For each
                $\sastate \satran{\sastate'} \sastateset
                 \in \satset$
                the run
                $\apply{\rew{\tuple{\cha, \set{\sat}}}}
                       {\stackv}$
                is an accepting run witnessing
                $\stackw \in \slang{\sastate}{\saauta}$.
            \end{itemize}
        \end{itemize}

        It remains to show that if
        $\stackw \in \slang{\sastate}{\saauta}$
        then the above algorithm always constructs a run that witnesses the fact.
        Observe that
        $\sastateset = \safinals_\opord$
        is the largest set for which
        $\sbrac{}{\opord}$
        is
        $\sastateset$-valid.
        Then, at each step of the above algorithm, we compute the largest
        $\sastateset$
        and set of transitions
        $\satset$
        for which the validity conditions can be maintained.
        Thus, all accepting runs are necessarily contained within the constructed run in the following sense.
        Let $\stackv$ be the result of the above algorithm.
        Moreover, let $\stacku$ be an accepting run witnessing
        $\stackw \in \slang{\sastate}{\saauta}$.
        Then the relationship
        $\saruncontained{\stacku}{\stackv}$
        holds where
        \[
            \begin{array}{rcl}
                \saruncontained{\annot{\tuple{\cha, \satset}}
                                      {\tuple{\opord, \idxi}}}
                               {\annot{\tuple{\cha, \satset'}}
                                      {\tuple{\opord, \idxi}}}
                &\iff&
                \satset \subseteq \satset'
                \\
                \saruncontained{\ccompose{\stacku'}{\opord}{\stacku''}}
                               {\ccompose{\stackv'}{\opord}{\stackv''}}
                &\iff&
                \saruncontained{\stacku'}{\stacku''}
                \land
                \saruncontained{\stackv'}{\stackv''} \ .
            \end{array}
        \]
        Thus, if a run $\stackv$ exists witnessing
        $\stackw \in \slang{\sastate}{\saauta}$
        then the above algorithm will find a witnessing run.
    \qed
}{
}


%% file: saturation.tex
\section{Saturation Algorithm}
\label{sec:saturation-alg}

Given a CPDS $\cpds$ and a stack automaton $\saauta_0$ with a state
$\sastate_\control \in \sastates_\cpdsord$ for each control state $\control$ in
$\cpds$, let
$\configset_f =
 \setcomp{\config{\control}{\stackw}}
         {\stackw \in \slang{\sastate_\control}{\saauta_0}}$.
We will write
$\prestar{\cpds}{\saauta_0}$
to denote
$\prestar{\cpds}{\configset_f}$.
We build a stack
automaton recognising $\prestar{\cpds}{\saauta_0}$.  We begin with $\saauta_0$
and iterate a saturation function denoted $\satstep$ --- which adds new
transitions to $\saauta_0$ --- until a `fixed point' has been reached.  That is,
we iterate $\saauta_{\idxi+1} = \apply{\satstep}{\saauta_\idxi}$ until
$\saauta_{\idxi+1} = \saauta_\idxi$.  As the number of states is bounded, we
eventually obtain this,  giving us the following theorem.

\begin{thm}
    Given an alternating CPDS $\cpds$ and a stack automaton $\saauta_0$, we can
    construct a stack automaton $\saauta$ accepting $\prestar{\cpds}{\saauta_0}$.
    That is
    $\config{\control}{\stackw} \in \prestar{\cpds}{\saauta_0}$
    iff
    $\stackw \in \slang{\sastate_\control}{\saauta}$.
    \qed
\end{thm}

The construction runs in $\cpdsord$-EXPTIME for alternating CPDS --- which is
optimal --- and can be improved to $(\cpdsord - 1)$-EXPTIME for non-alternating
CPDS when the initial automaton satisfies a certain notion of
\emph{non-alternation}, again optimal.  Correctness and complexity are discussed
in subsequent sections.

\subsection{Notation and Conventions}
\label{ssec:notations}

\paragraph{Initial States}
We slightly expand the definition of initial states.
That is, a state is \emph{initial} if it is of the form $\sastate_\control \in
\sastateset_{\cpdsord}$ for some control state $\control$ or if it is a state
$\sastate_{\opord} \in \sastateset_{\opord}$ for $\opord < \cpdsord$ such that
there exists a transition $\sastate_{\opord+1} \satran{\sastate_{\opord}}
\sastateset_{\opord+1}$ in $\sadelta_{\opord+1}$.  We make the assumption that
all initial states do not have any incoming transitions and that they are not
final\footnote{Hence automata cannot accept empty stacks from initial states.
This can be overcome by introducing a bottom-of-stack symbol.}.

\paragraph{Adding Transitions}
In the algorithm we will say we add transitions
$\satranfull{\sastate_\cpdsord}{\cha}{\sastateset_\branch}{\sastateset_1,
\ldots, \sastateset_\cpdsord}$ to the automaton.
By this, we mean we iterate from $\opord = \cpdsord$ down to $\opord = 2$ and add
$\sastate_\opord
 \satran{\sastate_{\opord-1}}
 \sastateset_\opord$
to $\sadelta_\opord$ if a transition between $\sastate_\opord$ and $\sastateset_\opord$ does not already exist, otherwise we use the existing transition and state $\sastate_{\opord-1}$.
Then, we add
$\sastate_1
 \satrancol{\cha}{\sastateset_\branch}
 \sastateset_1$
to $\sadelta_1$.

\paragraph{Justified Transitions}

When we add transitions via the saturation function we also add
\emph{justifications} to the new transitions that are not derived from
alternating transitions.  These justifications indicate the provenance of each
new transition.  This later permits counter example generation for CPDSs, as shown in
Section~\ref{sec:counter-egs}.

To each
$\sat = \brac{\sastate_1 \satrancol{\cha}{\sastateset_\branch} \sastateset_1}$
we will define the justification
$\apply{\tjust}{\sat}$ to be either $\unjust$ (indicating the transition is in
$\saauta_0$), a pair $\tuple{\cpdsruler,\idxi}$, a tuple
$\tuple{\cpdsruler,\sat',\idxi}$, $\tuple{\cpdsruler, \satset, \idxi}$ or a
tuple $\tuple{\cpdsruler, \sat', \satset, \idxi}$ where $\cpdsruler$ is a rule
of the CPDS, $\idxi$ is the number of iterations saturation
required to introduce the transition, $\sat'$ is an order-$1$ transition
(in $\sadelta_1$)
and $\satset$ is a set of such transitions.

\subsection{The Saturation Function}

We are now ready to give the saturation function $\satstep$ for a given $\cpds
= \tuple{\controls, \alphabet, \cpdsrules}$.  As described above, we apply this
function to $\saauta_0$ until a fixed point is reached.  First set
$\apply{\tjust}{\sat} = \unjust$ for all order-$1$ transitions of $\saauta_0$.  The
intuition behind the saturation rules can be quickly understood via a rewrite
rule $\cpdsrule{\control}{\cha}{\rew{b}}{\control'}$ which leads to the addition
of
$\satranfull{\sastate_{\control}}{\cha}{\sastateset_\branch}{\sastateset_1,
\dots, \sastateset_\cpdsord}$ whenever there already existed a transition with $\cpdsord$-expansion
$\satranfull{\sastate_{\control'}}{\chb}{\sastateset_\branch}{\sastateset_1,
\dots, \sastateset_\cpdsord}$.  Because the rewrite can change the control state
from $\control$ to $\control'$ and the top character from $\cha$ to $\chb$, we
must have an accepting run from $\sastate_\control$ with $\cha$ on top whenever
we had an accepting run from $\sastate_{\control'}$ with $\chb$ on top.  We
give examples and intuition of the more complex steps in
Section~\ref{sec:examples}, which may be read alongside the definition below.

\begin{defi}[The Saturation Function $\satstep$]
    Given an order-$\cpdsord$ stack automaton $\saauta_\idxi$ we define
    $\saauta_{\idxi+1} = \apply{\satstep}{\saauta_\idxi}$. The state-sets of
    $\saauta_{\idxi+1}$ are defined implicitly by the transitions which are
    those in $\saauta_\idxi$ plus, for each $\cpdsruler =
    \cpdsrule{\control}{\cha}{\genop}{\control'} \in \cpdsrules$,
    \begin{enumerate}
    \item
        when $\genop = \pop{\opord}$, for each order-$k$ state $\sastate_\opord$ with $\cpdsord$-expansion
        $\satranfullk{\sastate_{\control'}}{\sastate_\opord}{\sastateset_{\opord+1},
        \dots, \sastateset_\cpdsord}$ in $\saauta_\idxi$, add
        \[
            \satfull = \brac{
                \satranfull{\sastate_{\control}}{\cha}{\emptyset}{\emptyset,
                \ldots, \emptyset, \set{\sastate_\opord}, \sastateset_{\opord+1},
                \ldots, \sastateset_\cpdsord}
            }
        \]
        to $\saauta_{\idxi+1}$ and set $\apply{\tjust}{\tunexpand{\satfull}} =
        \tuple{\cpdsruler, \idxi+1}$ whenever $\tunexpand{\satfull}$ is not already in
        $\saauta_{\idxi+1}$,

    \item
        when $\genop = \push{\opord}$, for each order-$1$ transition $\sat$ with $\cpdsord$-expansion
        $\satranfull{\sastate_{\control'}}
                    {\cha}
                    {\sastateset_\branch}
                    {\sastateset_1, \ldots,\sastateset_\opord,\ldots, \sastateset_\cpdsord}$
        and set of order-$1$ transitions $\satset$ with strict $(\opord, \sastateset_\opord)$-expansion
        $\satranfull{\sastateset_\opord}
                    {\cha}
                    {\sastateset'_\branch}
                    {\sastateset'_1, \ldots, \sastateset'_\opord}$ in $\saauta_\idxi$,
        add to $\saauta_{\idxi+1}$ the transitions
        \[%
            \satfull' = \brac{%
                \satranfull{\sastate_{\control}}{\cha}{\sastateset_\branch%
                \cup \sastateset'_\branch}{%
                    \begin{array}{c}%
                        \sastateset_1 \cup \sastateset'_1, \ldots,%
                        \sastateset_{\opord-1} \cup \sastateset'_{\opord-1},\\ %
                        \sastateset'_\opord,\\ %
                        \sastateset_{\opord+1}, \ldots,%
                        \sastateset_\cpdsord%
                    \end{array}%
                }%
            }%
        \]%
        and set
        $\apply{\tjust}{\tunexpand{\satfull'}} =
         \tuple{\cpdsruler, \sat, \satset, \idxi+1}$
         if $\tunexpand{\satfull'}$ is not already in $\saauta_{\idxi+1}$,

    \item
        when $\genop = \collapse{\opord}$ for each order-$\opord$ state with $\cpdsord$-expansion
        $\satranfullk{\sastate_{\control'}}{\sastate_\opord}{\sastateset_{\opord+1},
        \dots, \sastateset_\cpdsord}$ in $\saauta_\idxi$, add to
        $\saauta_{\idxi+1}$ the transitions
        $\satfull = \brac{
            \satranfull{\sastate_{\control}}
                       {\cha}
                       {\set{\sastate_\opord}}
                       {\emptyset, \ldots, \emptyset,
                        \sastateset_{\opord+1}, \ldots, \sastateset_\cpdsord}
        }$
        if $\tunexpand{\satfull}$ does not already exist.  In all
        cases, if $\tunexpand{\satfull}$ is added, set
        $\apply{\tjust}{\tunexpand{\satfull}} = \tuple{\cpdsruler, \idxi+1}$,

    \item
        when $\genop = \cpush{\chb}{\opord}$ for all order-$1$ transitions $\sat$ with $\cpdsord$-expansion
        $\satfull = \brac{
            \satranfull{\sastate_{\control'}}
                       {\chb}
                       {\sastateset_\branch}
                       {\sastateset_1, \ldots, \sastateset_\cpdsord}
        }$
        and set of order-$1$ transitions $\satset$ with strict $(1, \sastateset_1)$-expansion
        $\satranfull{\sastateset_1}{\cha}{\sastateset'_\branch}{\sastateset'_1}$
        in $\saauta_\idxi$ with $\sastateset_\branch \subseteq
        \sastates_\opord$, add to $\saauta_{\idxi+1}$ the transitions
        \[
            \satfull' = \brac{
                \satranfull{\sastate_{\control}}{\cha}{\sastateset'_\branch}{\sastateset'_1,
                \sastateset_2, \ldots, \sastateset_\opord \cup
                \sastateset_\branch, \ldots, \sastateset_\cpdsord}
            } \ ,
        \]
        and set
        $\apply{\tjust}{\tunexpand{\satfull'}} =
        \tuple{\cpdsruler, \sat, \satset, \idxi+1}$
        if $\tunexpand{\satfull'}$ is not already in $\saauta_{\idxi+1}$,

    \item
        when $\genop = \rew{\chb}$ for each transition $\sat$ with $\cpdsord$-expansion
        $\satranfull{\sastate_{\control'}}
                    {\chb}
                    {\sastateset_\branch}
                    {\sastateset_1, \dots, \sastateset_\cpdsord}$
        in $\saauta_\idxi$, add to $\saauta_{\idxi+1}$ the transitions
        $\satfull = \brac{
            \satranfull{\sastate_{\control}}
                       {\cha}
                       {\sastateset_\branch}
                       {\sastateset_1, \dots, \sastateset_\cpdsord}}$,
        setting
        $\apply{\tjust}{\tunexpand{\satfull}} = (\cpdsruler, \sat, \idxi)$
        when $\tunexpand{\satfull}$ is not already in $\saauta_{\idxi+1}$.
    \end{enumerate}
    Finally, for every rule
    $\cpdsalttran{\control}{\controlset}$,
    let
    $\sastateset = \setcomp{\sastate_{\control'}}{\control' \in \controlset}$,
    then, for each set of order-$1$ transitions $\satset$ with strict $(\cpdsord, \sastateset)$-expansion
    $\satranfull{\sastateset}
                {\cha}
                {\sastateset_\branch}
                {\sastateset_1, \ldots, \sastateset_\cpdsord}$,
    add the transitions
    $\satfull = \satranfull{\sastate_\control}
                           {\cha}
                           {\sastateset_\branch}
                           {\sastateset_1, \ldots, \sastateset_\cpdsord}$
    and set
    $\apply{\tjust}{\tunexpand{\satfull}} =
     \tuple{\cpdsruler, \satset, \idxi+1}$
    if $\tunexpand{\satfull}$ is not already in $\saauta_{\idxi+1}$.
\end{defi}

From $\saauta_0$, we iterate $\saauta_{\idxi+1} =
\apply{\satstep}{\saauta_\idxi}$ until $\saauta_{\idxi+1} = \saauta_\idxi$.
Generally, as we show in Proposition~\ref{prop:sat-complexity}, we terminate in $\cpdsord$-EXPTIME.  When the CPDS does not use alternating transitions and $\saauta_0$ satisfies a
``non-alternating'' property (e.g. when we are only interested in reaching a
designated control state), we can restrict $\satstep$ to only add transitions
where $\sastateset_\cpdsord$ has at most one element, giving
$(\cpdsord-1)$-EXPTIME complexity.  In all cases saturation is linear in the
size of $\alphabet$.

\subsection{Examples of Saturation}
\label{sec:examples}

\begin{figure*}
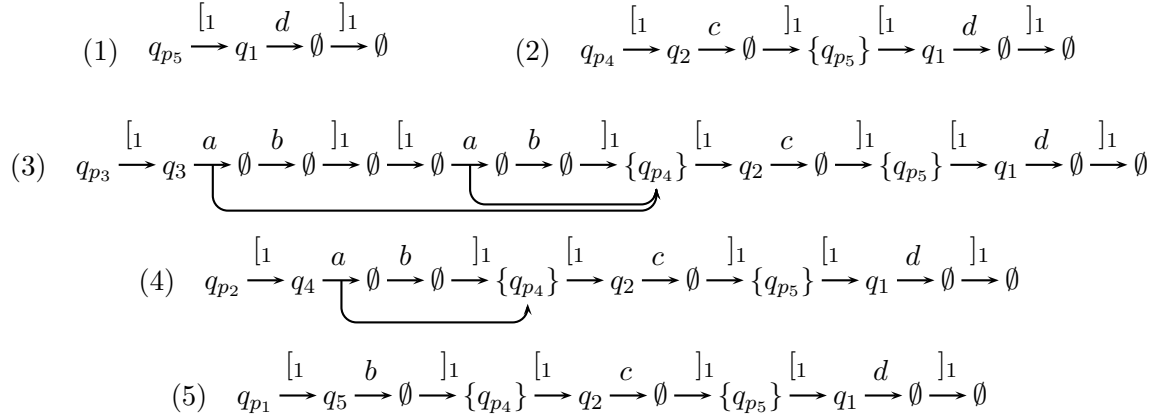

\begin{center}
    \newcommand\freenum[1]{\raisebox{2.5ex}{({#1})\ \ }}
    \begin{psmatrix}[nodealign=true,rowsep=4ex]
    \freenum{1}
    \begin{psmatrix}[nodealign=true,colsep=2ex,rowsep=1.5ex]
        \Rnode{N4}{$\sastate_{\control_5}$} &&
        \Rnode{N3}{$\sastate_1$} &&
        \Rnode{N2}{$\emptyset$} &&
        \Rnode{N1}{$\emptyset$} \\

        \psset{nodesep=.5ex,angle=-90,linearc=.2}
        \ncline{->}{N4}{N3}^{$\sopen{1}$}
        \ncline{->}{N3}{N2}^{$\chd$}
        \ncline{->}{N2}{N1}^{$\sclose{1}$}
    \end{psmatrix}
    \qquad\qquad
    \freenum{2}
    \begin{psmatrix}[nodealign=true,colsep=2ex,rowsep=1.25ex]
        \Rnode{N7}{$\sastate_{\control_4}$} &&
        \Rnode{N6}{$\sastate_2$} &&
        \Rnode{N5}{$\emptyset$} &&
        \Rnode{N4}{$\set{\sastate_{\control_5}}$} &&
        \Rnode{N3}{$\sastate_1$} &&
        \Rnode{N2}{$\emptyset$} &&
        \Rnode{N1}{$\emptyset$} \\

        \psset{nodesep=.5ex,angle=-90,linearc=.2}
        \ncline{->}{N7}{N6}^{$\sopen{1}$}
        \ncline{->}{N6}{N5}^{$\chc$}
        \ncline{->}{N5}{N4}^{$\sclose{1}$}
        \ncline{->}{N4}{N3}^{$\sopen{1}$}
        \ncline{->}{N3}{N2}^{$\chd$}
        \ncline{->}{N2}{N1}^{$\sclose{1}$}
    \end{psmatrix} \\

    \freenum{3}
    \begin{psmatrix}[nodealign=true,colsep=2ex,rowsep=1.25ex]
        \Rnode{N15}{$\sastate_{\control_3}$} &&
        \Rnode{N14}{$\sastate_3$} &\pnode{N1413}&
        \Rnode{N13}{$\emptyset$} &&
        \Rnode{N12}{$\emptyset$} &&
        \Rnode{N11}{$\emptyset$} &&
        \Rnode{N10}{$\emptyset$} &\pnode{N109}&
        \Rnode{N9}{$\emptyset$} &&
        \Rnode{N8}{$\emptyset$} &&
        \Rnode{N7}{$\set{\sastate_{\control_4}}$} &&
        \Rnode{N6}{$\sastate_2$} &&
        \Rnode{N5}{$\emptyset$} &&
        \Rnode{N4}{$\set{\sastate_{\control_5}}$} &&
        \Rnode{N3}{$\sastate_1$} &&
        \Rnode{N2}{$\emptyset$} &&
        \Rnode{N1}{$\emptyset$} \\

        \psset{nodesep=.5ex,angle=-90,linearc=.2}
        \ncline{->}{N15}{N14}^{$\sopen{1}$}
        \ncline{->}{N14}{N13}^{$\cha$}
        \ncline{->}{N13}{N12}^{$\chb$}
        \ncline{->}{N12}{N11}^{$\sclose{1}$}
        \ncline{->}{N11}{N10}^{$\sopen{1}$}
        \ncline{->}{N10}{N9}^{$\cha$}
        \ncline{->}{N9}{N8}^{$\chb$}
        \ncline{->}{N8}{N7}^{$\sclose{1}$}
        \ncline{->}{N7}{N6}^{$\sopen{1}$}
        \ncline{->}{N6}{N5}^{$\chc$}
        \ncline{->}{N5}{N4}^{$\sclose{1}$}
        \ncline{->}{N4}{N3}^{$\sopen{1}$}
        \ncline{->}{N3}{N2}^{$\chd$}
        \ncline{->}{N2}{N1}^{$\sclose{1}$}
        \ncbar[arm=1.8ex,nodesepA=0ex]{->}{N1413}{N7}
        \ncbar[arm=1.3ex,nodesepA=0ex]{->}{N109}{N7}
    \end{psmatrix} \\

    \freenum{4}
    \begin{psmatrix}[nodealign=true,colsep=2ex,rowsep=1.25ex]
        \Rnode{N11}{$\sastate_{\control_2}$} &&
        \Rnode{N10}{$\sastate_4$} &\pnode{N109}&
        \Rnode{N9}{$\emptyset$} &&
        \Rnode{N8}{$\emptyset$} &&
        \Rnode{N7}{$\set{\sastate_{\control_4}}$} &&
        \Rnode{N6}{$\sastate_2$} &&
        \Rnode{N5}{$\emptyset$} &&
        \Rnode{N4}{$\set{\sastate_{\control_5}}$} &&
        \Rnode{N3}{$\sastate_1$} &&
        \Rnode{N2}{$\emptyset$} &&
        \Rnode{N1}{$\emptyset$} \\

        \psset{nodesep=.5ex,angle=-90,linearc=.2}
        \ncline{->}{N11}{N10}^{$\sopen{1}$}
        \ncline{->}{N10}{N9}^{$\cha$}
        \ncline{->}{N9}{N8}^{$\chb$}
        \ncline{->}{N8}{N7}^{$\sclose{1}$}
        \ncline{->}{N7}{N6}^{$\sopen{1}$}
        \ncline{->}{N6}{N5}^{$\chc$}
        \ncline{->}{N5}{N4}^{$\sclose{1}$}
        \ncline{->}{N4}{N3}^{$\sopen{1}$}
        \ncline{->}{N3}{N2}^{$\chd$}
        \ncline{->}{N2}{N1}^{$\sclose{1}$}
        \ncbar[arm=1.5ex,nodesepA=0ex]{->}{N109}{N7}
    \end{psmatrix} \\

    \freenum{5}
    \begin{psmatrix}[nodealign=true,colsep=2ex,rowsep=1.25ex]
        \Rnode{N10}{$\sastate_{\control_1}$} &&
        \Rnode{N9}{$\sastate_5$} &&
        \Rnode{N8}{$\emptyset$} &&
        \Rnode{N7}{$\set{\sastate_{\control_4}}$} &&
        \Rnode{N6}{$\sastate_2$} &&
        \Rnode{N5}{$\emptyset$} &&
        \Rnode{N4}{$\set{\sastate_{\control_5}}$} &&
        \Rnode{N3}{$\sastate_1$} &&
        \Rnode{N2}{$\emptyset$} &&
        \Rnode{N1}{$\emptyset$} \\

        \psset{nodesep=.5ex,angle=-90,linearc=.2}
        \ncline{->}{N10}{N9}^{$\sopen{1}$}
        \ncline{->}{N9}{N8}^{$\chb$}
        \ncline{->}{N8}{N7}^{$\sclose{1}$}
        \ncline{->}{N7}{N6}^{$\sopen{1}$}
        \ncline{->}{N6}{N5}^{$\chc$}
        \ncline{->}{N5}{N4}^{$\sclose{1}$}
        \ncline{->}{N4}{N3}^{$\sopen{1}$}
        \ncline{->}{N3}{N2}^{$\chd$}
        \ncline{->}{N2}{N1}^{$\sclose{1}$}
    \end{psmatrix}
    \end{psmatrix}
    \vspace{-3ex}
    \caption{\label{fig:example-saturation}A sequence of saturation steps.}
\end{center}
\end{figure*}

As an example, consider a CPDS with the run
\begin{multline*}
    \config{\control_1}{
        \sbrac{
            \sbrac{\chb}{1}
            \sbrac{\chc}{1}
            \sbrac{\chd}{1}
        }{2}
    }
    \cpdatran{\cpush{\cha}{2}}
    \config{\control_2}{
        \sbrac{
            \sbrac{\rnode{A1}{\cha} \chb}{1}
            \pnode{S1} \sbrac{\chc}{1}
            \sbrac{\chd}{1}
        }{2}
    }
    \cpdatran{\push{2}}
    \config{\control_3}{
        \sbrac{
            \sbrac{\rnode{A2}{\cha} \chb}{1}
            \sbrac{\rnode{A22}{\cha} \chb}{1}
            \pnode{S2} \sbrac{\chc}{1}
            \sbrac{\chd}{1}
        }{2}
    }
    \\
    \cpdatran{\collapse{2}}
    \config{\control_4}{
        \sbrac{
            \sbrac{\chc}{1}
            \sbrac{\chd}{1}
        }{2}
    }
    \cpdatran{\pop{2}}
    \config{\control_5}{
        \sbrac{\sbrac{\chd}{1}}{2}
    }\ .
\end{multline*}
\psset{arcangleA=100,arcangleB=70,nodesep=1pt}%
\ncarc[ncurv=1.7]{->}{A1}{S1}
\ncarc[ncurv=1.3,arcangleB=75]{->}{A2}{S2}
\ncarc[ncurv=1.7]{->}{A22}{S2}
\vspace{-8ex}
\begin{center}
\end{center}
Figure~\ref{fig:example-saturation} shows the sequence of saturation steps,
beginning with an accepting run of the configuration
$\config{\control_5}{\sbrac{\sbrac{\chd}{1}}{2}}$
and finishing with an accepting run of
$\config{\control_1}{
    \sbrac{
        \sbrac{\chb}{1}
        \sbrac{\chc}{1}
        \sbrac{\chd}{1}
    }{2}
}$.
The individual steps are explained below.

\paragraph{Initial Automaton}

The top of Figure~\ref{fig:example-saturation} shows a stack
automaton containing the transitions $\sastate_{\control_5} \satran{\sastate_1}
\emptyset$ and $\sastate_1 \satrancol{\chd}{\emptyset} \emptyset$, which we
write $\satranfull{\sastate_{\control_5}}{\chd}{\emptyset}{\emptyset,
\emptyset}$.  This gives the run over
$\config{\control_5}{\sbrac{\sbrac{\chd}{1}}{2}}$.

\paragraph{Rule $\cpdsrule{\control_4}{\chc}{\pop{2}}{\control_5}$}

When the saturation step considers such a pop rule, it adds
$\satranfull{\sastate_{\control_4}}{\chc}{\emptyset}{\emptyset,
\set{\sastate_{\control_5}}}$.  This is added because we only require
the top order-$1$ stack (removed by $\pop{2}$) to have the top
character $\chc$ (hence $\emptyset$ is the next order-$1$ label), and after the
$\pop{2}$ the remaining stack needs to be accepted from $\sastate_{\control_5}$
(hence $\set{\sastate_{\control_5}}$ is the next order-$2$ label).  The new
transitions allow us to construct the next run over
$\config{\control_4}{
    \sbrac{
        \sbrac{\chc}{1}
        \sbrac{\chd}{1}
    }{2}
}$
in Figure~\ref{fig:example-saturation}.

\paragraph{Rule $\cpdsrule{\control_3}{\cha}{\collapse{2}}{\control_4}$}

Similarly to the pop rule above, the saturation step adds
$\satranfull{\sastate_{\control_3}}{\cha}{\set{\sastate_{\control_4}}}{\emptyset,
\emptyset}$.  The addition of these transitions allows us to construct the
pictured run over
$\config{\control_3}{
    \sbrac{
        \sbrac{\cha \chb}{1}
        \sbrac{\cha \chb}{1}
        \sbrac{\chc}{1}
        \sbrac{\chd}{1}
    }{2}
}$
(collapse links omitted), recalling that
$\emptyset \satran{\emptyset} \emptyset$, $\emptyset \satrancol{\cha}{\emptyset}
\emptyset$ and $\emptyset \satrancol{\chb}{\emptyset} \emptyset$ transitions are
always possible due to the empty initial set.  Note that the labelling of
$\set{\sastate_{\control_4}}$ comes from the collapse
link on the topmost $\cha$ character on the stack.

\paragraph{Rule $\cpdsrule{\control_2}{\cha}{\push{2}}{\control_3}$}

Consider the run from $\sastate_{\control_3}$ in
Figure~\ref{fig:example-saturation}.
The $2$-expansion of the head transition of the run accepting the first order-$1$ stack is
$\satranfull{\sastate_{\control_3}}{\cha}{\set{\sastate_{\control_4}}}{\emptyset,
\emptyset}$.  We also have $\emptyset \satran{\emptyset} \emptyset$ (trivially)
accepting the second order-$1$ stack.  Any $\push{2}$ predecessor of this stack
must have a top order-$1$ stack that could have appeared twice at the top of the
stack from $\sastate_{\control_3}$.  Thus, the saturation step combines the
initial order-$1$ transitions of first two order-$1$ stacks.
This results in
$\satranfull{\sastate_{\control_2}}{\cha}{\set{\sastate_{\control_4}} \cup
\emptyset}{\emptyset \cup \emptyset, \emptyset}$, which can be used to form the
shown run over
$\config{\control_2}{
    \sbrac{
        \sbrac{\cha \chb}{1}
        \sbrac{\chc}{1}
        \sbrac{\chd}{1}
    }{2}
}$
(collapse links omitted).

\paragraph{Rule $\cpdsrule{\control_1}{\chb}{\cpush{\cha}{2}}{\control_2}$}

The run from $\sastate_{\control_2}$ in Figure~\ref{fig:example-saturation}
has a head transition with $2$-expansion
$\satranfull{\sastate_{\control_2}}{\cha}{\set{\sastate_{\control_4}}}{\emptyset,
\emptyset}$ and the run continues with $\emptyset \satrancol{\chb}{\emptyset} \emptyset$.  Note that
the $\cpush{\cha}{2}$ gives a stack with $\cha \chb$ on top.  Moreover, the
collapse link on $\cha$ should point to the order-$1$ stack just below the
current top one.  Since the transition from $\sastate_{\control_2}$ requires
that the linked-to stack is accepted from $\sastate_{\control_4}$, we need this
requirement in the preceding stack (accepted from $\sastate_{\control_1}$ and
without the $\cha$ on top).  Thus, we move the target of the collapse link into
the order-$2$ destination of the new transitions.  That is,
for $\cpush{\cha}{2}$ we add
$\satranfull{\sastate_{\control_1}}{\chb}{\emptyset}{\emptyset, \emptyset \cup
\set{\sastate_{\control_4}}}$.  From this we can construct an accepting run
over
$\config{\control_1}{
    \sbrac{
        \sbrac{\chb}{1}
        \sbrac{\chc}{1}
        \sbrac{\chd}{1}
    }{2}
}$.


%% file: corrcomp.tex
\section{Correctness and Complexity}
\label{sec:correctness}
\label{sec:complexity}

In this section we show the complexity and correctness of saturation.  We prove
soundness by a witness generation algorithm.  This is an extension of the
witness generation given in ICFP 2013~\cite{BCHS13} to the case of alternating CPDSs.  In
ICALP 2012~\cite{BCHS12} we gave a more denotational proof of soundness which worked by
showing that all transitions added by saturation respect the ``meaning'' of the
transitions in the automata representing $\prestar{\cpds}{\saauta_0}$.  This
proof used a slightly different formulation of CPDS as Annotated Pushdown
Systems.  Although we believe this soundness proof to be more elegant, we do not
repeat it here for space reasons (since it would require the definition of
annotated pushdown systems).

\begin{thm}
    \label{thm:sat-correct}
    For a CPDS $\cpds$ and stack automaton $\saauta_0$, let $\saauta =
    \saauta_\idxi$ where $\idxi$ is the least index such that $\saauta_{\idxi+1}
    = \apply{\satstep}{\saauta_\idxi}$.  We have $\stackw \in
    \slang{\sastate_\control}{\saauta}$ iff $\config{\control}{\stackw} \in
    \prestar{\cpds}{\saauta_0}$.
    \qed
\end{thm}

The proof is given in the following sections.  Completeness is by a
straightforward induction over the ``distance'' to $\saauta_0$.  Soundness is
the key technical challenge.

\begin{prop}
	\label{prop:sat-complexity}
    The saturation construction for an alternating order-$\cpdsord$ collapsible PDS $\cpds$
    and an order-$\cpdsord$ stack automaton $\saauta_0$ runs in
    $\cpdsord$-EXPTIME, which is optimal.
\end{prop}
\proof
    Let $\exptower{0}{\numof} = \numof$ and $\exptower{\idxi+1}{\numof} =
    2^{\exptower{\idxi}{\numof}}$.  The number of states of $\saauta$ is bounded
    by $\exptower{(\cpdsord-1)}{\numof}$ where $\numof$ is the size of $\cpds$
    and $\saauta_0$:  each state in $\sastates_\opord$ was either in $\saauta_0$
    was added when a transition in $\sadelta_{\opord+1}$ was created.  Since
    the automata are alternating, the number of transitions of each order
    $\opord$ is exponential in the number of states of order $\opord$.
    Thus, there are potentially exponentially many order-$(\cpdsord-1)$ states, doubly exponentially many order-$(\cpdsord-2)$ states and so on.
    That is, there is an exponential blow up at each order except at
    order-$\cpdsord$.  Each iteration of the algorithm adds at least one new
    transition.  Only $\exptower{\cpdsord}{\numof}$ transitions can be added.
    Since reachability for alternating higher-order pushdown systems
    is complete for $\cpdsord$-EXPTIME~\cite{HO08}, our algorithm is optimal.
\qed

The complexity of reachability for non-alternating collapsible
PDS is in $(\cpdsord-1)$-EXPTIME.  The cause of the additional exponential blow
up is in the alternation of the stack automata.  However, for a suitable notion
of \emph{non-alternating} stack automata, our algorithm can be adapted to run in
$(\cpdsord-1)$-EXPTIME, when the CPDS is also non-alternating.  This is
discussed in the next section before the completeness and soundness proofs.

Finally, we remark the algorithm is PTIME for a fixed order and number of
control states.  If we obtained $\cpds$ from a higher-order recursion scheme, the number of control
states is given by the arity of the scheme~\cite{HMOS08,HMOS17} and the size of the
property automaton (giving $\lang_{\mathit{err}}$).  In practice, we expect the arity and
order to be small, and since simple reachability properties require small
automata, we expect the total number of control states to be small.

\subsection{Non-Alternation}

We introduce a notion of non-alternation at order-$\cpdsord$.  Note that the
automata are alternating both via transitions to $\sastateset$ with
$\cardinality{\sastateset} > 1$, and via collapse links.
Informally, a run is non-alternating at order-$\cpdsord$ over a stack
$\sbrac{\stackw_1 \ldots \stackw_\numof}{\cpdsord}$
if at most one transition from $\sadelta_\cpdsord$ is used to read each $\stackw_\idxi$.
In the graphical representation, each node with an outgoing edge labelled
$\sopen{\cpdsord-1}$
would be labelled by a set containing at most one state from
$\sastateset_\cpdsord$.
For the formal definition, we use the stack representation of runs, and the fact that $\stackw_\idxi$ being read by at most a single transition from $\sadelta_\cpdsord$ implies that at most a single transition from $\sadelta_1$ is used to read
$\apply{\ctop{1}}{\stackw_\idxi}$.
Conversely, if at most a single transition from $\sadelta_1$ is used to read
$\apply{\ctop{1}}{\stackw_\idxi}$
then, thanks to the unique $\cpdsord$-expansion, only at most one transition from $\sadelta_\cpdsord$ can be used to read $\stackw_\idxi$.

\begin{defi}[Non-Alternation at Order-$\cpdsord$]
    An order-$\cpdsord$ stack automaton $\saauta$ is \emph{non-alternating at order-$\cpdsord$} whenever, for all states $\sastate$ of $\saauta$ and all stacks
    $\sbrac{\stackw_1\ldots\stackw_\numof}{\cpdsord}
     \in
     \slang{\sastate}{\saauta}$,
    there is an accepting run
    $\sbrac{\stackv_1\ldots\stackv_\numof}{\cpdsord}$
    of $\saauta$ over $\stackw$ such that for all $\idxi$ we have
    $\apply{\ctop{1}}{\stackv_\idxi} = \tuple{\cha, \satset}$
    and
    $\cardinality{\satset} \leq 1$.
\end{defi}

Note, for example, that a stack automaton that does not follow collapse links,
and has no alternating transitions in $\sadelta_\cpdsord$, is trivially
non-alternating at order-$\cpdsord$.  Similarly, we may allow
$\satranfull{\sastate}{\cha}{\sastateset_\branch}{\emptyset, \ldots, \emptyset}$
when $\sastateset_\branch \subseteq \sastates_\cpdsord$ and
$\cardinality{\sastateset_\branch} \leq 1$.

We then define $\satstep'$ to be the saturation function $\satstep$ with the
additional constraint that transitions
$\satranfull{\sastate}{\cha}{\sastateset_\branch}{\sastateset_1, \ldots,
\sastateset_\cpdsord}$ are not added if $\cardinality{\sastateset_\cpdsord} > 1$.
Clearly saturation by $\satstep'$ remains sound, since it contains a subset of
the transitions of the automaton produced by saturation with $\satstep$.  Hence,
we only need to prove that the automaton remains complete.  We prove
completeness in conjunction with the completeness proof for the saturation
algorithm in general.  Intuitively, the automaton remains correct because a collapse
link at order-$n$ can only be used once, whereas, at lower orders, a
$\push{\opord}$ operation may make different copies of a link (with different
targets).  Hence, lower order links need alternation to keep track of the
different uses of the link throughout the run.

Henceforth, we will refer to non-alternation at
order-$\cpdsord$ as simply \emph{non-alternation}.

\subsection{Completeness}

We show that the automaton constructed by $\satstep$ (and $\satstep'$) is
complete for collapsible stacks.  The intuition behind the completeness proof is
well illustrated by the examples in Section~\ref{sec:examples}, hence we
encourage the reader to consult these examples when reading the proof.

\begin{lem}[Completeness of $\satstep$]
    \label{lem:completeness}
    Given a CPDS $\cpds$ and an order-$\cpdsord$ stack automaton
    $\saauta_0$, the automaton $\saauta$ constructed by saturation with
    $\satstep$ is such that $\config{\control}{\stackw} \in
    \prestar{\cpds}{\saauta_0}$ implies $\stackw \in
    \slang{\sastate_\control}{\saauta}$.  The result also holds for $\satstep'$
    when $\cpds$ and $\saauta_0$ are non-alternating.
\end{lem}
\proof
    The proof is by induction over $\ordinal$ such that
    $\config{\control}{\stackw} \in \pre{\ordinal}{\cpds}{\saauta_0}$.  We prove
    simultaneously during the induction that in the case of $\satstep'$ all
    $\config{\control}{\stackw} \in \prestar{\cpds}{\saauta_0}$ have a
    $\sastate_\control$-accepting run of $\saauta$ such that is non-alternating.

    In the base case, we have $\stackw \in \slang{\sastate_\control}{\saauta_0}$
    and the existence of a (non-alternating) run of $\saauta_0$, and thus a run in
    $\saauta$ comes directly from the (non-alternating) run of $\saauta_0$.

    Hence, inductively assume $\config{\control}{\stackw} \cpdstran
    \config{\control'}{\stackw'}$
     via a rule
    $\cpdsrule{\control}{\apply{\ctop{1}}{\stackw}}{\genop}{\control'}$
    and there is a (non-alternating) $\sastate_{\control'}$-accepting run $\stackv'$ of
    $\saauta$ over $\stackw'$.  Hence
    $\stackw' = \apply{\genop}{\stackw}$.
    We will construct an $\sastate_\control$-accepting (non-alternating) run $\stackv$ over $\stackw$.
    \begin{enumerate}
    \item
        When $\genop = \pop{\opord}$, let
        $\apply{\ctop{1}}{\stackv'} = \tuple{\cha, \set{\sat'}}$.
        Then, let
        $\satranfull{\sastate_\opord}
                    {\cha}
                    {\sastateset_\branch}
                    {\sastateset_1, \ldots, \sastateset_\opord}$
        be the $\opord$-expansion of $\sat'$ and let
        $\satranfullk{\sastate_{\control'}}
                     {\sastate_\opord}
                     {\sastateset_{\opord+1}, \ldots, \sastateset_\cpdsord}$
        be the $\cpdsord$-expansion of $\sastate_\opord$.
        We know, from the construction, that we have a transition $\sat$ with the $\cpdsord$-expansion
        \[
            \satranfull{\sastate_\control}
                       {\cha}
                       {\emptyset}
                       {\emptyset, \ldots, \emptyset,
                        \set{\sastate_\opord},
                        \sastateset_{\opord+1}, \ldots, \sastateset_\cpdsord} \ .
        \]
        We also know that $\stackv'$ is $\set{\sastate_\opord}$-valid (and non-alternating).
        In addition,
        $\stackw = \ccompose{\stacku_{\opord-1}}
                            {\opord}
                            {\stackw'}$
        for some order-$(\opord-1)$ stack $\stacku_{\opord-1}$.
        Hence, we build the run
        $\stackv = \ccompose{\stackv_{\opord-1}}
                            {\opord}
                            {\stackv'}$
        where $\stackv_{\opord-1}$ is the order-$(\opord-1)$ stack such that
        $\saprojection{\stackv_{\opord-1}} = \stacku_{\opord-1}$
        with
        $\apply{\ctop{1}}{\stackv_{\opord-1}} = \tuple{\cha, \set{\sat}}$
        (for some $\cha$)
        and all other characters
        $\tuple{\chb, \satset}$
        appearing at any position in $\stackv_{\opord-1}$ have
        $\satset = \emptyset$.

        One can verify that $\stackv$ is an accepting run over $\stackw$.
        The link-validity requirement is satisfied since $\stackv'$ was link-valid and all link constraints in $\stackv_{\opord-1}$ are empty.
        For $\sastateset$-validity, most cases either follow from the validity of $\stackv'$ or from the fact that $\sastateset$ is $\emptyset$.
        The only non-empty new constraint is $\set{\sastate_\opord}$, but we have already noted the required stack, $\stackv'$, is $\set{\sastate_\opord}$-valid.
        For $\satstep'$, it is immediate to verify that $\stackv$ is non-alternating as $\sastateset_\cpdsord$ is unchanged, and no new transition uses the collapse links.

    \item
        When $\genop = \push{\opord}$, let
        $\stackw =
         \ccompose{\stacku_{\opord-1}}
                  {\opord}
                  {\stackw''}$.
        We know that
        \[
            \stackw' =
            \ccompose{\stacku_{\opord-1}}
                     {\opord}
                     {\ccompose{\stacku_{\opord-1}}
                               {\opord}
                               {\stackw''}}
          \ .
        \]
        By induction we have an accepting run $\stackv'$ of $\stackw'$.
        Suppose
        $\apply{\ctop{1}}{\stackw'} = \tuple{\cha, \set{\sat'}}$
        and let
        $\satranfull{\sastate_{\control'}}
                    {\cha}
                    {\sastateset_\branch}
                    {\sastateset_1,
                     \ldots,
                     \sastateset_\opord,
                     \ldots,
                     \sastateset_\cpdsord}$
        be the $\cpdsord$-expansion of $\sat'$.
        Moreover, suppose
        $\apply{\ctop{1}}{\stackv'} = \tuple{\cha, \satset}$
        and the $(\opord, \sastateset_\opord)$-expansion of $\satset$ is
        $\satranfull{\sastateset_\opord}
                    {\cha}
                    {\sastateset'_\branch}
                    {\sastateset'_1, \ldots, \sastateset'_\opord}$.
        Note, this expansion ``reads'' the second copy of $\stacku_\opord$.
        It exists because otherwise the run would not meet the validity constraints.

        From the construction we added a transition $\sat$ with $\cpdsord$-expansion
        \[
            \satranfull{\sastate_{\control}}{\cha}{\sastateset_\branch \cup
            \sastateset'_\branch}{\sastateset_1 \cup \sastateset'_1, \ldots,
            \sastateset_{\opord-1} \cup \sastateset'_{\opord-1},
            \sastateset'_\opord, \sastateset_{\opord+1}, \ldots,
            \sastateset_\cpdsord}
            \ .
        \]

        Let
        $\stackv' = \ccompose{\stackv^1_{\opord-1}}
                             {\opord}
                             {\ccompose{\stackv^2_{\opord-1}}
                                       {\opord}
                                       {\stackv''}}$.
        Note
        $\saprojection{\stackv^1_{\opord-1}} =
         \saprojection{\stackv^2_{\opord-1}} =
         \stacku_{\opord-1}$.
        We define
        $\stackv_{\opord-1} = \stackv^1_{\opord-1} \cup \stackv^2_{\opord-1}$
        where
        \[
            \begin{array}{rcl}
                \annot{\tuple{\cha, \satset_1}}{\tuple{\opord', \idxi}}
                \cup
                \annot{\tuple{\cha, \satset_2}}{\tuple{\opord', \idxi}}
                &=&
                \annot{\tuple{\cha, \satset_1 \cup \satset_2}}
                      {\tuple{\opord', \idxi}}
                \\
                \brac{\ccompose{\stackv^1}{\opord'}{\stackv^2}}
                \cup
                \brac{\ccompose{\stackv^3}{\opord'}{\stackv^4}}
                &=&
                \ccompose{\brac{\stackv^1 \cup \stackv^3}}
                         {\opord'}
                         {\brac{\stackv^2 \cup \stackv^4}} \ .
            \end{array}
        \]
        The run
        $\stackv =
         \apply{\rew{\tuple{\cha, \set{\sat}}}}
               {\ccompose{\stackv_{\opord-1}}{\opord}{\stackv''}}$
        is an accepting run over $\stackw$.
        This can be verified as follows.
        The topmost transition $\sat$ requires the following validity constraints.
        For
        $\opord' \in \set{1, \ldots, \opord-1}$
        we require
        $\apply{\ctop{\opord'+1}}
               {\apply{\pop{\opord'}}{\stackv''}}$
        is
        $(\sastateset_{\opord'} \cup \sastateset'_{\opord'})$-valid,
        which follows from the validity of
        $\stackv^1_{\opord-1}$
        and
        $\stackv^2_{\opord-1}$.
        At order-$\opord$, $\sastateset'_\opord$-validity follows from validity of $\stackv''$.
        Similarly for orders
        $\opord' \in \set{\opord+1, \ldots, \cpdsord}$
        and
        $\sastateset_{\opord'}$-validity.
        Link validity follows by a similar argument.
        All other required $\sastateset$-validity and link constraints follows from the fact that they held in either $\stackv^1_{\opord}$, $\stackv^2_{\opord}$, or $\stackv''$.

        For $\satstep'$, from the above observation, we know that the new run is non-alternating
        (that is, taking the union of $\sastateset_\branch$ and $\sastateset'_\branch$ does not introduce alternation).
        We also remark that, while this is
        \emph{always} the case for order-$\cpdsord$ links, it is not always the
        case for order-$\opord$ links with $\opord < \cpdsord$, and hence,
        alternation is needed at orders lower than $\cpdsord$.

    \item
        When $\genop = \collapse{\opord}$, the case is similar to $\pop{\opord}$.
        Let
        $\apply{\ctop{1}}{\stackv'} = \tuple{\cha, \set{\sat'}}$.
        Then, let
        $\satranfull{\sastate_\opord}
                    {\cha}
                    {\sastateset_\branch}
                    {\sastateset_1, \ldots, \sastateset_\opord}$
        be the $\opord$-expansion of $\sat'$ and let
        $\satranfullk{\sastate_{\control'}}
                     {\sastate_\opord}
                     {\sastateset_{\opord+1}, \ldots, \sastateset_\cpdsord}$
        be the $\cpdsord$-expansion of $\sastate_\opord$.
        We know, from the construction, that we have a transition $\sat$ with the $\cpdsord$-expansion
        \[
            \satranfull{\sastate_\control}
                       {\cha}
                       {\set{\sastate_\opord}}
                       {\emptyset, \ldots, \emptyset,
                        \sastateset_{\opord+1}, \ldots, \sastateset_\cpdsord} \ .
        \]
        We also know that $\stackv'$ is $\set{\sastate_\opord}$-valid (and non-alternating).
        In addition,
        \[
            \stackw =
            \ccompose{\stacku^1_{\opord-1}}
                     {\opord}
                     {\ccompose{\cdots}
                               {\opord}
                               {\ccompose{\stacku^\numof_{\opord-1}}
                                         {\opord}
                                         {\stackw'}}} \ .
        \]
        That is, $\stackw'$ was obtained by performing a collapse on $\stackw$.
        We build the run
        $\stackv =
         \ccompose{\stackv^1_{\opord-1}}
                  {\opord}
                  {\ccompose{\cdots}
                            {\opord}
                            {\ccompose{\stackv^\numof_{\opord-1}}
                                      {\opord}
                                      {\stackv'}}}$
        where $\stackv^1_{\opord-1}$ is the order-$(\opord-1)$ stack such that
        $\saprojection{\stackv^1_{\opord-1}} =
         \stacku^1_{\opord-1}$
        with
        $\apply{\ctop{1}}{\stackv^1_{\opord-1}} = \tuple{\cha, \set{\sat}}$
        (for some $\cha$)
        and all other characters
        $\tuple{\chb, \satset}$
        appearing at any position in $\stackv^1_{\opord-1}$ have
        $\satset = \emptyset$.
        Similarly, for all
        $\idxi \in \set{2,\ldots,\numof}$
        we have $\stackv^\idxi_{\opord-1}$ is the order-$(\opord-1)$ stack such that
        $\saprojection{\stackv^\idxi_{\opord-1}} =
         \stacku^\idxi_{\opord-1}$
        and all characters
        $\tuple{\chb, \satset}$
        appearing at any position in $\stackv^\idxi_{\opord-1}$ have
        $\satset = \emptyset$.

        One can verify that $\stackv$ is an accepting run over $\stackw$.
        The link-validity requirement is satisfied since $\stackv'$ was link-valid and all new link constraints are empty except at the top of
        $\stackv^1_{\opord-1}$.
        Here we require the link target to be
        $\set{\sastate_\opord}$-valid.
        Since the link target is $\stackw'$, we have already noted the required validity.
        For $\sastateset$-validity, all cases either follow from the validity of $\stackv'$ or from the fact that $\sastateset$ is $\emptyset$.
        For $\satstep'$, it is immediate to verify that $\stackv$ is non-alternating because as $\stackv'$ is non-alternating and the new parts of the stack do not have alternation at order-$\cpdsord$.

    \item
        When $\genop = \cpush{\chb}{\opord}$, then for the appropriate $\numof$, we know that
        $\stackw' = \apply{\cpush{\chb}{\opord}}{\stackw}$
        is
        $\ccompose{\chb^{\tuple{\opord, \numof}}}{1}{\stackw}$.
        Let
        $\apply{\ctop{1}}{\stackv'} = \tuple{\chb, \set{\sat'}}$
        where $\sat'$ has the $\cpdsord$-expansion
        $\satranfull{\sastate_{\control'}}
                    {\chb}
                    {\sastateset_\branch}
                    {\sastateset_1, \ldots, \sastateset_\cpdsord}$.
        Moreover, let
        $\apply{\ctop{1}}{\apply{\pop{1}}{\stackv'}} = \tuple{\chb, \satset}$
        and let
        $\sastateset_1
         \satrancol{\cha}{\sastateset'_\branch}
         \sastateset'_1$
        be the $(1, \sastateset_1)$-expansion of $\satset$.
        From the construction we know that we added a transition $\sat$ with $\cpdsord$-expansion
        \[
            \satranfull{\sastate_{\control}}
                       {\cha}
                       {\sastateset'_\branch}
                       {\sastateset'_1,
                        \sastateset_2,
                        \ldots,
                        \sastateset_\opord \cup \sastateset_\branch,
                        \ldots,
                        \sastateset_\cpdsord} \ .
        \]
        We note that
        $\stackv =
         \apply{\rew{\tuple{\cha, \set{\sat}}}}
               {\apply{\pop{1}}{\stackv'}}$
        is an accepting run over $\stackw$.
        This follows using arguments similar to the previous cases.
        The main difference is
        $(\sastateset_\opord \cup \sastateset_\branch)$-validity of
        $\apply{\pop{\opord}}{\stackv}$.
        This follows from a combination of link-validity and $\sastateset_\opord$-validity in $\stackv'$.

        For $\satstep'$, to see that the run is non-alternating, we observe that
        even though the new transition has
        $\sastateset_\opord \cup \sastateset_\branch$
        on the right, the run
        $\apply{\pop{1}}{\stackv}$
        contains, in the characters the same sets of transitions as in $\stackv'$, which was non-alternating.

    \item
        When $\genop = \rew{\chb}$ let
        $\apply{\ctop{1}}{\stackv'} = \tuple{\chb, \set{\sat'}}$
        where $\sat'$ has the $\cpdsord$-expansion
        $\satranfull{\sastate_{\control'}}
                    {\chb}
                    {\sastateset_\branch}
                    {\sastateset_1, \ldots, \sastateset_\cpdsord}$.
        From the construction we know that we have a transition $\sat$ with $\cpdsord$-expanstion
        $\satranfull{\sastate_\control}
                    {\cha}
                    {\sastateset_\branch}
                    {\sastateset_1, \ldots, \sastateset_\cpdsord}$,
        from which we get an accepting, (non-alternating) run of
        $\stackw$ as required.
        That is, the run
        $\apply{\rew{\tuple{\cha, \set{\sat}}}}{\stackv'}$.
    \end{enumerate}
    Hence, for every $\config{\control}{\stackw} \in \prestar{\cpds}{\saauta_0}$
    we have $\stackw \in \slang{\sastate_\control}{\saauta}$, and when we use
    $\satstep'$, and $\cpds$ and $\saauta_0$ are non-alternating, the run is
    non-alternating.

    In the alternating case we may have a branching transition
    $\config{\control}{\stackw} \cpdstran \configset$
    (where $\configset$ is a set of configurations)
    via a rule
    $\cpdsalttran{\control}{\controlset}$.
    In this case, for all
    $\control' \in \controlset$
    we have a run $\stackv_{\control'}$ by induction.
    Let
    $\apply{\ctop{1}}{\stackv_{\control'}} = \tuple{\cha, \set{\sat_{\control'}}}$.
    Let
    $\satset = \setcomp{\sat_{\control'}}{\control' \in \controlset}$
    and
    $\sastateset =
     \setcomp{\sastate}
             {\sastate
              \satrancol{\cha}{\sastateset_\branch}
              \sastateset' \in \satset}$.
    Then, take
    $\satranfull{\sastateset}
                {\cha}
                {\sastateset_\branch}
                {\sastateset_1, \ldots, \sastateset_\cpdsord}$,
    the $(\cpdsord, \sastateset)$-expansion of $\satset$.
    By construction, we have a transition $\sat$ with $\cpdsord$-expansion
    \[
        \satranfull{\sastate_\control}
                   {\cha}
                   {\sastateset_\branch}
                   {\sastateset_1, \ldots, \sastateset_\cpdsord} \ .
    \]
    It can be seen that
    $\apply{\rew{\tuple{\cha, \set{\sat}}}}
           {\bigcup\limits_{\control' \in \controlset}
                \stackv_{\control'}}$
    is an accepting run over $\stackw$.
    Hence, for every $\config{\control}{\stackw} \in \prestar{\cpds}{\saauta_0}$
    we have $\stackw \in \slang{\sastate_\control}{\saauta}$.
\qed

\subsection{Soundness by Witness Generation}
\label{sec:counter-egs}

Take a CPDS $\cpds$, a stack automaton $A_{0}$, and a configuration
$\config{\control}{\stackw}$ of
$\cpds$ belonging to $\prestar{\cpds}{A_{0}}$.
In this section we describe an algorithm that constructs a tree of rules of $\cpds$
such that, each branch, when applied from  $\config{\control}{\stackw}$ leads to
a configuration in $\lang({\saauta_{0}})$.  When an alternating rule
$\cpdsalttran{\control}{\controlset}$ is applied, the tree has a child for each
control state appearing in $\controlset$.  Otherwise, each node has a single
child.

The algorithm is a natural one and the full details are given in the
sequel.  We describe it informally here by means of the example in
Figure~\ref{fig:example-saturation}, described in Section~\ref{sec:examples}.
In this case, since there is no alternation, we construct a single-branch tree,
i.e., a trace.

To construct a trace from
$\config{\control_1}{\sbrac{\sbrac{\chb}{1}\sbrac{\chc}{1}\sbrac{\chd}{1}}{2}}$ to
$\config{\control_5}{\sbrac{\sbrac{\chd}{1}}{2}}$ we first note that, when adding the head
transition of the pictured run from $\sastate_{\control_1}$, the saturation
step marked that the transition was added due to the rule
$\cpdsrule{\control_1}{\chb}{\cpush{\cha}{2}}{\control_2}$.  If we apply this
rule to $\config{\control_1}{\sbrac{\sbrac{\chb}{1}\sbrac{\chc}{1}\sbrac{\chd}{1}}{2}}$ we obtain
$\config{\control_2}{\sbrac{\sbrac{\cha \chb}{1}\sbrac{\chc}{1}\sbrac{\chd}{1}}{2}}$ (collapse
links omitted).  Furthermore, the justifications added during the saturation
step tell us which transitions to use to construct the pictured run from
$\sastate_{\control_2}$.  Hence, we have completed the first step of counter
example extraction and moved one step closer to the target configuration.  To
continue, we consider the initial transition of the run from
$\sastate_{\control_2}$.  Again, the justifications added during saturation
tell us which CPDS rule to apply and which stack automaton transitions to use to
build an accepting run of the next configuration.  Thus, we follow the
justifications back to a run of $\saauta_0$, constructing a complete trace on
the way.

The main technical difficulty lies in proving that the reasoning outlined above
leads to a terminating algorithm.  For example, we need to prove that following
the justifications does not result us following a loop indefinitely.  Since the
stack may shrink and grow during a run, this is a non-trivial property.  To
prove it, we require a subtle relation on runs over higher-order collapsible
stacks.

\subsubsection{A Well-Founded Relation on Stack Automaton Runs}

We define a well-founded relation over runs of the stack automaton
$\saauta$ constructed by saturation from $\cpds$ and $\saauta_0$.
We can define by induction a relation
$\runindexrel_{k}$ on the order-$\opord$ runs of $\saauta$. Note that this is
not an order relation as it is not always transitive.  There are several cases
to $\runindexrel_{\opord}$.
\begin{enumerate}
    \item For $\opord=1$ and order-$1$ runs $\stackr$ and $\stackr'$, we say $\stackr' \runindexrel_{1} \stackr$ if for some
          $i \geq 0$, $\stackr$ contains strictly fewer transitions in
          $\sadelta_{1}$ justified at step $i$ of the saturation than $\stackr'$ and that for all
          $j>i$ they both contain the same number of transitions in
          $\sadelta_{1}$ justified at step $j$.

    \item For $\opord>1$, we say $\stacku = \sbrac{ \stacku_{\numof} \ldots
          \stacku_{1}}{\opord} \runindexrel_{k} \stackv=\sbrac{
          \stackv_{\numof'} \ldots \stackv_{1}}{\opord}$ if
          \begin{enumerate}
              \item $\numof'< \numof$ and $\stacku_{i}=\stackv_{i}$ for $i \in
                    \set{1,\ldots,\ell'-1}$ and either $\stacku_{\ell'}=\stackv_{\ell'}$ or
                    $\stacku_{\ell'} \runindexrel_{\opord-1} \stackv_{\ell'}$,
                    or

              \item $\numof' \geq \numof$ and $\stacku_{i}=\stackv_{i}$ for $i
                    \in \set{1,\ldots,\ell-1}$ and $u_{\numof} \runindexrel_{\opord-1}
                    v_{i}$ for all $i \in \set{\numof,\ldots,\numof'}$.
          \end{enumerate}
\end{enumerate}

The proof of the following lemma is given in Section~\ref{subsubsection:correctness-Algorithm}.

\begin{lem}
    \label{lemma:well-founded}
    For all $\opord \in \set{1,\ldots,\cpdsord}$, the relation $\runindexrel_{\opord}$ is
    well-founded. Namely there is no infinite sequence $ \stackr_{0}
    \runindexrel_{\opord} \stackr_{1} \runindexrel_{\opord} \stackr_{2}
    \runindexrel_{\opord} \cdots$.
    \qed
\end{lem}

It is possible to show that by following the justifications, from stack
$\stackw$ to a $\stackw'$, we always have $\stackw \runindexrel_{\cpdsord}
\stackw'$.  Since this relation is well-founded, witness generation
always terminates.

\subsubsection{Witness Trees}

We define what it means to be a witness of $\config{\control}{\stackw} \in
\prestar{\cpds}{\saauta_0}$.  Without alternation, we simply require a trace of
rules which take $\config{\control}{\stackw}$ to some configuration in
$\langof{\saauta_0}$.  In the presence of alternating transitions
$\cpdsalttran{\control}{\controlset}$ we need to account for each possible next
configuration.  Hence, we require finite trees rather than sequences.

A $\treelabset$-labelled \emph{finite tree} is a tuple $\tuple{\treedom,
\treelab}$ where $\treedom \subset \naturals^\ast$ is a tree domain that is both
prefix- and younger-sibling-closed.  That is, for all $\treenode \idxi \in
\treedom$ with $\treenode \in \naturals^\ast$ and $\idxi \in \naturals$ we have
$\treenode \in \treedom$ and moreover, for all $\idxj < \idxi$ we have
$\treenode \idxj \in \treedom$.  Furthermore, $\treelab : \treedom \rightarrow
\treelabset$ is a tree labelling for a set $\treelabset$ of labels.  A leaf node
is a node $\treenode \in \treedom$ such that there is no $\idxi$ with $\treenode
\idxi \in \treedom$.  Otherwise, $\treenode$ is an internal node.

\begin{defi}[Witness Trees]
    For a CPDS $\cpds$, configuration $\config{\control}{\stackw}$ and stack automaton
    $\saauta_0$, a witness tree is a $\treelabset$-labelled finite tree
    $\tuple{\treedom, \treelab}$ where $\treelabset$ contains labels of the form
    $\nodelab{\configc}$ and $\nodelab{\configc, \cpdsruler}$ with $\configc$ a
    configuration of $\cpds$ and $\cpdsruler$ a rule of $\cpds$.  Moreover, for all
    $\treenode \in \treedom$ we have
    \begin{itemize}
    \item
        if $\treenode = \varepsilon$ then $\apply{\treelab}{\treenode} =
        \nodelab{\configc}$ or $\apply{\treelab}{\treenode} =
        \nodelab{\configc, \cpdsruler}$ for some $\cpdsruler$ and $\configc =
        \config{\control}{\stackw}$, and

    \item
        if $\treenode$ is an internal node then $\apply{\treelab}{\treenode} =
        \nodelab{\config{\control_1}{\stackw_1}, \cpdsruler}$ for some $\control_1$,
        $\stackw_1$, and $\cpdsruler$, and
        \begin{itemize}
        \item
            if $\cpdsruler = \cpdsrule{\control_1}{\cha}{\genop}{\control_2}$ then
            $\treenode 0$ is the only child of $\treenode$ and
            $\apply{\treelab}{\treenode 0} =
            \nodelab{\config{\control_2}{\stackw_2}}$ or
            $\apply{\treelab}{\treenode 0} =
            \nodelab{\config{\control_2}{\stackw_2}, \cpdsruler'}$ with
            $\stackw_2 = \apply{\genop}{\stackw_1}$
            and
            $\apply{\ctop{1}}{\stackw_1} = \cha$, and

        \item
            if $\cpdsruler = \cpdsaltrule{\control_1}{\controlset}$
            with
            $\controlset = \set{\control'_0, \ldots, \control'_\numof}$
            then $\treenode$ has $\numof$ children and for each
            $\idxi \in \set{0, \ldots, \numof}$
            we have
            $\apply{\treelab}{\treenode \idxi} =
            \nodelab{\config{\control'_\idxi}{\stackw_1}}$ or
            $\apply{\treelab}{\treenode \idxi} =
            \nodelab{\config{\control'_\idxi}{\stackw_1}, \cpdsruler'}$, and
        \end{itemize}

    \item
        if $\treenode$ is a leaf node then $\apply{\treelab}{\treenode} =
        \nodelab{\configc}$ for some $\configc \in \langof{\saauta_0}$.
    \end{itemize}
\end{defi}

The following proposition gives us the required property of witness trees that
allows us to use them to prove soundness.

\begin{prop}
    For a CPDS $\cpds$, stack automaton $\saauta_0$ and configuration $\configc$
    of $\cpds$, if there is a witness tree for $\configc$, then $\configc \in
    \prestar{\cpds}{\saauta_0}$.
\end{prop}
\proof
    A straightforward induction beginning at the leaves of the witness tree.
\qed

\begin{algorithm}[H]
    \caption{\label{alg:counter-eg-extract}Counter-example Extraction}
    \begin{algorithmic}
        \REQUIRE{
            A stack-automaton $\saauta$ generated by saturating $\saauta_{0}$ and a configuration
            $\config{\control_{0}}{\stackw} \in \langof{\saauta}$.
        }
        \ENSURE{
            A witness tree from
            $\config{\control_{0}}{\stackw}$.
        }
        \STATE{Fix $\boldsymbol{\stackr}$ to be a trimmed accepting run of $\saauta$ over
        $\stackw$ from $\sastate_{\control_0}$}
        \RETURN{$\apply{\getcountereg}{\boldsymbol{\stackr}}$}
    \end{algorithmic}
\end{algorithm}

\begin{algorithm}[H]
    \caption{\label{alg:counter-eg-extract-rec}$\apply{\getcountereg}{\boldsymbol{\stackr}}$}
    \begin{algorithmic}
        \REQUIRE{A trimmed accepting run of $\saauta$ over a stack $\stacku$ from a control state $\control$.}
        \ENSURE{A witness tree for $\config{\control}{\stacku}$.}
        \IF{the head transition $\sat$ of $\boldsymbol{\stackr}$ is justified by $\unjust$}
            \RETURN{$\nodelab{\config{\control}{\stacku}}$}
        \ELSE
            \STATE{Let $\cpdsruler$ be
                  the CPDS rule appearing in the justification of $\sat$.}
            \STATE{Let $\nodelabel = \nodelab{\config{\control}{\stacku},
                                              \cpdsruler}$.}
            \IF{$\cpdsruler = (\control, \cha, \pop{\opord}, \control')$
                for some $1 \leq \opord \leq \cpdsord$}
               \STATE{The transition $\sat$ has $\cpdsord$-expansion
                      $\satranfull{\sastate_\control}
                                  {\cha}
                                  {\emptyset}
                                  {\emptyset, \dots, \emptyset,
                                   \set{\sastate_\opord},
                                   \sastateset_{\opord +1}, \dots, \sastateset_\cpdsord}$}
                \STATE{Suppose
                       $\apply{\ctop{1}}{\apply{\pop{\opord}}{\boldsymbol{\stackr}}} =
                        \tuple{\chb, \satset}$}
                \STATE{Pick
                       $\sastateset_{\opord},
                        \dots,
                        \sastateset_1,
                        \sastateset_{\branch}$
                       such that there is some
                       $\sat' \in \satset$
                       with the $\cpdsord$-expansion
                       $\satranfull{\sastate_{\control'}}
                                   {\chb}
                                   {\sastateset_{\branch}}
                                   {\sastateset_1, \dots, \sastateset_\opord,
                                    \sastateset_{\opord +1}, \dots, \sastateset_\cpdsord}$}
                \RETURN{$\treeap{\nodelabel}
                                {\apply{\getcountereg}
                                       {\apply{\rew{\tuple{\cha,\set{\sat'}}}}
                                              {\apply{\pop{\opord}}
                                                     {\boldsymbol{\stackr}}}}}$}
            \ELSIF{$\cpdsruler = (\control, \cha, \collapse{\opord}, \control')$
                   for $\opord \in \set{2, \ldots, \cpdsord}$}
                \STATE{The transition $\sat$ has the $\cpdsord$-expansion
                       $\satranfull{\sastate_\control}
                                   {\cha}
                                   {\set{\sastate_\opord}}
                                   {\emptyset, \dots, \emptyset,
                                    \sastateset_{\opord +1}, \dots, \sastateset_\cpdsord}$}
                \STATE{Suppose
                       $\apply{\ctop{1}}
                              {\apply{\collapse{\opord}}
                                     {\boldsymbol{\stackr}}} =
                        \tuple{\chb, \satset}$}
                \STATE{Pick
                       $\sastateset_{\opord},
                        \dots,
                        \sastateset_1,\sastateset_{\branch}$
                       such that there is
                       $\sat' \in \satset$
                       with $\cpdsord$-expansion
                       $\satranfull{\sastate_{\control'}}
                                   {\cha}
                                   {\sastateset_{\branch}}
                                   {\sastateset_1, \dots, \sastateset_{\opord},
                                    \sastateset_{\opord +1}, \dots, \sastateset_\cpdsord}$}
                \RETURN{$\treeap{\nodelabel}
                                {\apply{\getcountereg}
                                       {\apply{\rew{\tuple{\chb, \set{\sat'}}}}
                                              {\apply{\collapse{\opord}}
                                                     {\boldsymbol{\stackr}}}}}$}
            \ELSIF{$\cpdsruler = (\control, \cha, \rew{\chb}, \control')$
                   for some $\chb \in \alphabet$}
                \STATE{$\apply{\tjust}{\sat}$ must be of form $(\cpdsruler, t',i)$}
                \RETURN{$\treeap{\nodelabel}
                                {\apply{\getcountereg}
                                       {\apply{\rew{\tuple{\chb, \set{t'}}}}
                                              {\boldsymbol{\stackr}}}}$}
            \ELSIF{$\cpdsruler = (\control, \cha, \push{\opord}, \control')$}
                \STATE{$\apply{\tjust}{\sat}$
                       must be of the form
                       $(\cpdsruler, \sat', \satset , \idxi)$}
                \RETURN{$\treeap{\nodelabel}
                                {\apply{\getcountereg}
                                       {\apply{\rew{\tuple{\cha, \set{\sat'}}}}
                                              {\apply{\push{\opord}}
                                                     {\apply{\rew{\tuple{\cha, \satset}}}
                                                            {\boldsymbol{\stackr}}}}}}$}
            \ELSIF{$\cpdsruler = (\control, \cha, \cpush{\chb}{\opord}, \control')$}
                \STATE{$\apply{\tjust}{\sat}$
                       must be of the form
                       $(\cpdsruler, \sat', \satset , \idxi)$}
                \RETURN{$\treeap{\nodelabel}
                                {\apply{\getcountereg}
                                       {\apply{\rew{\tuple{\chb, \set{\sat'}}}}
                                              {\apply{\push{\opord}}
                                                     {\apply{\rew{\tuple{\cha, \satset}}}
                                                            {\boldsymbol{\stackr}}}}}}$}
            \ELSIF{$\cpdsruler = \cpdsalttran{\control}{\controlset}$}
                \STATE{$\apply{\tjust}{\sat}$
                       must be of the form
                       $(\cpdsruler, \satset, \idxi)$}
                \STATE{$\controlset$ is of the form
                       $\set{\control_1, \ldots, \control_\numof}$}
                \STATE{For each $\idxj$ we have a transition
                       $\sat_\idxj \in \satset$
                       with $\cpdsord$-expansion
                       $\satranfull{\sastate_{\control_\idxj}}
                                   {\cha}
                                   {\sastateset^\idxj_\branch}
                                   {\sastateset^\idxj_1,
                                    \ldots,
                                    \sastateset^\idxj_\cpdsord}$}
                \RETURN{$\treeap{\nodelabel}
                                {\apply{\getcountereg}
                                       {\apply{\rew{\tuple{\cha, \set{\sat_1}}}}
                                              {\boldsymbol{\stackr}}},
                                 \ldots,
                                 \apply{\getcountereg}
                                       {\apply{\rew{\set{\tuple{\cha, t_\numof}}}}
                                                   {\boldsymbol{\stackr}}}}$}
            \ENDIF
        \ENDIF
    \end{algorithmic}
\end{algorithm}

\subsubsection{The Algorithm}

Algorithm~\ref{alg:counter-eg-extract} and
Algorithm~\ref{alg:counter-eg-extract-rec} shows how we construct counter
examples from a given initial configuration
$\config{\control_{0}}{\stacku_{0}} \in \langof{\saauta}$.

An important notion in these algorithms is that of a \emph{trimmed} stack.
Intuitively, a trimmed run contains only useful transitions.
In particular, notice that the definition of a run may permit a substack
$\stackr =
 \ccompose{\annot{\tuple{\cha, \satset}}{\tuple{\opord, \idxi}}}
          {1}
          {\stackr'}$
where there is some
$\sat = \brac{
    \sastate
    \satrancol{\cha}{\sastateset_\branch}
    \sastateset
 }
 \in \satset$
and $\stackr'$ is not $\sastateset$-valid.
This may occur when $\sat$ is not needed to prove validity of the stack containing $\stackr$.
In other words, $\sat$ is redundant.
A trimmed stack does not contain such redundant transitions.

More formally, a run $\stackr$ is \emph{trimmed} if the following holds.
Take any sequence $o_{1}, \ldots, o_{j}$ of $\pop{}$ operations producing a subrun $\stackr'=o_{j}(\ldots o_{1}(\stackr)\ldots)$ and any transition
$\sastate_1 \satrancol{\cha}{\sastateset_\branch}{\sastateset_1}$
appearing in
$\ctop{1}(\stackr')$.
Let $\opord$ be the smallest index such that $\pop{\opord}$ appears in the sequence $o_{1}, \ldots, o_{j}$.
Since we used no $\pop{\opord'}$ with $\opord' < \opord$ the topmost stacks in $\stackr'$ up to order-$\opord$ are also topmost in $\stackr$ and must be read from initial states.
Thus we can obtain the $\opord$-expansion
\[
    \satranfull{\sastate_\opord}
               {\cha}
               {\sastateset_\branch}
               {\sastateset_1,\ldots,\sastateset_\opord}
\]
of the transition.
We require, for all $\opord' \in \set{1,\ldots,\opord}$, that
$\ctop{\opord'+1}(\pop{\opord}(\stackr'))$ is $Q_{\opord'}$-valid.

In the algorithms, variable $\boldsymbol{\stackr}$ contains a run of $\saauta$ which is
$\sastate_{\control}$-accepting for some state $\control$.  The initial value
of $\boldsymbol{\stackr}$, denoted $\stackr_{0}$, is an accepting run for the
initial configuration $\config{\control_{0}}{\stacku_{0}}$.  We construct
a witness tree recursively, with each recursive call building a
different branch of the witness tree.  At the beginning of each recursive call,
let $\stackr$ be the value of $\boldsymbol{\stackr}$ which we assume to be an
accepting run for a configuration $\config{\control}{\stacku}$.
Moreover, let $t$ denote the head transition of $\stackr$.

Each recursive call returns a witness tree from $\config{\control}{\stacku}$.
Moreover and crucially for termination, each recursive call with argument value
$\stackr'$ is such that $\stackr \runindexrel_{\cpdsord} \stackr'$.  As
$\runindexrel_{\cpdsord}$ is well-founded, the recursive calls eventually reach
the base case after a finite number of calls with the final $t$ justified by $\unjust$.
It will then be possible to prune the run $\stackr$ obtained in the base case to form a run that consists
entirely of transitions already belonging to $\saauta_{0}$.
This is by the assumption
that initial states at every order of $\saauta$ have no incoming transitions
and we only added transitions to the initial states of $\saauta_0$ (and to new states not in $\saauta_0$).
Thus, after the first transition of the run, we only use transitions from non-initial states of $\saauta$, which were necessarily already present in $\saauta_0$.
It follows that the configurations reached at the leaves of the witness tree
belongs to $\lang({\saauta_{0}})$.

\subsubsection{Correctness of the Algorithm}\label{subsubsection:correctness-Algorithm}

In this section, we establish the correctness of Algorithm~\ref{alg:counter-eg-extract} and give omitted proofs.
We start with the proof of Lemma~\ref{lemma:well-founded}.

\begin{proof}
For $\opord=1$, consider for any order-1 run $\stackr$ the tuple $|\stackr|=(n_{m},\ldots,n_{0})$ where $m$ is the step at which the saturation
algorithm terminates and for all $i \in \set{0,\ldots,m}$, $n_{i}$ is the number of occurrences  in $\stackr$ of transitions in $\sadelta_{1}$  justified at step $i$. The relation $\runindexrel_{1}$ can be equivalently defined as $\stackr \runindexrel_{1} \stackr'$ if $|\stackr'|$ is lexicographically smaller than $|\stackr|$. It immediately follows that $\runindexrel_{1}$ is well-founded.

For $\opord+1>1$ assuming the property holds for $\runindexrel_{k}$. Suppose for contradiction that $\runindexrel_{k+1}$ is not well-founded. Then there
must be an infinite chain of runs of the form:
\begin{equation*}
 \stackr_1 \runindexrel_{k+1} \stackr_2 \runindexrel_{k+1} \stackr_3
        \runindexrel_{k+1} \cdots
\end{equation*}
Now pick an index $i$ such that for every $j > i$ it is the case that $\stackr_j$ is at least as long (w.r.t the number of order-$(k-1)$ stacks) as the run $\stackr_i$
(infinitely many such indices must clearly exist since comparing runs by their lengths is a well-founded relation). If $\stackr_{i}= \ccompose{\stacku}{k+1}{\stackr_{i}'}$, it is a straightforward induction to see that for every $j > i$
$\stackr_{j}$ is of the form $\stackr_{j}'' \stackr_{i}'$ with $\stacku \runindexrel_{\opord}^{+} \stackv$ for all order-$\opord$ runs $v$ occurring in $\stackr_{j}''$ where $\runindexrel_{\opord}^{+}$ designates the transitive closure of $\runindexrel_{\opord}$.

So in particular if we pick infinitely many positions in the chain $i_\ell$ such
that the run $\stackr_{i_\ell}= \ccompose{\stacku_{i_{\ell}}}{k+1}{\stackr'_{i_\ell}}$ is at least as long as the sequence
$\stackr_j$ for all $j > i_\ell$ it must be the case that:
\begin{equation*}
        \stacku_{i_1} \;\;{\runindexrel_{k}}^{+} \;\; \stacku_{i_2} {\runindexrel_{k}}^+ \;\;
                        \stacku_{i_3} \;\; {\runindexrel_{k}}^{+} \cdots
\end{equation*}
This in turn contradicts the fact that $\runindexrel_{k}$ is well-founded.
\end{proof}

The next lemma describes two sufficient conditions condition
for $\stackr \runindexrel_{k} \stackr'$ to hold.

\begin{lem}
\label{lem:technical-runrel}
The following properties hold:
\begin{enumerate}
\item Let $w$ and $w'$ be two order-$n$ runs such that for some
      $k \in \set{1, \ldots, n-1}$,
      $\ctop{k+1}(w) \runindexrel_{k} \ctop{k+1}(w')$ and $\pop{k+1}(w)=\pop{k+1}(w')$ then $\stackr \runindexrel_{n} \stackr'$.
\item Let $w$ be an order-$k$ run and let $T$ be a set of transitions that is smaller than some transition appearing in $\ctop{1}(w)$.
That is,
$\apply{\ctop{1}}{w} = \tuple{\cha, \satset'}$
and there is some $\sat \in \satset'$ such that
$\sbrac{\set{\sat}}{1} \runindexrel_{1} \sbrac{\set{T}}{1}$.
Then, we have
$\stackr \runindexrel_{k} \apply{\rew{\tuple{\cha, T}}}{w}$
for any $\cha$.
\end{enumerate}
\end{lem}
\proof
    For the first property, we will show by induction on $k'$ that for all $k'
    \in \set{k,\ldots,n}$, $\ctop{k'+1}(w) \runindexrel_{k'} \ctop{k'+1}(w')$. The case
    $k'=k$ is assumed to hold in the hypothesis. Assume that the property holds
    for $k' < \cpdsord$.
    We show that it holds for $k' + 1$.
    We have
    $\ctop{k'+2}(w)=\ccompose{\ctop{k'+1}(w)}{(k'+1)}{\ctop{k'+2}(\pop{k'+1}(w))}$
    and
    $\ctop{k'+2}(w')=\ccompose{\ctop{k'+1}(w')}{(k'+1)}{\ctop{k'+2}(\pop{k'+1}(w'))}$.
    Observe that $\pop{k'+1}(w)=\pop{k'+1}(w')=u$. This is by assumption for
    $k=k'$ and if $k'>k$ then
    $\pop{k'+2}(w)=\pop{k'+2}(\pop{k+1}(w))=\pop{k'+2}(\pop{k+1}(w'))=\pop{k'+2}(w')$.
    Hence  $\ctop{k'+2}(w)=\ccompose{\ctop{k'+1}(w)}{(k'+1)}{u}$ and
    $\ctop{k'+2}(w')=\ccompose{\ctop{k'+1}(w')}{(k'+1)}{u}$ with $\ctop{k'+1}
    \runindexrel_{k'} \ctop{k'+1}(w')$. By definition of
    $\runindexrel_{k'+1}$, we have $\ctop{k'+2} \runindexrel_{k'+1}
    \ctop{k'+2}(w')$.

    For the second property, we have
    $\apply{\ctop{2}}{w}
     \runindexrel_{1}
     \apply{\ctop{2}}
           {\apply{\rew{\tuple{\cha, T}}}{w}}$
    (by definition of $\runindexrel_{1}$) and
    $\pop{2}(w)=\pop{2}(w')$. Hence by the first-property $\stackr \runindexrel_{k}
    \apply{\rew{\tuple{\cha, T}}}{w}$.
\qed

We now prove Algorithm~\ref{alg:counter-eg-extract} is correct.

\begin{prop}
\label{prop:correctness-counter-eg-extract}
Algorithm~\ref{alg:counter-eg-extract} is correct.
\end{prop}
\proof
    The initial value of $\boldsymbol{\stackr}$, denoted $\stackr_{0}$, is an
    accepting run for the initial configuration
    $\config{\control_{0}}{\stacku_{0}}$. An updated value of
    $\boldsymbol{\stackr}$ is passed at each recursive call. We denote by
    $\stackr$ the value of $\boldsymbol{\stackr}$ at the beginning of each
    call.

    We are going to prove by induction on the depth of recursion that $\stackr$
    is always a trimmed $\sastate_{\control}$-accepting run on some stack $\stacku$.
    Furthermore, for each recursive call we have $w \runindexrel_{\cpdsord} w'$
    where $w'$ is the value passed to the call.

    In the base case, before any calls, we had assumed $\stackr$ to be a trimmed run.
    Next, assume that the property holds for $\stackr$, and let us prove it
    for each $\stackr'$ appearing in a recursive call.  By the induction
    hypothesis, $\stackr$ is a trimmed $\sastate_{\control}$-accepting run
    on a stack $\stacku'$. This implies that its head transition $t$ has a $\cpdsord$-expansion
    \[
        \satranfull{\sastate_{\control}}
                   {\cha}
                   {\sastateset_\branch}
                   {\sastateset_1,\ldots,\sastateset_\cpdsord} \ .
    \]
    Hence its justification contains a transition of the CPDS of the form
    $(\control,\cha,o,\control')$ or $\cpdsalttran{\control}{\controlset}$.
    In the first case, we reason by case distinction on the operation $o$.

    \textbf{Case $o=\rew{b}$ for some $b \in \Sigma$.}
    The transition $t$ has a justification of the form $J(t)=(r,t',i)$ with $t'$ having $\cpdsord$-expansion
    $\sastate_{\control'} \satrancol{\cha}{Q_{\branch}}(Q_{1}, \dots, \sastateset_\cpdsord)$.
    Note that $t'$ was introduced before $t$.

    The run $w'$ is equal to
    $\apply{\rew{\tuple{\chb, \set{t'}}}}{\boldsymbol{\stackr}}$.
    It is clear that $w'$ is a trimmed $\sastate_{\control'}$-accepting run on the
    stack $\rew{b}(\stacku)$. By the second property of Lemma~\ref{lem:technical-runrel}, $\stackr \runindexrel_{n} \stackr'$.

    \textbf{Case $o=\pop{k}$ for some $k \in \set{1,\ldots,n}$.}
     The transition $t$ has $\cpdsord$-expansion
    \[
        \satranfull{\sastate_\control}
                   {\cha}
                   {\emptyset}
                   {\emptyset, \dots, \emptyset,
                    \set{\sastate_\opord},
                    \sastateset_{\opord +1}, \dots, \sastateset_\cpdsord}.
    \]
    As $w$ is $\sastate_{\control}$-accepting, it follows that
    for all $j \in \set{k+1,\ldots,n}$, $\ctop{j+1}(\pop{j}(w))$ is $Q_{j}$-valid
    and that $\ctop{k+1}(\pop{k}(\stackr))$ is $\set{q_\opord}$-valid.
    Since $t$ was introduced when processing a pop operation, the state $\sastate_\opord$ has a $\cpdsord$-expansion
    $\satranfullk{\sastate_{\control'}}
                 {\sastate_\opord}
                 {\sastateset_{\opord+1}, \ldots, \sastateset_\cpdsord}$.
    Then, by unfolding the notion of $\set{q_\opord}$-validity, we obtain that $\ctop{1}(\pop{k}(w))$ contains for some
    $\sastateset_1, \ldots, \sastateset_\opord$
    at least one transition $t'$ with $\cpdsord$-expansion
    \[
        \satranfull{\sastate_{\control'}}
                   {\cha}
                   {\sastateset_\branch}
                   {\sastateset_1, \dots, \sastateset_\opord,
                    \sastateset_{\opord +1}, \dots, \sastateset_\cpdsord} \ .
    \]
    Let $t'$ be the transition of this form picked by the algorithm.
    As $w$ is trimmed it follows that for all $j \in \set{1,\ldots,k}$, $\ctop{j+1}(\pop{j}(\pop{k}(w)))$ is $Q_{j}$-valid.

    We have
    $w'= \apply{\rew{\tuple{\cha,\set{\sat'}}}}
               {\apply{\pop{\opord}}
                      {\boldsymbol{\stackr}}}$.
    Recall $w$ is link-valid and trimmed.
    Observe $w'$ is a subrun of $w$ in the following sense:
    $\pop{k}(\stackr)$ is a substack of $w$ and $t'$ is contained in $\ctop{1}(w)$.
    Thus, $w'$ is also link-valid and trimmed. To prove that it is also $\sastate_{\control'}$-valid it
    is enough to show that for all $i \in \set{1,\ldots,n}$, we have $\ctop{i+1}(\pop{i}(\stackr'))$ is $Q_{i}$-valid.
    For $i \in \set{k+1,\ldots,n}$, we have seen  that $\ctop{i+1}(\pop{i}(\stackr'))=
    \ctop{i+1}(\pop{i}(\stackr))$ is $Q_{i}$-valid.
    For $i \in \set{1,\ldots,k}$, we have seen that $\ctop{i+1}(\pop{i}(\stackr'))=
    \ctop{i+1}(\pop{i}(\pop{k}(\stackr))$ is $Q_{i}$-valid.

    It only remains to show that $\stackr \runindexrel_{n} \stackr'$.
    By the first  property of Lemma~\ref{lem:technical-runrel}, it is
    enough to show that $\ctop{k+1}(w) \runindexrel_{k} \ctop{k+1}(w')$ (as
    $\pop{k+1}(w')=\pop{k+1}(w)$ if $k<n$). First consider the case
    when $k=1$. That
    $\stackr \runindexrel_{n} \stackr'$
    follows from the fact that the set of order-1 transitions appearing in $\ctop{2}(w')$ is strictly included in $\ctop{2}(w)$. Now assume
    that $k>1$. The run $\ctop{k+1}(w)$ can be written as $\ccompose{u}{k+1}{\ccompose{u'}{k+1}{v}}$  and $\ctop{k+1}(w')=\ccompose{\rew{\tuple{\cha, \set{t'}}}(u')}{k+1}{v}$.
    To show
    $\ctop{k+1}(w) \runindexrel_{k} \ctop{k+1}(w')$
    there are two cases according to the definition of $\runindexrel_k$.
    If $\rew{\tuple{\cha, \set{t'}}}(u') = u'$ then we get
    $\ctop{k+1}(w) \runindexrel_{k} \ctop{k+1}(w')$.
    Otherwise, $\set{t'}$ is a strict subset of the set of transitions in $\ctop{1}(u')$ giving
    $\ctop{2}(u') \runindexrel_1 \ctop{2}(\rew{\tuple{\cha, \set{t'}}}(u'))$
    and since
    $\pop{2}(\rew{\tuple{\cha, \set{t'}}})(u') = \pop{2}(u')$
    we get from Lemma~\ref{lem:technical-runrel}
    $u' \runindexrel_{k-1} \rew{\tuple{\cha, \set{t'}}}(u')$.
    Then by definition of $\runindexrel_{k}$, $\ctop{k+1}(w) \runindexrel_{k} \ctop{k+1}(w')$.

    \textbf{Case $o=\collapse{k}$ for some $k \in \set{2,\ldots,n}$.} This case is similar to the $\pop{k}$ case.

    \textbf{Case $o=\push{k}$ for some $k \in \set{2,\ldots,n}$.}
    The transition $t$ has $\cpdsord$-expansion
    \[
        \satranfull{\sastate_{\control}}{\cha}{\sastateset_\branch
        \cup \sastateset'_\branch}{\sastateset_1 \cup \sastateset'_1,
        \ldots, \sastateset_{\opord-1} \cup \sastateset'_{\opord-1},
        \sastateset'_\opord, \sastateset_{\opord+1}, \ldots,
        \sastateset_\cpdsord}
    \]
    with $J(t)=\tuple{\cpdsruler, \sat', \satset,
                  \idxi+1}$ where
    $\sat'$ has $\cpdsord$-expansion
    \[
        \brac{\satranfull{\sastate_{\control'}}{\cha}{\sastateset_\branch}{\sastateset_1,
                  \ldots,\sastateset_\opord,\ldots, \sastateset_\cpdsord}}
    \]
    and $T$ is a set of transitions with strict $(\opord, \sastateset_\opord)$-expansion
    $\satranfull{\sastateset_\opord}{\cha}{\sastateset'_\branch}{\sastateset'_1,
                  \ldots, \sastateset'_\opord}$.

    The run $w'$ is equal to
    $\apply{\rew{\tuple{\cha, \set{\sat'}}}}
           {\apply{\push{\opord}}
                  {\apply{\rew{\tuple{\cha, \satset}}}
                         {\boldsymbol{\stackr}}}}$.
    Let $\stackr = \ccompose{\stacku}{k}{\stackv}$. The run $w'$ is then equal to
    $\ccompose{\rew{\tuple{\cha, \set{t'}}}(u)}
              {k}
              {\ccompose{\rew{\tuple{\cha, T}}(u)}{k}{\stackv}}$.

    Let us first show that $w'$ is $\set{\sastate_{\control'}}$-valid. For this it is enough
    to show that:
    \begin{itemize}
    \item for all $k' \in \set{k+1,\ldots,n}$, $\ctop{k'+1}(\pop{k'}(w'))=\ctop{k'+1}
    (\pop{k'}(w))$ is $Q_{k'}$-valid. This immediately follows from the fact that
    $w$ is $\sastate_\control$-accepting with head transition $t$.
    \item
    $\ctop{k+1}(\pop{k}(w'))=\rew{\tuple{\cha, T}}(u)=\ctop{k+1}(\rew{\tuple{\cha, T}}(w))$ is
    $Q_{k}$-valid. As $T$ has the strict $(\opord, \sastateset_\opord)$-expansion
    $\satranfull{\sastateset_\opord}
                {\cha}
                {\sastateset'_\branch}
                {\sastateset'_1, \ldots, \sastateset'_\opord}$,
    it enough to show that for all $k' \in \set{1,\ldots,k}$, we have
    $\ctop{k'+1}(\pop{k'}(\rew{\tuple{\cha, T}}(w)))=\ctop{k'+1}(\pop{k'}(w))$
    is $Q_{k'}$-valid.
    This immediately follows from the fact that
    $w$ is $\sastate_{\control}$-accepting with head transition $t$.
    \item for all $k' \in \set{1,\ldots,k-1}$,  $\ctop{k'+1}(\pop{k'}(w'))=\ctop{k'+1}
    (\pop{k'}(w))$ is $Q_{k'}$-valid. This immediately follows from the fact that
    $w$ is $\sastate_\control$-accepting with head transition $t$.
    \end{itemize}

    We now show that $w'$ is link-valid. We only need to check the validity
    for the substack $\ccompose{\rew{\tuple{\cha, T}}(u)}{k}{\stackv}$ and the substacks
    of the form $\ccompose{u'}{k}{\ccompose{\rew{\tuple{\cha, T}}(u)}{k}{\stackv}}$ where $u'$
    is a substack of $\rew{\tuple{\cha, \set{t'}}}(u)$. Let us first consider
    the stack $\ccompose{\rew{\tuple{\cha, T}}(u)}{k}{\stackv}$ and let $h$ be a transition
    in $T$ with $\cpdsord$-expansion
    \[
    \satranfull{\sastate_{h}}{\cha}{\sastateset_\branch^{h}}{\sastateset_1^{h},
                 \ldots, \sastateset_\cpdsord^{h}}.
    \]
    We have that $\sastateset_{\branch}^{h}$ is a subset of $Q'_{\branch}$. Let $k'$ be the order of the link on top of $\ccompose{\rew{\tuple{\cha, T}}(u)}{k}{\stackv}$.
    As $w$ is link-valid, we know that $\ctop{k'+1}(\collapse{k'}(w))=\ctop{k'+1}(\collapse{k'}(\ccompose{\rew{\tuple{\cha, T}}(u)}{k}{\stackv}))$ is $Q_{\branch} \cup Q_{\branch}'$-valid hence it is also $Q_{\branch}^{h}$-valid.
    We now move on to the case of $x=\ccompose{\rew{\tuple{\cha, \set{t'}}}(u)}{k}{\ccompose{\rew{\tuple{\cha, T}}(u)}{k}{\stackv}}$.
    Let $k'$ be the order of the link on top of $x$. We have that
    $\ctop{k'+1}(\collapse{k'}(x))=
    \ctop{k'+1}(\collapse{k'}(w))$. By link-validity of $w$, it is the case that $\ctop{k'+1}(\collapse{k'}(w))$ is $Q_{\branch} \cup Q_{\branch}'$-valid and in particular $Q_{\branch}$-valid.

    Finally let $u'$ be a strict substack of $\rew{\tuple{\cha, \set{t'}}}(u)$.
    Let $k'$ be the order of the link appearing on top of $x=\ccompose{u'}{k}{\ccompose{\rew{\tuple{\cha, T}}(u)}{k}{\stackv}}$ and let $h$ be a transition attached to the top of $x$ with $\cpdsord$-expansion
    \[
    \satranfull{\sastate_{h}}{\cha}{\sastateset_\branch^{h}}{\sastateset_1^{h},
                 \ldots, \sastateset_\cpdsord^{h}}.
    \]

    We have that
    $\ctop{k'+1}(\collapse{k'}(x))=
    \ctop{k'+1}(\collapse{k'}(w))$. By link-validity of $w$, it is the case that $\ctop{k'+1}(\collapse{k'}(w))$ is $Q_{\branch}^{h}$-valid.

    It now remains to show that $w'$ is trimmed. The only interesting case
    is that of the substack $\ccompose{\rew{\tuple{\cha, T}}(u)}{k}{\stackv}$ which is
    reached by a $\pop{k}$ operation. Any transition $h \in T$ has $\cpdsord$-expansion
    \[
    \satranfull{\sastate_{h}}{\cha}{\sastateset_\branch^{h}}{\sastateset_1^{h},
                 \ldots, \sastateset_\cpdsord^{h}}
    \]
    with for all $k' \in \set{1,\ldots,k}$, $Q_{k'}^{h} \subseteq Q_{k'}'$. Hence it is
    enough for us to show that for all $k' \in \set{1,\ldots,k}$, $\ctop{k'+1}(\pop{k'}(\ccompose{\rew{\tuple{\cha, T}}(u)}{k}{\stackv}))=\ctop{k'+1}(\pop{k'}(w))$ is $Q'_{k}$-valid. This immediately follows from the fact that
    $w$ is $q$-accepting with head transition $t$.

    It only remains to show that $\stackr \runindexrel_{n} \stackr'$. First observe that
    $u \runindexrel_{k-1} \rew{\tuple{\cha, \set{t'}}}(u)$ and $u \runindexrel_{k-1} \rew{\tuple{\cha, T}}(u)$
    as in both cases $t$ is replaced by one or several transitions with a smaller
    timestamp (cf. second property of Lemma~\ref{lem:technical-runrel}). By definition of $\runindexrel_{k}$, we have $\ctop{k+1}(w) \runindexrel_{k} \ctop{k+1}(w')$. The first property of Lemma~\ref{lem:technical-runrel} then
    implies that $ \stackr \runindexrel_{k} \stackr'$.

    \textbf{Case $o=\cpush{b}{k}$ for some $\chb \in \Sigma$ and $k \in \set{2,\ldots,n}$.} This case is similar to the $\push{k}$ case.

    This concludes the case where the justification contains a transition of
    the form $(\control,\cha,o,\control')$.  When it is of the form
    $\cpdsalttran{\control}{\controlset}$ then the transition $\sat$ has $\cpdsord$-expansion
    $
        \sastate_{\control}
        \satrancol{\cha}{Q_{\branch}}(Q_{1}, \dots, \sastateset_\cpdsord)
    $
    with a justification of the form $\apply{\tjust}{\sat} =
    \tuple{\cpdsruler,\satset, \idxi}$ with $\satset$ having the strict $(\cpdsord, \sastateset)$-expansion
    $
        \satranfull{\sastateset}
                   {\cha}
                   {\sastateset_\branch}
                   {\sastateset_1, \ldots, \sastateset_\cpdsord}
    $
    for
    $\sastateset = \setcomp{\sastate_{\control'}}{\control' \in \controlset}$.
    Thus for each $\control' \in \controlset$ we have some transition $\sat'$ with $\cpdsord$-expanion
    $\satranfull{\sastate_{\control'}}
                {\cha}
                {\sastateset'_\branch}
                {\sastateset'_1, \ldots, \sastateset'_\cpdsord}$
    and
    $\sastateset'_\opord \subseteq \sastateset_\opord$
    for all
    $\opord \in \set{1, \ldots, \cpdsord}$.
    Note that $\sat'$ was introduced before $\sat$.
    The run $\stackr'$ in the corresponding recursive call is equal to
    $\rew{\tuple{\cha, \set{\sat'}}}(\boldsymbol{\stackr})$.  It is clear that
    $\stackr'$ is a trimmed $\sastate_{\control'}$-accepting run on the stack
    $\stacku$. By the second property of Lemma~\ref{lem:technical-runrel},
    $\stackr \runindexrel_{n} \stackr'$.

    In all cases, the recursive call is made with a smaller stack.  Since
    $\runindexrel_\cpdsord$ is well-founded, we eventually reach the base case
    of the recursion.  Thus the algorithm terminates.

    That the algorithm returns a witness tree can be proven by induction from
    the leaves of the recursion back to the beginning of the algorithm.  It is
    immediate in the base case, since a justification of $\unjust$ implies that
    the configuration is accepted by $\saauta_0$.  When a rule of the form
    $\cpdsrule{\control}{\cha}{\genop}{\control'}$ leads to the recursive call
    we know by induction that we obtain a witness tree for
    $\config{\control'}{\apply{\genop}{\stacku}}$.  By adding $\nodelabel$ as
    the root of this tree, we immediately get a witness tree for
    $\config{\control}{\stacku}$.  The remaining case is when a rule
    $\cpdsalttran{\control}{\controlset}$ is used.  For each $\control' \in
    \controlset$ we obtain a witness tree for $\config{\control'}{\stacku}$.
    By constructing the tree with $\nodelabel$ at the root and children from
    each of the recursive calls, we have a witness tree for
    $\config{\control}{\stacku}$ as required.
\qed

%% file: forward.tex
\section{Initial Forward Analysis}
\label{section:optimisations} \label{sec:contrib-first}

It is generally completely impractical to compute
$\prestar{\cpds}{\saauta_0}$ in full (most non-trivial examples considered
in our experiments would time-out).  For our saturation algorithm to be usable
in practice, it is therefore essential that the search space is restricted,
which we achieve by means of an initial forward analysis of the CPDS.  In
short, we compute an over-approximation of all reachable configurations, and
try to restrict our backwards reachability analysis to only include
configurations in this over-approximation.

In this section we make two assumptions about the reachability problem.
First we assume we are given an initial configuration, which for simplicity has the form
$\config{\control_0}{\sbrac{\cdots\sbrac{\cha_0}{1}\cdots}{\cpdsord}}$
(we can always adjust the CPDS to construct a different stack during the initial moves).
Next, we distinguish an \emph{error state}  $\qerror$ and  we are
interested only in whether $\cpds$ can reach a configuration of the form
$\config{\qerror}{\stackw}$.  That is, the set of target configurations is
$\setcomp{\config{\qerror}{\stackw}}
         {\stackw \in \stacksalpha{\cpdsord}{\alphabet}
          \text{ and $\apply{\ctop{1}}{\stackw}$ is defined}}$.
The condition on $\ctop{1}$ ensures a stack automaton can be defined to accept the set.
This suffices to capture the same safety (reachability) properties of recursion
schemes as \trecs{}~\cite{K09b}.

We fix a stack-automaton $\errorconfigs$ recognising all
error configurations (those with the state $\qerror$).  We write $\poststarraw$
for the set of configurations reachable by $\cpds$ from the initial
configuration $\configc_0$.
More formally, $\poststarraw$ is the
smallest set such that $\configc_0 \in \poststarraw$ and
\[
    \poststarraw \supseteq \setcomp{\config{\control'}{\stackw'}}{%
        \begin{array}{l}%
            \exists
            \config{\control}{\stackw} \in \poststarraw
            \;\textrm{with\;}
            \brac{
                \begin{array}{l}
                    \config{\control}{\stackw} \cpdstran%
                    \config{\control'}{\stackw'} \ \lor
                    \\ %
                    \config{\control}{\stackw} \cpdstran
                    \configset\;\text{and}\; \config{\control'}{\stackw'} \in
                    \configset
                \end{array}
            }
        \end{array}%
    } \ .%
\]%
This set cannot be represented precisely by a stack
automaton~\cite{BM04} (for instance using $\push{2}$, we can create
$\sbrac{\sbrac{a^{n}}{1}\sbrac{a^{n}}{1}}{2}$ from $\sbrac{\sbrac{a^{n}}{1}}{2}$
for any  $n \geq 0$). We summarise our approach then give details in
Sections~\ref{sec:guarded}, \ref{sec:approx-graph} and~\ref{sec:extract}.
Note that the handling of alternating transitions in $\poststarraw$ allows
us to treat these transitions in our approximation algorithms in the same
way as if the choice were non-deterministic rather than alternating.

Ideally
we would compute only $\prestar{\cpds}{\errorconfigs} \cap \poststarraw$.  Since
this cannot be represented by an automaton, we instead compute a sufficient approximation $T$
(ideally a \emph{strict} subset of  $\prestar{\cpds}{\errorconfigs}$) where:
\begin{equation*}
  \prestar{\cpds}{\errorconfigs} \; \cap \; \poststarraw\; \subseteq \; T \;
  \subseteq \; \prestar{\cpds}{\errorconfigs}.
\end{equation*}
The initial configuration will belong to $T$ iff it can reach a configuration
recognised by $\errorconfigs$.  Computing such a $T$ is much more feasible.

The first step in obtaining a sufficient approximation $T$ is to compute an over-approximation of $\poststarraw$ (Section~\ref{sec:approx-graph}).
For this we use a \emph{summary algorithm} \cite{SP81}
(that happens to be precise at order-$1$).
From this over-approximation we extract a further
over-approximation of the set of CPDS rules that may be used on a
run to $\qerror$.  Let $\cpds'$ be the (smaller) CPDS containing only these
rules.  I.e., we remove all rules that we know cannot appear on a run to
$\qerror$.  We could thus take $T = \prestar{\cpds'}{\errorconfigs}$ (computable
by saturation for $\cpds'$) since it satisfies the conditions
above. This is what we meant by \emph{`pruning'} the CPDS (1a on
page \pageref{enum:contributions})

However, we further improve performance by computing an even smaller $T$ (1b in the list
on page \pageref{enum:contributions}).  We
extract contextual information from our over-approximation of $\poststarraw$
about how pops and collapses might be used during a run to $\qerror$ (Section~\ref{sec:extract}).
Our $\cpds'$ is then restricted to a model $\cpds''$ that `guards' its rules by
these contextual constraints.
Taking $T = \prestar{\cpds''}{\errorconfigs}$ we
have a $T$ smaller than $\prestar{\cpds'}{\errorconfigs}$, but still satisfying
our sufficient conditions.  In fact, $\cpds''$ will be a \emph{`guarded CPDS'}
(Section~\ref{sec:guarded}).
Computing
$\prestar{\cpds''}{\errorconfigs}$
precisely for a guarded CPDS is likely to be impractical
(as described at the end of Section~\ref{sec:guarded}).
Instead, we adjust saturation to compute $T$ such that $\prestar{\cpds''}{\errorconfigs} \;
\subseteq \; T \; \subseteq \; \prestar{\cpds'}{\errorconfigs}$.  This set will
thus also satisfy our sufficient conditions.

\subsection{Guarded Destruction}
\label{sec:guarded}

An \emph{order-$\cpdsord$ \emph{guarded} CPDS ($\cpdsord$-GCPDS)} is an
$\cpdsord$-CPDS where conventional $\pop{\opord}$ and $\collapse{\opord}$
operations are replaced by \emph{guarded operations} of the form
$\popg{\opord}{\alphaset}$ and $\collapseg{\opord}{\alphaset}$ where $\alphaset
\subseteq \alphabet$. These operations may only be fired if the resulting stack
has a member of $\alphaset$ on top. That is, for
$\genop \in \setcomp{\collapse{\opord}}{\opord \in \set{2,\ldots,\cpdsord}}$
or
$\genop \in \setcomp{\pop{\opord} }{\opord \in \set{1,\ldots,\cpdsord}}$:
\begin{equation*}
    \genop^\alphaset(\stacku) =
	\begin{cases}
	  \genop(\stacku) & \textrm{ if } \genop(\stacku) \textrm{ defined and } \ctop{1}(\genop(\stacku)) \in \alphaset\\
	  \textrm{undefined} & \textrm{ otherwise . }
	\end{cases}
\end{equation*}
Note, we do not guard the other stack operations since these themselves
guarantee the symbol on top of the new stack (e.g. when a transition
$(\control, \cha, \push{2}, \control')$ fires it must always result in a stack with
$\cha$ on top, and $(\control, \cha, \cpush{k}{\chb}, \control')$ produces a stack
with $\chb$ on top).
Observe also that guarded operations cannot empty the topmost stack.
This is reasonable since, by our assumptions on stack automata, once the topmost stack is empty, there is no hope of reaching a target configuration.

For a GCPDS $\cpds$, we write $\trivialise{\cpds}$ for the \emph{trivialisation} of $\cpds$: the ordinary CPDS obtained by replacing each
$\popg{\opord}{S}$ (resp. $\collapseg{\opord}{S}$) in the rules of $\cpds$ with
$\pop{\opord}$ (resp. $\collapse{\opord}$).
Non-trivial guards
reduce the size of the stack-automaton constructed by avoiding
additions that are only relevant for unreachable (and hence uninteresting)
configurations in the pre-image.  Thus, we improve performance.

We modify the saturation algorithm to use `guarded' saturation steps for pop and
collapse rules.
Note, the justifications remain unchanged, as do the other saturation steps.
\begin{enumerate}
        \item when $\genop = \popg{\opord}{\alphaset}$, for each order-$\opord$ state $\sastate_\opord$ with $\cpdsord$-expansion
              $\satranfullk{\sastate_{\control'}}{\sastate_\opord}{\sastateset_{\opord+1},
              \dots, \sastateset_\cpdsord}$ in $\saauta$ such that
              there is an order-$1$ transition with $\opord$-expansion $\satranfull{\sastate_k}{\chb}{\_}{\_, \dots, \_}$ in $\saauta$
              such that $\chb \in \alphaset$, add the transitions
              $\satranfull{\sastate_{\control}}{\cha}{\emptyset}{\emptyset,
              \ldots, \emptyset, \set{\sastate_\opord}, \sastateset_{\opord+1},
              \ldots, \sastateset_\cpdsord}$ to $\saauta'$,

	    \setcounter{enumi}{2} 
        \item when $\genop = \collapseg{\opord}{\alphaset}$, for each order-$\opord$ state with $\cpdsord$-expansion
              $\satranfullk{\sastate_{\control'}}{\sastate_\opord}{\sastateset_{\opord+1},
              \dots, \sastateset_\cpdsord}$ in $\saauta$  where
              there is an order-$1$ transition with $\opord$-expansion
              $\satranfull{\sastate_k}{\chb}{\_}{\_, \dots, \_}$ in $\saauta$
              with $\chb \in \alphaset$, add the transitions
              $\satranfull{\sastate_{\control}}{\cha}{\set{\sastate_\opord}}{\emptyset,
              \ldots, \emptyset, \sastateset_{\opord+1}, \ldots,
              \sastateset_\cpdsord}$ to $\saauta'$.
\end{enumerate}
E.g., suppose that an ordinary (non-guarded) $2$-CPDS has rules $(\control_1,
\chc, \collapse{2}, \control)$ and $(\control_2, \chd, \collapse{2},
\control')$.  The \emph{original} saturation algorithm would process these rules
to add the transitions:
$\satranfull{\sastate_{\control_1}}{\chc}{\set{\sastate_\control}}{\emptyset, \emptyset}$
and
$\satranfull{\sastate_{\control_2}}{\chd}{\set{\sastate_{\control'}}}{\emptyset,  \emptyset}$.

Now suppose that the saturation algorithm has produced two transitions with $\cpdsord$-expansions
$\satranfull{\sastate_\control}{\cha}{\_}{\_, \_}$ and
$\satranfull{\sastate_{\control'}}{\chb}{\_}{\_,  \_}$.  If a \emph{G}CPDS had,
for example, the rules $(\control_1, \chc, \collapse{2}^{\set{\cha}}, \control)$
and $(\control_2, \chd, \collapse{2}^{\set{\chb}}, \control')$, then these same
two transitions would be added by the modified saturation algorithm.  On the
other hand, the rule $(\control_1, \chc, \collapse{2}^{\set{\cha}}, \control)$
and the rule $(\control_2, \chd, \collapse{2}^{\set{\cha}}, \control')$ would only result
in the first of the two transitions being added.

\begin{lem}
    \label{lemma:guardedCorrect}
    The revised saturation algorithm applied to $\errorconfigs$ (for a
    \emph{G}CPDS $\cpds$) gives a stack automaton recognising $T$ such that
    $\prestar{\cpds}{\errorconfigs} \; \subseteq \; T \; \subseteq \;
    \prestar{\trivialise{\cpds}}{\errorconfigs}$
    \qed
\end{lem}
\proof
  We can see that $T \; \subseteq \; \prestar{\trivialise{\cpds}}{\errorconfigs}$ since every time we can add a transition
  during the modified saturation algorithm we could have added the corresponding guard-free rule in the original algorithm, and the original algorithm
  is already known to be sound.

  Checking that
  $\prestar{\cpds}{\errorconfigs} \; \subseteq \; T$ is an easy modification of the completeness proof for the original
  algorithm in Lemma~\ref{lem:completeness}. This works by induction on the length of a path from a configuration in
  $\prestar{\cpds}{\errorconfigs}$ to one in $\errorconfigs$. Suppose we have a stack-automaton $\saauta$ recognising
  a configuration $\config{\control'}{\stacku'}$ together with a rule $(\control, \cha, \genop^S, \control')$ of $\cpds$ where $\genop$ is
  either a \emph{pop} or a \emph{collapse} operation. Suppose that $\config{\control}{\stacku}$ can reach $\config{\control'}{\stacku'}$
  in a single step via this rule. By definition it must then be the case that $\ctop{1}(\stacku') = \chb$ for some $\chb \in S$
  (and also that $\stacku' = \genop(\stacku)$). But then the run recognising $\config{\control'}{\stacku'}$ must have a head transition with $\opord$-expansion
  $\satranfull{\sastate_\opord}{\chb}{\sastateset_\branch}{\sastateset, \ldots, \sastateset_\opord}$
  and $\sastate_\opord$ has the $\cpdsord$-expansion
  \hbox{$\satranfullk{\sastate_{\control'}}{\sastate_\opord}{\sastateset_{\opord+1}, \dots, \sastateset_\cpdsord}$}.
  By taking this
  $\sastate_{\opord}$
  we can see that applying the step for the operation $\genop^S$ in the revised saturation algorithm will
  create a stack-automaton recognising  $\stacku$.
\qed

The reason that the algorithm may result in a stack-automaton recognising configurations that do not belong to
$\prestar{\cpds}{\errorconfigs}$ (albeit still in $\prestar{\trivialise{\cpds}}{\errorconfigs}$)
is as follows.
To take account of the guards, we check before adding a transition because of $\sastate_\opord$ with $\cpdsord$-expansion
$\satranfullk{\sastate_{\control'}}
             {\sastate_\opord}
             {\sastateset_{\opord+1}, \ldots, \sastateset_\cpdsord}$
that there is a transition with $\opord$-expansion
$\satranfull{\sastate_k}{\chb}{\_}{\_, \dots, \_}$
with $\chb$ appearing in the guard of the rule being processed.
However, since there may also exist a transition with $\opord$-expansion
$\satranfull{\sastate_k}{\chb'}{\_}{\_, \dots, \_}$ with $\chb \neq \chb'$
then we will also accepts stacks that are the predecessors of stacks where the guard is not satisfied (i.e.\ there is a $\chb'$ where there should be a $\chb$).
We could
obtain a precise algorithm by taking order-$\cpdsord$ stack-automaton states of the form
$\controls \; \times \; \alphabet$  so that they represent the top stack-character of a configuration
as well as its control-state. However, since $\alphabet$ is usually large compared to $\controls$ and since the worst-case size of
the stack-automaton is $\cpdsord$-exponential in the number of order-$\cpdsord$ states this would potentially come at a
large practical cost and in any case destroy fixed-parameter tractability.
We leave it for future work to investigate how this potential for  accuracy could be balanced with the
inevitable cost.

\begin{rem}
    The above modification to the naive saturation algorithm can also
    be easily incorporated into the efficient fixed point algorithm described in
    Section~\ref{sec:fastalgorithm}.
\end{rem}

\subsection{Approximate Reachability Graphs}
\label{sec:approx-graph}
We now describe the summary algorithm used to obtain an over-approximation of $\poststarraw$
and thus compute the GCPDS $\cpds''$ mentioned previously.  For simplicity, we assume that a stack symbol uniquely
determines the order of any link that it emits (which is the case for a CPDS obtained
from a HORS).
This condition is easily satsified by creating a copy of each stack character for each order in $\set{1, \ldots, \cpdsord}$ and adjusting the CPDS accordingly.
We first describe the approximate reachability graph, and then the approximate summary algorithm.

\subsubsection{The Approximate Reachability Graph}

We begin with an informal description before the formal definition.
Informally, an \emph{approximate reachability graph} for $\cpds$ is a structure
$\tuple{H, E, B}$ describing an over-approximation of the reachable configurations of $\cpds$.
\begin{itemize}
\item
    The set of nodes of the graph $H$ consists of \emph{heads} of the CPDS, where a head is a pair
    $(\control, \cha) \in \controls \times \alphabet$
    and describes configurations of the form $\config{\control}{\stacku}$ where $\ctop{1}(\stacku) = \cha$.

\item
    The set $E$ contains directed edges  $((\control, \cha), r, (\control', \cha'))$
    labelled by rules of $\cpds$. Such edges over-approximate the transitions that $\cpds$ might
    make using a rule $r$ from a configuration described by $(\control, \cha)$ to one
    described by $(\control', \cha')$.
      For example, suppose that $\cpds$ is order-$2$ and has, amongst others, the rules
      $r_1 = (\control_1, \chb, \push{2}, \control_2)$, $r_2 = (\control_2, \chb, \cpush{c}{2}, \control_3)$
      and $r_3 = (\control_3, \chc, \pop{1}, \control_4)$
      so that it can perform transitions:
      \begin{center}
      \scriptsize
      \begin{psmatrix}
       $\bigconfig{\control_1}{
       \stack{
          \stack{
            \begin{array}{c}
              b \\
              a
            \end{array}
          }
          }
       }$
       $\xrightarrow{\; r_1 \; }$
       $\bigconfig{\control_2}{
       \stack{
          \stack{
            \begin{array}{c}
              b \\
              a
            \end{array}
          }
          \;
          \stack{
            \begin{array}{c}
              b \\
              a
            \end{array}
          }
        }
       }$
       \end{psmatrix}
       \begin{psmatrix}
       $\xrightarrow{\; r_2\; }$
       $\bigconfig{\control_3}{
       \stack{
          \stack{
            \begin{array}{c}
              \rnode{C}{c} \\
              b \\
              a
            \end{array}
          }
          \;
          \stack{
            \begin{array}{c}
              \phantom{c}\\
              \rnode{B}{b} \\
              a
            \end{array}
         }
        }
       }$
       $\xrightarrow{\; r_3\; }$
       $\bigconfig{\control_4}{
       \stack{
          \stack{
            \begin{array}{c}
              \phantom{c} \\
              b \\
              a
            \end{array}
          }
          \;
          \stack{
            \begin{array}{c}
              \phantom{c}\\
              {b} \\
              a
            \end{array}
         }
        }
       }$
       \ncarc{->}{C}{B}
      \end{psmatrix}
    \end{center}
    where the first configuration mentioned here is reachable.
    We should then have  edges
    $((\control_1, b), r_1 , (\control_2, b))$,
    $((\control_2, b), r_2, (\control_3, c))$ and $((\control_3, c), r_3,
    (\control_4, b))$ in $E$.  We denote the configurations above $C_1, C_2, C_3$
    and $C_4$ respectively, with respective stacks $s_1, s_2, s_3, s_4$.

\item
    Finally, $B$ is a map assigning each head $h$ in the graph
    a set $B(h)$ of \emph{stack descriptors}, which are $(\cpdsord + 1)$-tuples
    $(h_\cpdsord, \dots, h_1, h_c)$ of heads. In the following, we refer to $h_{\opord}$ as the
    order-$\opord$ component and $h_{c}$ the collapse component.
    We give a rough description of $h_\opord$ before explaining an example.
    After applying a $\pop{\opord}$ operation, a previously created order-$(\opord-1)$ stack will be exposed.
    This newly exposed stack was either in the initial stack or created during the run.
    The pair $h_\opord$ describes
    at which head the new exposed order-$(\opord-1)$ stack resulting from a $\pop{\opord}$ operation (applied to a configuration with head $h$) may have been created.
    The pair $h_c$ does likewise for a $\collapse{}$ operation.
    (We will use $\bot$ in place of a head to indicate when $\pop{\opord}$ or $\collapse{}$
    leads to an empty stack.)

    Consider $C_3 = \config{\control_3}{s_3}$ from the example above. This has control-state
    $\control_3$ and top stack symbol $c$ and so is associated with the head
    $(\control_3, c)$. Thus $B((\control_3, c))$ should contain the stack-descriptor
    $((\control_1, b), (\control_2, b), (\control_1, b))$, which
    describes $s_3$. The first (order-{2}) component is because $\ctop{2}(s_3)$ was created by a
    $\push{2}$ operation from a configuration with head $(\control_1, b)$.  The second (order-1) component
    is because the top symbol was created via an order-$1$ push from $(\control_2, b)$.
    Finally, the order-$2$ link from the top of $s_3$ points to a stack occurring
    on top of a configuration at the head $(\control_1, b)$, giving rise to the
    final (collapse) component describing the collapse link.

    Tracking this information allows the summary algorithm to process the rule $r_3$
    to obtain a description of $C_4$ from the description of $C_3$. Since this rule
    performs a $\pop{1}$, it can look at the order-1 component of the stack descriptor
    to see the head $(\control_2, b)$, telling us that $\pop{1}$
    results in $b$ being on top of the stack. Since the rule $r_3$ moves into control-state
    $\control_4$, this tells us that  the new head should be $(\control_4, b)$.
    It also tells us that certain pieces of information in $B((\control_2, b))$
    are relevant to the description of $\ctop{2}(s_4)$  contained in
    $B((\control_4, b))$. First notice that this situation only occurs for the $\pop{\opord}$
    and $\collapse{\opord}$ operations. To keep track of these correlations, we will introduce in Section~\ref{sec:approx-summary} another component $U$ of the graph.
\end{itemize}

More formally, let us fix an ordinary order-$\cpdsord$ CPDS with rules $\cpdsrules$ and initial configuration
 $c_0 = \config{\control_0}{\sbrac{\cdots\sbrac{\cha_0}{1}\cdots}{\cpdsord}}$.
 A \emph{head} is an element $(\control, \cha) \in \controls \times \alphabet$ and should be viewed as describing
 \emph{stacks} $\stacku$ such that there is a \emph{reachable} configuration of the form
 $\config{\control}{\stacku}$ where $\ctop{1}(\stacku) = \cha$. Formally we define:
 \begin{equation*}
    \sembr{(\control, \cha)} = \setcomp{\stacku \in \stacksalpha{\cpdsord}{\alphabet}}{\ctop{1}(\stacku) = \cha \textrm{ and }
	  \config{\control}{\stacku} \in \poststarraw}
 \end{equation*}

A \emph{stack descriptor} is an $(\cpdsord + 1)$-tuple $\tuple{h_\cpdsord, \dots, h_1, h_c}$ where for each $1 \leq i \leq n$,
each of $h_i$ and $h_c$ is either a head or $\bot$.
We write
$\stackdes = {\brac{(\controls\times\alphabet) \cup \set{\bot}}^{\cpdsord + 1}}$ for the set of stack descriptors and it will also be useful
 to have $\stackdesk{\opord} = {\brac{(\controls\times\alphabet) \cup \set{\bot}}}^{\cpdsord - \opord}$ for the set of \emph{order-$\opord$
 stack-descriptor prefixes}. Note
 that $\stackdesk{\cpdsord} = \set{\tuple{}}$---\emph{i.e.} consists only
 of the empty tuple.
Assuming a map
$B : (\controls\times\alphabet)\rightarrow2^\stackdes$
a stack descriptor
\[
    \semdesc{B}{\tuple{h_\cpdsord, \dots, h_1, h_c}}
\]
is inductively defined and describes a set of stacks which contains
$\stacku \in \stacksalpha{\cpdsord}{\alphabet}$
iff
\begin{itemize}
\item
    for every
    $\opord \in \set{1, \ldots, \cpdsord}$,
    \begin{itemize}
    \item
        if
        $h_\opord = \bot$
        then
        $\apply{\ctop{\opord+1}}{\pop{\opord}(\stacku)} = \sbrac{}{\opord}$,
        and
    \item
        otherwise
        $\ctop{1}(\pop{\opord}(\stacku)) = \chb_\opord$
        where
        $h_\opord = (\_, \chb_\opord)$
        and
        \[
            \pop{\opord}(\stacku)
            \in
            \semdesc{B}{\tuple{h_\cpdsord, \dots, h_{\opord+1},
                               h_\opord', \dots, h_1', h_c'}}
        \]
        for some
        $(\_, \dots, \_, h_\opord', \dots, h_1', h_c') \in B(h_\opord)$, and
    \end{itemize}
\item
    for some
    $\opord \in \set{2, \ldots, \cpdsord}$,
    \begin{itemize}
    \item
        if
        $h_c = \bot$
        the topmost symbol
        $\ctop{1}(\stacku)$
        has link
        $\tuple{\opord, 0}$, and
    \item
        otherwise
        $\ctop{1}(\collapse{\opord}(\stacku)) = \chb_c$
        where
        $h_c = (\_, \chb_c)$
        and
        \[
            \collapse{\opord}(\stacku)
            \in
            \semdesc{B}{\tuple{h_\cpdsord, \dots, h_{\opord+1},
                               h_\opord', \dots, h_1', h_c'}}
        \]
        for some
        $(\_, \dots, \_, h_\opord', \dots, h_1', h_c') \in B(h_c)$.
    \end{itemize}
\end{itemize}
We now define an approximate reachability graph.
\begin{defi}
  An \emph{approximate reachability graph} for the CPDS $\cpds$ is a triple $\tuple{H, E, B}$ such that
  \begin{enumerate}[(i)]
  \item
    $H \; \subseteq \; \controls \times \alphabet$ is a set of heads such that
    $\config{\control}{\stacku} \in \poststarraw$ implies that $(\control, \ctop{1}(\stacku)) \in H$ when $\ctop{1}(\stacku)$ is defined,

  \item
    $E  \subseteq  H \times \cpdsrules \times H$ is a set of triples such that if $\config{\control}{\stacku} \in \poststarraw$ with $\ctop{1}(\stacku)$ defined and
    \begin{enumerate}
    \item
      $\cpdsruler = \cpdsrule{\control}{\ctop{1}(\stacku)}{\genop}{\control'} \in \cpdsrules$ for which $\ctop{1}(\genop(\stacku))$ is defined,
      then it is the case that $((\control, \ctop{1}(\stacku)), \cpdsruler, (\control', \ctop{1}(\genop(\stacku)))) \in E$, and
    \item
      $\cpdsruler = \cpdsaltrule{\control}{\controlset} \in \cpdsrules$
      then $((\control, \ctop{1}(\stacku)), \cpdsruler, (\control',
      \ctop{1}(\stacku))) \in E$ for all $\control' \in \controlset$,
    \end{enumerate}

  \item
    $B$ is a map
    $B : H \rightarrow \stackdes$ such that for every $h \in H$ we have $\sembr{h} \; \subseteq \; \setcomp{\semdesc{B}{d}}{d \in B(h)}$.
  \end{enumerate}
\end{defi}

A non-trivial approximate reachability graph is computed using an algorithm that works \emph{forwards} (while saturation works backwards), and which resembles a summary algorithm in the spirit of Sharir and Pnueli~\cite{SP81}.

\subsubsection{The Approximate Summary Algorithm}
\label{sec:approx-summary}

The construction of the approximate reachability graph is described in
Algorithms \ref{alg:sumMain}, \ref{alg:addSD}, \ref{alg:procHeadDesc} and
\ref{alg:addSum}.  The main work is done in the function
$\mathrm{ProcessHeadWithDescriptor}$. In particular, this is where summary edges
are added for the $\pop{\opord}$ and $\collapse{\opord}$ operations.

The approximate summary algorithm computes an approximate reachability graph $\tuple{H, E, B}$ `as accurately
as possible based on an order-$1$ approximation'.
In order to do this, the algorithm builds up an object $\tuple{H, E, B, U}$ where the additional component $U$ is a set of \emph{approximate higher-order summary edges}.
A summary edge describes how information contained in stack descriptors should be shared between heads.
We will first describe the role of edges in $E$ before describing summary edges.

An edge
$(h, r, h') \in E$
means that there is a transition from a configuration with head $h$ to a configuration with head $h'$.
This means that whenever we add a descriptor to $h$ we may need to add descriptors to $h'$ as we may apply $r$ to the stacks represented by the new descriptor.
The simplest case is when $r$ is a rewrite rule.
Any descriptor $(h_\cpdsord, \ldots, h_1, h_c)$ added to $h$ must be propagated to $h'$ as rewriting the top character results in a stack with the same descriptor.
In Algorithms \ref{alg:addSD} and \ref{alg:procHeadDesc} we will refer to this process as respecting $E$.

An \emph{order-$\opord$ summary edge} from a head $h$ to a head $h'$ is a triple of the form
\[
    (h, \tuple{h'_\cpdsord, \dots, h'_{\opord + 1}} , h')
\]
where each $h_i'$ is a head.  That is, a triple in $H  \times  \stackdesk{\opord}  \times  H$.
Such a summary edge is added when processing either a $\pop{\opord}$ or a $\collapse{\opord}$ operation on an order-$\opord$ link.
Intuitively such a summary edge means that if
$\tuple{h_\cpdsord, \dots, h_{\opord + 1}, h_\opord, \dots, h_1, h_c} \in B(h)$,
then we have
$\tuple{h'_\cpdsord, \dots, h'_{\opord + 1}, h_\opord, \dots, h_1, h_c} \in B(h')$.
In Algorithms \ref{alg:addSD} and \ref{alg:addSum} this addition is what we mean by respecting summary edges.
When $\cpdsord = \opord = 1$
(so that $h_c$ is also unnecessary since there would be no links)
note that $(h, \tuple{}, h')$ behaves like a summary edge in a standard order-$1$ summary algorithm \cite{SP81},
which is complete
at order-$1$.

To continue  our example, the $r_3$ rule (which performs a
$\pop{1}$ operation) from $C_3$ to $C_4$ means  $U$ should contain an
order-$1$ summary edge $((\control_2, b), \tuple{(\control_1, b)} ,
(\control_4, b))$. Since $\pop{1}$ is an order-1 operation, we have
$\pop{2}(s_3) = \pop{2}(s_4)$.  Hence $(\control_1, b)$ (the order-$2$
component of the stack descriptor for $s_3$) should also be the first
component of a stack descriptor for $s_4$.  However, since $\ctop{1}(s_4)$
was created at a configuration with head $(\control_2, b)$, the order-1 and
collapse components of such a stack descriptor for $s_4$ should be inherited
from a stack descriptor in $B((\control_2, b))$. In general if we go from a
configuration $\config{\control}{s}$ with head $h$ to a configuration $\config{\control'}{s'}$
with head $h'$ by the $\pop{\opord}$ operation or $\collapse{\opord}$ on an
order-$\opord$ link, we have that $\pop{\opord+1}(s)=\pop{\opord+1}(s')$ and
hence we have a summary edge $(h, \tuple{h'_\cpdsord, \dots, h'_{\opord +
1}} , h')$

The algorithm is presented as Algorithm \ref{alg:sumMain}.

\begin{algorithm}
 \caption{\label{alg:sumMain} The Approximate Summary Algorithm}
 \begin{algorithmic}
 \REQUIRE {An $\cpdsord$-CPDS with rules $\cpdsrules$ and heads $\controls \times \alphabet$ and initial configuration
	    $\config{\control_0}{\sbrac{\cdots\sbrac{\cha_0}{1}\cdots}{\cpdsord}}$}
  \ENSURE {The creation of a structure $\tuple{H, E, B, U}$ where $\tuple{H, E, B}$ is an approximate reachability graph
	    and $U$ is a set of approximate higher-order summary edges.}
  \STATE{Set $H = \set{(\control_0, \cha_0)}$ and set $E$, $B$ and $U$ to be empty}
  \STATE{Call {AddStackDescriptor($(\control_0, \cha_0), \tuple{\bot, \dots, \bot, \bot}$)}}
  \RETURN{Done, $\tuple{H, E, B, U}$ will now be as required}
 \end{algorithmic}
\end{algorithm}


\begin{algorithm}
    \caption{\label{alg:addSD}{AddStackDescriptor($h, \tuple{h_\cpdsord,  \dots, h_1, h_c}$)}}
    \begin{algorithmic}
        \REQUIRE {A head $h \in H$ and a stack descriptor
        $\tuple{h_\cpdsord, \dots, h_1, h_c}$}

        \ENSURE {$\tuple{h_\cpdsord,  \dots, h_1, h_c} \in B(h)$, that $E$ is respected and that any further additions to $B(h')$ for each $h' \in H$ necessary to respect summary edges are made.}
	\IF {$\tuple{h_\cpdsord,  \dots, h_1, h_c} \in B(h)$}
	  \RETURN {Done (Nothing to do)}
	\ENDIF

	\STATE {Add $\tuple{h_\cpdsord, \dots, h_1, h_c}$ to $B(h)$}
	\STATE {Call {ProcessHeadWithDescriptor($h, \tuple{h_\cpdsord, \dots, h_1, h_c}$)}}
	\FOR {$h' \in H$ such that $(h, \tuple{h'_\cpdsord, \dots, h'_{\opord + 1}} , h') \in U$}
	  \STATE {Call {AddStackDescriptor($h', \tuple{h'_\cpdsord, \dots, h'_{\opord + 1}, h_\opord, \dots, h_1, h_c}$)}}
	\ENDFOR

        \RETURN {Done}
    \end{algorithmic}
\end{algorithm}


\begin{algorithm}
  \caption{\label{alg:procHeadDesc} { ProcessHeadWithDescriptor($h, \tuple{h_\cpdsord, \dots, h_1, h_c}$)}}
  \begin{algorithmic}
    \REQUIRE {A head $h = (\control, \cha) \in H$ and a stack descriptor $\tuple{h_\cpdsord, \dots, h_1, h_c} \in B(h)$}
    \ENSURE {
        All necessary modifications are made to the graph to respect $E$ after adding
        $\tuple{h_\cpdsord, \dots, h_1, h_c} \in B(h)$.
        In addition, new summary edges are created when required.
    }
    \FOR {$\genop$ and $\control'$ such that $\cpdsruler = \cpdsrule{\control}{\cha}{\genop}{\control'} \in \cpdsrules$}
	      \IF {$\genop$ of form $\rew{\chb}$}
		  \STATE {Add $(\control', \chb)$ to $H$
                  and
                  $\tuple{ (\control, \cha), \cpdsruler, (\control', \chb) }$ to $E$}
		  \STATE {Call { AddStackDescriptor($ (\control', \chb),  \tuple{h_\cpdsord, \dots, h_1, h_c}$)}}
	      \ELSIF {$\genop$ of form  $\cpush{\chb}{\opord}$}
		  \STATE {Add $(\control', \chb)$ to $H$
                  and
                  $\tuple{ (\control, \cha), \cpdsruler, (\control', \chb) }$ to $E$}
		  \STATE {Call {AddStackDescriptor($ (\control', \chb),  \tuple{h_\cpdsord, \dots, h_2, (\control, \cha), h_\opord}$)}}
	      \ELSIF {$\genop$ of form $\push{\opord}$}
		  \STATE {Add $(\control', \cha)$ to $H$
                  and
                  $\tuple{ (\control, \cha), \cpdsruler, (\control', \cha) }$ to $E$}
		  \STATE {Call {AddStackDescriptor($ (\control', \cha),  (h_\cpdsord, \dots, h_{\opord +1}, (\control, \cha), h_{\opord -1}, \dots, h_1, h_c)$)}}
	       \ELSIF {$\genop$ of form $\pop{\opord}$ with $h_\opord = (\control_\opord, \cha_\opord)$ where $\cha_\opord \neq \bot$}
		  \STATE {Add $(\control', \cha_\opord)$ to $H$
                  and
                  $\tuple{ (\control, \cha), \cpdsruler, (\control', \cha_\opord) }$ to $E$}
		  \STATE {Call {AddSummary($ (\control_\opord, \cha_\opord), \tuple{h_\cpdsord, \dots, h_{\opord + 1}} , (\control', \cha_\opord)$)}}
		\ELSIF {$\genop$ of form $\collapse{\opord}$ with $h_c = (\control_c, \cha_c)$ where $\cha_c \neq \bot$}
		  \STATE {Add $(\control', \cha_c)$ to $H$
                  and
                  $\tuple{ (\control, \cha), \cpdsruler, (\control', \cha_c) }$ to $E$}
		  \STATE {Call {AddSummary($ (\control_c, \cha_c), \tuple{h_\cpdsord, \dots, h_{\opord + 1}} , (\control', \cha_c)$)}}
	      \ENDIF
      \ENDFOR

      \FOR {$\cpdsruler = \cpdsalttran{\control}{\controlset} \in \cpdsrules$, and $\control' \in \controlset$}
		  \STATE {Add $(\control', \cha)$ to $H$
                  and
                  $\tuple{ (\control, \cha), \cpdsruler, (\control', \cha) }$ to $E$}
		  \STATE {Call { AddStackDescriptor($ (\control', \cha),  \tuple{h_\cpdsord, \dots, h_1, h_c}$)}}
      \ENDFOR

      \RETURN {Done}
  \end{algorithmic}
\end{algorithm}


\begin{algorithm}
    \caption{\label{alg:addSum}{AddSummary($h, \tuple{h_\cpdsord',  \dots, h_{\opord + 1}}, h'$)}}
    \begin{algorithmic}
        \REQUIRE {An approximate higher-order summary edge $\tuple{h, \tuple{h_\cpdsord',  \dots, h'_{\opord + 1}}, h'}$}

        \ENSURE {$\tuple{h, \tuple{h_\cpdsord',  \dots, h_{\opord + 1}'}, h'} \in U$ and that all necessary stack descriptors
		    are added to the appropriate $B(h'')$ for $h'' \in H$ so that all summary edges (including the new one) are respected.}
	\IF {$\tuple{h, \tuple{h_\cpdsord',  \dots, h_{\opord + 1}'}, h'} \in U$}
	  \RETURN {Done (Nothing to do)}
	\ENDIF

	\STATE {Add $\tuple{h, \tuple{h_\cpdsord',  \dots, h_{\opord + 1}'}, h'}$ to U}

	\FOR {$\tuple{h_\cpdsord, \dots, h_{\opord +1}, h_\opord, \dots, h_1, h_c} \in B(h)$}
	  \STATE {{AddStackDescriptor($h', \tuple{h'_\cpdsord, \dots, h'_{\opord + 1}, h_\opord, \dots, h_1, h_c}$)}}
	\ENDFOR

        \RETURN {Done}
    \end{algorithmic}
\end{algorithm}


\begin{lem}
  Algorithm \ref{alg:sumMain} terminates and the resulting structure $\tuple{H, E, B, U}$ gives
  an approximate reachability graph $\tuple{H, E, B}$.
\end{lem}
\proof
  For termination note that the respective procedures in Algorithms \ref{alg:addSD} and \ref{alg:addSum}
  will immediately return if the stack-descriptor (respectively summary) that they are
  called with is already contained in a particular set. If it does not belong to this
  set, then it is added. Since there are only finitely many possible arguments for these
  functions, they can thus only be called finitely many times without immediately returning.
  From this fact it is easy to see that the entire algorithm must always terminate.

  Now we show that $\tuple{H, E, B}$ is an approximate reachability graph.
  Recursively define
  $\postdepth{0} = \set{c_0}$ and
  $$\postdepth{i+1} = \postdepth{i} \cup
                       \setcomp{c}{\exists c' \in \postdepth{i} \textrm{ s.t. }
                                   c' \cpdstran c \textrm{ or }
                                   c' \cpdstran \configset \textrm{ with } c \in \configset}.$$
  That is $\postdepth{i}$ is the set of configurations that can be reached from the initial configuration in at most $i$ steps.
  For a head $(\control, \cha) \in \controls\times\alphabet$, define
  $$\sembr{(\control, \cha)}_i = \setcomp{\config{\control}{\stacku}}{\config{\control}{\stacku} \in \postdepth{i} \textrm{ and }
	\ctop{1}(\stacku) = \cha}.$$

  We can now define an \emph{$i$-partial approximate reachability graph} to be a version of an approximate reachability
  graph defined for `reachability up to depth $i$'.
  \begin{defi}
  An \emph{$i$-partial approximate reachability graph} for the CPDS $\cpds$ is a triple $\tuple{H, E, B}$ such that
  \begin{enumerate}[(i)]
  \item
    $H \; \subseteq \; \controls \times \alphabet$ is a set of heads such that
    $\config{\control}{\stacku} \in \postdepth{i}$ implies that $(\control, \ctop{1}(\stacku)) \in H$ when $\ctop{1}(\stacku)$ is defined,
  \item
    $E  \subseteq  H \times \cpdsrules \times H$ is a set of triples such that if $i > 0$ and $\config{\control}{\stacku} \in \postdepth{i-1}$ with $\ctop{1}(\stacku)$ defined and
    \begin{enumerate}
    \item
      $\cpdsruler = \cpdsrule{\control}{\ctop{1}(\stacku)}{\genop}{\control'} \in \cpdsrules$ for which $\ctop{1}(\genop(\stacku))$ is defined,
      then it is the case that $((\control, \ctop{1}(\stacku)), \cpdsruler, (\control', \ctop{1}(\genop(\stacku)))) \in E$,
     \item
       $\cpdsruler = \cpdsaltrule{\control}{\controlset} \in \cpdsrules$
       then $((\control, \ctop{1}(\stacku)), \cpdsruler, (\control',
       \ctop{1}(\stacku))) \in E$ for all $\control' \in \controlset$,
     \end{enumerate}

  \item
    $B$ is a map
    $B : H \rightarrow \stackdes$ such that for every $h \in H$ we have $\sembr{h}_i \; \subseteq \; \setcomp{\sembr{d}}{d \in B(h)}$.
  \end{enumerate}
  \end{defi}
  Observe that a structure $\tuple{H, E, B}$ is an approximate reachability graph if and only if
  it is an $i$-partial approximate reachability graph for every $i \geq 0$.

  Now observe that the algorithm monotonically grows the sets making up $\tuple{H, E, B, U}$
 (it only adds to the sets, it never removes from them). We may thus argue by induction
 to show that the $\tuple{H, E, B}$ after termination is an $i$-partial approximate reachability
 graph for every $i \geq 0$ (and hence an approximate reachability graph).
 First note that the opening statements of Algorithm \ref{alg:sumMain}
 (including the call to add $(\bot, \dots, \bot, \bot)$ as a stack descriptor to
 $B(\control_0, \cha_0)$) guarantees that $\tuple{H, E, B}$ is a $0$-partial approximate reachability
 graph.

 Now suppose $\tuple{H, E, B}$ is an $i$-partial approximate reachability graph.
 We show it is also an $(i+1)$-partial approximate reachability graph. Let
 $\config{\control}{\stacku} \in \postdepth{i}$ and let either
 \begin{enumerate}
 \item
   $r = (\control, \cha, \genop, \control') \in \cpdsrules$
   be such that $\genop(\stacku)$ is defined and $\ctop{1}(\stacku) = \cha$ so that
   $(\control', \genop(\stacku)) \in \postdepth{i+1}$, or
 \item
   $r = \cpdsalttran{\control}{\controlset} \in \cpdsrules$ so that
   $\config{\control'}{\stacku} \in \postdepth{i+1}$ for all $\control' \in \controlset$.
 \end{enumerate}
 Let $\cha' = \ctop{1}(\genop(\stacku))$. It suffices to show that
 (i) $h ' = (\control', \cha') \in H$, (ii) $e = ((\control, \cha), r, (\control', \cha')) \in E$
 and that
 (iii) some $d' = (h_\cpdsord', \dots, h_1', h_c')
	  \in B(\control', \cha')$ with $\genop(\stacku) \in \semdesc{B}{d'}$.

By the induction hypothesis (that the structure is an $i$-partial approximate reachability graph)
we must have $h = (\control, \ctop{1}(\stacku)) \in H$ and
$d = (h_\cpdsord, \dots, h_1, h_c)
	  \in B(h)$ such that $\stacku \in \semdesc{B}{d}$.
Inspection of the algorithm shows that the addition of $d$ to $B(h)$
is only possible if $\mathrm{AddStackDescriptor}(h, d)$
was called at some point during its execution. However, this also implies
that $\mathrm{ProcessHeadWithDescriptor}(h, d)$ must have
been called.

Note also that when $\genop$ is a rewrite operation or $r$ is an alternating rule we must have
$\pop{j}(\genop(\stacku)) = \pop{j}(\stacku)$
and $\collapse{j}(\genop(\stacku)) = \collapse{j}(\stacku)$
for all $j$.
When $\genop = \push{\opord}$ for $\opord \geq 2$ we must have
$\ctop{j+1}(\pop{j}(\genop(\stacku))) = \ctop{j+1}(\pop{j}(\stacku))$
and $\ctop{j+1}(\collapse{j}(\genop(\stacku))) = \ctop{j+1}(\collapse{j}(\stacku))$
for all $j \neq \opord$ and $\pop{\opord}(\genop(\stacku)) = \stacku$.
When $\genop = \cpush{\chb'}{\opord}$ we must have
$\pop{j}(\genop(\stacku)) = \pop{j}(\stacku)$ for all $j \geq 2$,
but $\pop{1}(\genop(\stacku)) = \stacku$
and $\collapse{\opord}(\genop(\stacku)) = \pop{\opord}(\stacku)$.

Thus if $\genop$ is any operation other than $\pop{\opord}$
or $\collapse{\opord}$ it can be seen that the function
$\mathrm{AddStackDescriptor}(h', d')$ must be called for a $d'$
such that $\stacku \in \semdesc{B}{d'}$. Also, $e$ is added to $E$. Since the algorithm never deletes elements from sets,
this ensures that $\tuple{H, E, B}$ must satisfy the constraints (i), (ii) and (iii)
above.

Now consider the case when $\genop$ is either $\pop{\opord}$ or $\collapse{\opord}$.
Suppose again that $\ctop{1}(\genop(\stacku)) = \cha'$. Since $u \in \semdesc{B}{d}$ we must
have:
\begin{itemize}
  \item For some control-state $\control^-$ we have: $h_\opord = (\control^-, \cha')$ if $\genop = \pop{\opord}$ and
	$h_c = (\control^-, \cha')$ if $\genop = \collapse{\opord}$ such that\dots
  \item \dots there exists $(\_, \dots, \_, h_\opord', \dots, h_1', h_c') \in B((\control^-, \cha'))$
	such that \\ $\genop(u) \in \semdesc{B}{(h_\cpdsord, \dots, h_{\opord+1}, h_\opord', \dots, h_1', h_c')}$.
\end{itemize}
Thus a suitable $d'$ is $d' = (h_\cpdsord, \dots, h_{\opord+1}, h_\opord', \dots, h_1', h_c')$.

The call to $\mathrm{ProcessHeadWithDescriptor}(h, d)$
guarantees that (i) $h' = (\control', \cha') \in H$ and (ii) $e = ((\control, \cha), r, (\control', \cha')) \in E$.
It just remains to check that $d' \in B((\control', \cha'))$.

Note that the above call must also ensure
 a call to $$\mathrm{AddSummary}((\control^-, \cha'), (h_\cpdsord, \dots, h_{\opord+1}), (\control', \cha')) \ .$$
We are thus guaranteed the existence of a summary edge
$s = ((\control^-, \cha'), (h_\cpdsord, \dots, h_{\opord +1}), (\control', \cha')) \in U$
(although it
may have been added at an earlier point in the algorithm).
There are two cases to consider:
\begin{itemize}
  \item  If the summary edge $s$ was created \emph{after} a stack-descriptor
      $(\_, \dots, \_, h_\opord', \dots, h_1', h_c')$ was added to  $B((\control^-, \cha'))$,
      then the call to $\mathrm{AddSummary}$ creating $s$ must add $d'$ to
      $B((\control', \cha'))$.
   \item If the summary edge $s$ was created \emph{before} a stack-descriptor
   $d^- = (\_, \dots, \_, h_\opord', \dots, h_1', h_c')$ was added to  $B((\control^-, \cha'))$,
   then the $\mathrm{AddStackDescriptor}((\control^-, \cha'))$ call
   creating this stack-descriptor must result in $d'$ being added to $B((\control', \cha'))$.
\end{itemize}
Either way, (iii) must also be satisfied.
\qed

\subsubsection{A Remark On Complexity}
The approximate summary algorithm runs in time polynomial in the size of the CPDS (see below). Since
the graph constructed must also be of polynomial size, it follows that the rules for the
guarded CPDS $\cpds'$ can also be extracted in polynomial time. Since the raw saturation
algorithm is also PTIME when the number of control-states is fixed, it follows that the
\cshore{} algorithm as a whole -- including the forwards approximation and saturation -- runs in PTIME when the number of control-states is fixed.

We sketch here how to see that the approximate summary algorithm runs in polynomial time (when, as is standard, the order $\cpdsord$ is fixed).
First note that an approximate reachability graph can contain at most $|Q|\cdot|\Sigma|$ heads
and at most $|Q|\cdot|\Sigma|\cdot|\cpdsrules|\cdot|Q|\cdot|\Sigma|$ edges (recalling that $\cpdsrules$ is
the set of CPDS rules). Moreover the maximum size of the function $B$ (when viewed as a relation defined by
$\setcomp{(h, d) \in (Q \times \Sigma) \times (Q \times \Sigma)^{\cpdsord+1}}{d \in B(h)}$)
is $|Q|.|\Sigma|.(|Q|.|\Sigma|)^{\cpdsord +1}$. The maximum number of summary edges is
$\sum_{i = 1}^\cpdsord |Q|\cdot|\Sigma|\cdot(|Q|\cdot|\Sigma|)^{\cpdsord -i}\cdot|Q|\cdot|\Sigma|$.
It follows that the size of the structure $(H, E, B, U)$ constructed by algorithm
is at most polynomial in the size of the original CPDS. Moreover, since the algorithm
only \emph{adds} to the structure and never removes elements previously added, it
will perform at most polynomially many additions. Let $Z$ be this polynomial bound
on the size of the structure.

Moreover, recall that the procedures for adding summaries and heads/stack-descriptors are guarded.
I.e. the procedure only processes the new object if it had not already been added; if
it had already been added, the procedure will return after constant time.

So we consider the cases when the created object is new.
For each \emph{new} head/stack-descriptor pair,
$\mathrm{ProcessHeadWithDescriptor}$ will check it against every rule
and for each rule may attempt to create a new object. Disregarding the result of the calls
to create \emph{new} objects (with calls to create old objects returning in constant time),
the run-time of this procedure will thus be bounded by
$O(|\cpdsrules|)$. Likewise each time a \emph{new} stack descriptor is added,
$\mathrm{AddStackDescriptor}$ will compare it against
existing summary edges and so run in time $O(Z)$.

Similarly the run-time of a call to
$\mathrm{AddSummary}$ on a new summary edge (disregarding run-times
to calls from this procedure that create \emph{new} objects) is $O(Z)$ since
the new summary edge will, at worst, be compared against every possible
stack-descriptor.

Thus creating a new object takes at most $O(Z\cdot|\cpdsrules|)$ time and new objects
are created only during the call to a procedure that itself is creating a new object.
Thus the overall run-time is bounded by $O(Z\cdot Z\cdot|\cpdsrules|)$ and so is polynomial.

\subsection{Extracting the Guarded CPDA}
\label{sec:extract}

Let $\approxgraph = \tuple{H, E, B}$ be an approximate reachability graph for $\cpds$. Let $\heads{\errorconfigs}$
be the set of heads of error configurations, \emph{i.e.}
$\heads{\errorconfigs} = \setcomp{(\qerror, \cha)}{\cha \in \alphabet}$.
We do a simple backwards reachability computation on the finite graph $\approxgraph$ to compute
$\backrules{\approxgraph}$, defined to be the smallest set satisfying:
\[
    \begin{array}{rcl}%
      \backrules{\approxgraph}%
      &=&%
      \setcomp{e \in E}{e = (h, \cpdsruler, h') \in E  \textrm{ for some } h' \in \heads{\errorconfigs}} \ \cup%
      \\%
      & &%
      \setcomp{e \in E}{e = (h, \cpdsruler, h') \in E \textrm{ for some } (h', \_, \_) \in \backrules{\approxgraph}}%
    \end{array}
\]
The CPDS rules occurring in the triples in $\backrules{\approxgraph}$
can be used to define a pruned CPDS that is safe if and only if the original also is.
However, the approximate reachability graph provides enough information to construct a guarded CPDS whose guards are non-trivial.
It is clear that the following set $\backrulesG{\approxgraph}$ of \emph{guarded} rules can be computed:
\[
    \begin{array}{c}%
        \setcomp{\cpdsalttran{\control}{\controlset}} %
                {(\_, \cpdsalttran{\control}{\controlset}, \_) \in \backrules{\approxgraph}} %
        \\ \cup \\ %
        \setcomp{\cpdsrule{\control}{\cha}{\genop'}{\control'}}%
                        {\begin{array}{l}  (\_, \cpdsrule{\control}{\cha}{\genop}{\control'}, \_) \in \backrules{\approxgraph}%
                            \textrm{ and } \\ \genop' =%
                              \begin{cases}%
                                  \genop^S & \textrm{if } \genop \textrm{ is a pop or a collapse and }S \textrm{ is }  \\ %
                                  & \setcomp{\chb \in \alphabet}%
                                  {\begin{array}{l}((\control, \cha), \cpdsruler, (\control', \chb)) \in E\end{array}} \\ %
                                  & \textrm{with } r = \cpdsrule{\control}{\cha}{\genop}{\control'} \\ %
                                  \genop & \textrm{if } \genop \textrm{ is a rewrite or push}%
                              \end{cases}%
                        \end{array}}%
    \end{array}%
\]

These rules define a GCPDS on which \cshore finally performs saturation.
\begin{lem}
\label{lemma:GCPDSextract}
  The GCPDS $\cpds'$ defined using the rules $\backrulesG{\approxgraph}$ satisfies:
  \begin{equation*}
      \poststarraw \; \cap \; \prestar{\cpds}{\errorconfigs} \; \subseteq \; \prestar{\cpds'}{\errorconfigs}
	  \; \subseteq \; \prestar{\cpds}{\errorconfigs}
  \end{equation*}
\end{lem}
\proof
 $ \prestar{\cpds'}{\errorconfigs}
	  \; \subseteq \; \prestar{\cpds}{\errorconfigs}$
is trivial since $\trivialise{\cpds'}$ is a subset of the rules for $\cpds$.

Now suppose that $\config{\control}{\stacku} \in \poststarraw \; \cap \; \prestar{\cpds}{\errorconfigs}$.
By (i) in the definition of approximate reachability graphs it must be the case that
$(\control, \ctop{1}(\stacku)) \in H$ (since $\config{\control}{\stacku} \in \poststarraw$).

Since $\config{\control}{\stacku} \in \prestar{\cpds}{\errorconfigs}$ we must also
have $\config{\control}{\stacku} \in \pre{\ordinal}{\cpds}{\errorconfigs}$.
That is, $\config{\control}{\stacku}$ reaches
$\errorconfigs$ in $\ordinal$ steps.  We induct over $\ordinal$.

When $\ordinal = 0$ we have $\stacku \in \slang{\sastate_\control}{\errorconfigs}$ and the result is immediate.
Otherwise, for $(\ordinal + 1)$, there are two cases.
When $\config{\control}{\stacku} \cpdstran \configset \subseteq \pre{\ordinal}{\cpds}{\errorconfigs}$ via a rule
$r = \cpdsalttran{\control}{\controlset}$
we have by induction
$\configset \subseteq \prestar{\cpds'}{\errorconfigs}$
and by (i) and (ii) in the definition of approximate
reachability graph, $h = (\control, \ctop{1}(\stacku)) \in H$ and  $(h, r, h') \in E$ for every $h' = (\control', \ctop{1}(\stacku))$ with $\control' \in \controlset$.
In the second case we have
$\config{\control}{\stacku} \cpdstran \config{\control'}{\genop(\stacku)}$
by a rule
$r = \cpdsrule{\control}{\cha}{\genop}{\control'}$
and by induction
$\config{\control'}{\genop(\stacku)} \in \prestar{\cpds'}{\errorconfigs}$.
Furthermore, by (i) and (ii) in the definition of approximate reachability graphs we have
$h = (\control, \ctop{1}(\stacku)) \in H$ and
$(h, r, h') \in E$ where
$h' = (\control, \ctop{1}(\genop(\stacku)))$.
Thus one can verify
$(\_, r, \_) \in \backrules{\approxgraph}$.

Thus when $r$ is alternating or $\genop$ is neither a \emph{pop} nor \emph{collapse} operation
$r' = r$ will itself occur as a rule of $\cpds'$.
Otherwise
$r' = (\control, \cha, \genop^S, \control')$ will be in $\cpds'$
where
$\apply{\ctop{1}}{\stacku} \in S$.
Thus applying $r'$ to $\config{\control}{\stacku}$ witnesses
$\config{\control}{\stacku} \in \prestar{\cpds'}{\errorconfigs}$,
as required.
\qed


%% file: fastalg.tex
\section{Efficient Fixed Point Computation}
\label{sec:fastalgorithm}

We introduce an efficient method of computing the fixed point in
Section~\ref{sec:saturation-alg},  inspired by Schwoon\etal's algorithm for
alternating (order-$1$) pushdown systems~\cite{SSE06}.  Rather than checking all
CPDS rules at each iteration, we fully process \emph{all} consequences of each
new transition at once.  New transitions are kept in a set $\transtodo$
(implemented as a stack), processed, then moved to a set $\transdone$, which
forms the transition relation of the final stack automaton.  We assume \wlogen
that a character's link order is determined by the character.  This is true for
all CPDSs obtained from HORSs.

In the case of rewrite, pop and collapse rules, new transitions only depend ultimately on a single existing transition or state,
hence processing the consequences of new transitions is straightforward.  The
key difficulty is in the push rules and the alternating rules, for which new transitions depend on
\emph{sets} of existing transitions.  For example, given a rule
$\cpdsrule{\control}{\cha}{\push{\opord}}{\control'}$,
processing a transition with $\cpdsord$-expansion
$\satranfull{\sastate_{\control'}}{\cha}{\sastateset_\branch}{\sastateset_1,
\ldots,\sastateset_\opord,\ldots, \sastateset_\cpdsord}$ `once and once only'
must somehow include adding transitions whenever there is a set of
transitions with strict $(\opord, \sastateset_\opord)$-expansion
$\satranfull{\sastateset_\opord}{\cha}{\sastateset'_\branch}{\sastateset'_1,
\ldots, \sastateset'_\opord}$ in $\saauta_\idxi$ either now \emph{or in the
future}.
We discuss new notation required for our implementation of this before describing our algorithms.

\subsection{Notation}

To solve the problem of adding transitions depending on sets of transitions we use the notions below.

We first introduce some notation for referring to sets of transitions from $\sastateset_\opord$.
When $\opord = 1$ we write
$\sastateset_1 \satrancol{\cha}{\sastateset_\branch} \sastateset'_1$
to indicate the property that there exists some
$\satset \subseteq \sadelta_1$
with strict $(1, \sastateset_1)$-expansion
$\satranfull{\sastateset_1}{\cha}{\sastateset_\branch}{\sastateset'_1}$.
When $\opord > 1$ we write
$\sastateset_\opord \satran{\sastateset_{\opord-1}} \sastateset'_\opord$
to denote the existence of a set of transitions
$\satset \subseteq \sadelta_\opord$
such that
$\sastateset_\opord =
 \setcomp{\sastate}
         {\sastate \satran{\sastate'} \sastateset \in \satset}$
and
$\sastateset_{\opord-1} =
 \setcomp{\sastate'}
         {\sastate \satran{\sastate'} \sastateset \in \satset}$
and
$\sastateset'_\opord =
 \bigcup\limits_{\sastate \satran{\sastate'} \sastateset \in \satset}
    \sastateset$.
In both cases, we will say that $\satset$ \emph{witnesses} the existence of
$\sastateset_1 \satrancol{\cha}{\sastateset_\branch} \sastateset'_1$
or
$\sastateset_\opord \satran{\sastateset_{\opord-1}} \sastateset'_\opord$.

When $\gentr$ is processed, we create a \emph{trip-wire},
consisting of a \emph{source} and a \emph{target}.  A \emph{target} collects
transitions from a given set of states (such as $\sastateset_\opord$ above),
whilst a \emph{source} describes how such a collection could be used to form a
new transition according to a \emph{push} saturation step.
For the purposes of the following definition, let
$\sastates_0 = \set{\bot}$.

\begin{defi}
    An \emph{order-$\opord$ source} for $\opord \geq 1$ is a tuple
    $(\srcorig_\opord, \sastate_{\opord-1}, \cha, \sastateset_\opord)$ in
    $\brac{\controls \times \sastateset_{\opord+1} \times \cdots \sastateset_\cpdsord}
     \times
     \brac{\sastates_{\opord -1}\cup\set{\bot}}
     \times
     \alphabet \times 2^{\sastates_\opord}$.  An
    \emph{order-$\opord$ target} is a tuple
    \[
        \begin{cases} %
            (\sastateset_\opord, \countdown_\opord, \sastateset_{\mathit{lbl}}, %
            \sastateset_\opord') \in 2^{\sastates_\opord} \times
            2^{\sastates_\opord} \times 2^{\sastates_{\opord -1}} \times %
            2^{\sastates_\opord} %
            & %
            \text{if $\opord \geq 2$,} %
            \\ %
            (\sastateset_1, \countdown_1, \cha, \sastateset_{\branch}, %
            \sastateset_1') \in \bigcup_{k'=2}^\cpdsord\left(2^{\sastates_1} \times %
            2^{\sastates_1} \times \Sigma \times 2^{\sastates_{\opord'}} \times %
            2^{\sastates_1}\right) %
            & %
            \text{if $\opord = 1$.} %
        \end{cases} %
    \]
\end{defi}
The set $\countdown_\opord$ is a \emph{countdown} containing states in
$\sastateset_\opord$ still awaiting a transition. We always have
$\countdown_\opord \subseteq \sastateset_\opord$ and $\left(\sastateset_\opord
\setminus \countdown_\opord \right) \satran{\sastateset_{\mathit{lbl}}}
\sastateset'_\opord$.  Likewise, an order-$1$ target $(\sastateset_1,
\countdown_1, \cha, \sastateset_{\branch}, \sastateset_1)$ will satisfy
$(\sastateset_1 \setminus \countdown_1) \satrancol{\cha}{\sastateset_{\branch}}
\sastateset_1'$.  A target is \emph{complete} if $\countdown_\opord = \emptyset$
or $\countdown_1 = \emptyset$.
We say a source
$(\srcorig_\opord, \sastate_{\opord -1}, \cha, \sastateset_\opord)$
\emph{matches}
complete targets of the form
$(\sastateset'_\opord,
  \emptyset,
  \sastateset_{\mathit{lbl}},
  \sastateset_\opord')$
or
$(\sastateset_\opord, \emptyset, \cha, \sastateset_{\branch}, \sastateset_1')$.

A
\emph{trip-wire} of order-$\opord$ is an order-$\opord$ source-target pair which can take two forms:
$((\_, \_, \_, \sastateset_\opord), (\sastateset_\opord, \_, \_, \_))$ when $\opord \geq 2$
or $((\_, \_, \cha, \sastateset_\opord), (\sastateset_\opord, \_, \cha, \_, \_))$ when $\opord = 1$.
When the target in a trip-wire is \emph{complete}, the action specified by its source is triggered,
which we now sketch.

An order-$\opord$ \emph{source} for $\opord \geq 2$ describes how an
order-$(\opord - 1)$ source should be created from a complete target, propagating
the computation to the level below, and an order-$1$ source describes how new
transitions should be created from a complete target.  That is, when we
have a source
$\tuple{\srcorig_\opord, \_, \cha, \sastateset_\opord}$
(we hide the second component for simplicity of description)
and an associated target
$\tuple{\sastateset_\opord,
        \emptyset,
        \sastateset_{\mathit{lbl}},
        \sastateset'_\opord}$
this means we have found a set of transitions witnessing
$\sastateset_\opord \satran{\sastateset_{\mathit{lbl}}} \sastateset'_\opord$
and should now look for transitions from $\sastateset_{\mathit{lbl}}$.
Hence the algorithm creates a new source and target for the order-$(\opord-1)$ state-set
$\sastateset_{\mathit{lbl}}$.
When this process reaches order-$1$, new transitions are created.
This results in the construction of the $\satfull'$ from a \emph{push} saturation step.
In particular, given the complete trip-wire
$((\srcorig_1, \bot, \cha, \sastateset_1),
  (\sastateset_1, \emptyset, \cha, \sastateset_\branch, \sastateset'_1))$
where
$\srcorig_1 = \tuple{\control, \sastateset_2, \ldots, \sastateset_\cpdsord}$
we add
$\satranfull{\sastate_{\control}}
            {\cha}
            {\sastateset_\branch}
            {\sastateset'_1, \sastateset_2, \ldots, \sastateset_\cpdsord}$.

\begin{algorithm}
    \caption{\label{alg:main}Computing $\prestar{\cpds}{\saauta_0}$}
    \begin{algorithmic}
        \STATE {Let $\transdone = \emptyset$, $\transtodo = \bigcup_{\opord \in \set{1, \ldots, \cpdsord}}
               \sadelta_\opord$, $\sources[\opord] =
               \emptyset$, $\targs[\opord] = \set{\tuple{\emptyset, \emptyset,
               \emptyset, \emptyset}}$ for each $\cpdsord \geq \opord
               > 1$ and $\targs[1] = \setcomp{\tuple{\emptyset, \emptyset, \cha,
               \emptyset, \emptyset}}{\cha \in \alphabet}$.}

        \FOR {$\cpdsruler =
             \cpdsrule{\control}{\cha}{\pop{\cpdsord}}{\control'} \in
             \cpdsrules$}
            \STATE
            {$\apply{\addworklist}{\satranfull{\sastate_\control}{\cha}{\emptyset}{\emptyset,
            \dots, \emptyset, \set{\sastate_{\control'}}}, r}$}
        \ENDFOR

        \FOR {$\cpdsruler =
             \cpdsrule{\control}{\cha}{\collapse{\cpdsord}}{\control'} \in
             \cpdsrules$}
            \STATE
            {$\apply{\addworklist}{\satranfull{\sastate_\control}{\cha}{\set{\sastate_{\control'}}}{\emptyset,
            \dots, \emptyset}, r}$}
        \ENDFOR

        \FOR {$\cpdsruler = \cpdsalttran{\control}{\controlset} \in \cpdsrules$
              and
              $\cha \in \alphabet$}
            \STATE
            {$\apply{\createtrip}{\control, \bot, \cha, \sastateset, \cpdsruler}$
             where
             $\sastateset = \setcomp{\sastate_{\control'}}
                                    {\control' \in \controlset}$

            }
        \ENDFOR

        \WHILE {$\exists \gentr \in \transtodo$}
            \STATE {$\apply{\updatebyrules}{\gentr}$}
            \STATE{$\apply{\updatetrip}{\gentr}$}
            \STATE{Move $t$ from $\transtodo$ to $\transdone$}
        \ENDWHILE
    \end{algorithmic}
\end{algorithm}


\begin{algorithm}
    \caption{\label{alg:update-cpdsrules}$\apply{\updatebyrules}{\sat}$}
    \begin{algorithmic}
        \REQUIRE {A transition $\sat$ to be processed against $\transdone$}

        \IF {$\sat$ is an order-$\opord$ transition
             $\sastate_\opord \satran{\sastate_{\opord-1}} \sastateset_\opord$
             for some
             $\opord \in \set{2, \ldots, \cpdsord}$}
            \STATE{Suppose $\sastate_\opord$ has $\cpdsord$-expansion
                   $\satranfullk{\sastate_{\control'}}
                                {\sastate_\opord}
                                {\sastateset_{\opord+1}, \ldots, \sastateset_\cpdsord}$}
            \FOR {$\control \in \controls$ and $\cha \in \alphabet$ such that
                  $r = \cpdsrule{\control}{\cha}{\pop{\opord-1}}{\control'}
                        \in \cpdsrules$}

                \STATE {$\apply{\addworklist}
                               {\satranfull{\sastate_\control}
                                           {\cha}
                                           {\sastateset_\branch}
                                           {\emptyset, \dots, \emptyset,
                                            \set{\sastate_{\opord-1}},
                                            \sastateset_{\opord},
                                            \dots,
                                            \sastateset_\cpdsord}, r}$}
            \ENDFOR

            \FOR {$\control \in \controls$ and $\cha \in \alphabet$ such that
                  $r = \cpdsrule{\control}{\cha}{\collapse{\opord-1}}{\control'}
                        \in \cpdsrules$}
                \STATE {$\apply{\addworklist}
                               {\satranfull{\sastate_\control}
                                           {\cha}
                                           {\set{\sastate_{\opord-1}}}
                                           {\emptyset, \ldots, \emptyset,
                                            \sastateset_{\opord},
                                            \ldots,
                                            \sastateset_\cpdsord}, r}$}
            \ENDFOR

            \FOR {$\control \in \controls$ and $\cha \in \alphabet$ such that $r
                 = \cpdsrule{\control}{\cha}{\push{\opord}}{\control'} \in
                 \cpdsrules$}
                \STATE {$\apply{\createtrip}
                               {\tuple{\control,
                                       \sastateset_{\opord+1},
                                       \ldots,
                                       \sastateset_\cpdsord},
                                \sastate_{\opord-1},
                                \cha,
                                \sastateset_\opord,
                                (r, \sat)}$}
            \ENDFOR

        \ELSIF{$\sat$ is an order-$1$ transition with $\cpdsord$-expansion
              $\satranfull{\sastate_{\control'}}
                          {\chb}
                          {\sastateset_\branch}
                          {\sastateset_1, \ldots, \sastateset_\cpdsord}$}

            \FOR {$\control \in \controls$ and $\cha \in \alphabet$ such that $r
                 = \cpdsrule{\control}{\cha}{\rew{\chb}}{\control'} \in
                 \cpdsrules$}
                \STATE {$\apply{\addworklist}{ \satranfull{\sastate_\control}
                       {\cha} {\sastateset_\branch} {\sastateset_1, \ldots,
                       \sastateset_\cpdsord}, (r, \sat)}$}
            \ENDFOR

            \FOR {$\control \in \controls$ and $\cha \in \alphabet$ such that $r
                 = \cpdsrule{\control}{\cha}{\cpush{\chb}{\opord}}{\control'}
                 \in \cpdsrules$}
                \STATE {Let $\srcorig_1 = \tuple{\control,
                                                  \sastateset_2,
                                                  \ldots,
                                                  \sastateset_{\opord - 1},
                                                  \sastateset_\opord \cup \sastateset_\branch,
                                                  \sastateset_{\opord + 1},
                                                  \ldots,
                                                  \sastateset_\cpdsord}$}
                \STATE {$\apply{\createtrip}
                               {\srcorig_1, \bot, \cha, \sastateset_1, (r, \sat)}$}
            \ENDFOR
        \ENDIF
    \end{algorithmic}
\end{algorithm}


\begin{algorithm}
    \caption{\label{alg:updatetrip}
             $\apply{\updatetrip}
                    {\satfull =
                     \brac{\satranfull{\sastate_\control}
                                      {\cha}
                                      {\sastateset_\branch}
                                      {\sastateset_1, \dots, \sastateset_\cpdsord}}}$}
    \begin{algorithmic}
        \FOR{
            order-$\opord$
            $t_\opord\in \extractshort(\satfull)$
            where
            $t_\opord = \sastate_\opord
                        \satran{\sastate_{\opord-1}}
                        \sastateset_\opord$
            or
            $t_\opord = \sastate_\opord
                        \satrancol{\cha}{\sastateset_\branch}
                        \sastateset_\opord$
        }
            \FOR{$\mathit{targ} \in \targs[\opord]$ with $\mathit{targ} = (\_,
                \countdown_{\opord'}, \_, \_)$ or $(\_, \countdown_{\opord'},
                \cha, \_, \_)$ and $\sastate_\opord \in \countdown_{\opord'}$}
                \STATE{$\apply{\proctarg}{\mathit{targ}, t_\opord}$}
            \ENDFOR
        \ENDFOR
    \end{algorithmic}
\end{algorithm}


\begin{algorithm}
    \caption{\label{alg:create-tripwire} $\apply{\createtrip}{\srcorig_\opord,
            \sastate_{\opord -1}, \cha, \sastateset_\opord, \mathit{jus}}$}
    \begin{algorithmic}
        \IF{$(\srcorig_\opord, {\sastate_{\opord -1}}, \cha, \sastateset_\opord)
           \notin \sources[\opord]$}
            \STATE{Add $\mathit{src} = (\srcorig_\opord, {\sastate_{\opord
                  -1}}, \cha, \sastateset_\opord)$ to $\sources[\opord]$, set
                  $\tjust(\mathit{src}) = \mathit{jus}$}
            \STATE{Let $\mathit{targ} = (\sastateset_\opord,
                  \sastateset_\opord, \emptyset, \emptyset)$ if $\opord > 1$ or
                  $(\sastateset_\opord, \sastateset_\opord, \cha, \emptyset,
                  \emptyset)$ if $\opord = 1$}
            \IF{$\mathit{targ} \in \targs[\opord]$}
                \FOR{each \emph{complete} target $targ$ matching $src$}
                    \STATE{$\apply{\procsrccomptarg}{src, targ}$}
                \ENDFOR
            \ELSE
                \STATE{$\apply{\addtarget}{\mathit{targ}, \opord}$}
                \STATE{set $\tjust(\mathit{targ}) = \emptyset$}
            \ENDIF
        \ENDIF
    \end{algorithmic}
\end{algorithm}


\begin{algorithm}
    \caption{\label{alg:proc-tran} $\apply{\proctarg}{\mathit{targ}, t}$}
    \begin{algorithmic}
        \STATE{Suppose $\left\{\begin{array}{ll}
                   t = \sastate_\opord \satran{\sastate_{\opord-1}} \sastateset_\opord''
                   \text{ and } \mathit{targ} = (\sastateset_\opord,
                   \countdown_\opord, \sastateset_{\mathit{lbl}},
                   \sastateset_\opord') & \text{if $\opord \geq 2$} \\

                   t = \sastate_1 \satrancol{\cha}{\sastateset_\branch}
                   \sastateset_1'' \text{ and } \mathit{targ} = (\sastateset_1,
                   \countdown_1, \cha, \sastateset_{\mathit{lbl}},
                   \sastateset_1') & \text{if $\opord = 1$}
               \end{array}\right.$}

        \STATE{Let $\mathit{targ}' = \left\{\begin{array}{ll}
                  (\sastateset_\opord, \countdown_\opord \setminus
                  \set{\sastate_\opord}, \sastateset_{\mathit{lbl}} \cup
                  \set{\sastate_{\opord-1}},
                  \sastateset_\opord' \cup \sastateset_\opord'') & \text{if
                  $\opord \geq 2$} \\

                  (\sastateset_1, \countdown_1 \setminus \set{\sastate_1}, \cha,
                  \sastateset_{\mathit{lbl}} \cup \sastateset_\branch,
                  \sastateset_1' \cup \sastateset_1'') & \text{if $\opord = 1$}
               \end{array}\right.$}

        \IF{$\sastate_\opord \in \countdown_\opord$ and
           $\mathit{targ}'\notin\targs[\opord]$}
            \STATE{$\apply{\addtarget}{\mathit{targ}', \opord}$; if $k = 1$, set
                  $\tjust(\mathit{targ}') = \tjust(\mathit{targ}) \cup
                  \set{t}$}
            \IF{$\countdown_\opord \setminus \set{\sastate_\opord} = \emptyset$}
                \FOR{
                    each source
                    $\mathit{src} \in \sources[\opord]$
                    that matches $\mathit{targ}'$
                }
                    \STATE{$\apply{\procsrccomptarg}{\mathit{src},
                          \mathit{targ}'}$}
                \ENDFOR
            \ENDIF
        \ENDIF
    \end{algorithmic}
\end{algorithm}


\begin{algorithm}
    \caption{\label{alg:proc-src-complete-target}
            $\apply{\procsrccomptarg}{\mathit{src}, \mathit{comp\_targ}}$}
    \begin{algorithmic}
    \REQUIRE{An order-$\opord$ source of the form $\mathit{src} =
            (\srcorig_\opord, \sastate_{\opord -1}, \cha, \sastateset_\opord)$
            with
            $\srcorig = \tuple{\control,
                               \sastateset_{\opord+1}, \ldots, \sastateset_\cpdsord}$
            and an order-$\opord$ complete target of the form
            $\mathit{comp\_targ} = (\sastateset_\opord, \emptyset,
            \sastateset_{\mathit{lbl}}, \sastateset_\opord')$ when $\opord \geq
            2$ and $(\sastateset_1, \emptyset, \cha, \sastateset_{\mathit{lbl}},
            \sastateset_1')$ when $\opord = 1$}

        \IF{$\opord \geq 2$}
            \STATE{Let
                   $\srcorig_{\opord-1} =
                    \tuple{\control,
                           \sastateset'_\opord,
                           \sastateset_{\opord+1}, \ldots, \sastateset_\cpdsord}$}
            \STATE{$\apply{\createtrip}
                          {\srcorig_{\opord-1},
                           \bot,
                           \cha,
                           \sastateset_{\mathit{lbl}} \cup S,
                           \tjust(\mathit{src})}$
                   where
                   $S = \setcomp{\sastate_{\opord - 1}}
                                 {\sastate_{\opord - 1} \neq \bot}$}
        \ELSIF{$\opord = 1$}
            \STATE{Suppose $\tjust(\mathit{comp\_targ}) = T$}
            \STATE{$\apply{\addworklist}{\sastate_\control
                   \satrancol{\cha}{\sastateset_{\mathit{lbl}}}(\sastateset_1',
                   \sastateset_2, \dots, \sastateset_\cpdsord), \mathit{jus}}$
                   where
                   $\mathit{jus} = \begin{cases}
                                        (r, t, T) & \tjust(\mathit{src}) = (r, t) \\
                                        (r, T) & \tjust(\mathit{src}) = r \\
                                    \end{cases}$
                  }

        \ENDIF
    \end{algorithmic}
\end{algorithm}


\begin{algorithm}
    \caption{\label{alg:add-to-work-list}
             $\apply{\addworklist}{\satfull, \mathit{jus}}$}
    \begin{algorithmic}
        \REQUIRE{A $\cpdsord$-expansion $\satfull$ and justification $\mathit{jus}$.}

        \FOR {$\sat \in \extractshort(\satfull)$
              such that
              $\sat \notin \transdone \cup \transtodo$}
            \STATE{Add $\sat$ to $\transtodo$ and set
                   $\tjust(\sat) =
                    \tuple{\mathit{jus}, \sizeof{\transtodo \cup \transdone}}$
                  if $\sat$ is order-1}
        \ENDFOR
    \end{algorithmic}
\end{algorithm}


\begin{algorithm}
    \caption{\label{alg:add-target} $\apply{\addtarget}{\mathit{targ}, \opord}$}
    \begin{algorithmic}
        \IF{$\mathit{targ} \notin \targs[\opord]$}
            \STATE{Add $\mathit{targ}$ to $\targs[\opord]$}
            \FOR{$t' \in \transdone$}
                \STATE{$\apply{\proctarg}{\mathit{targ}, t'}$}
            \ENDFOR
        \ENDIF
    \end{algorithmic}
\end{algorithm}

\subsection{Algorithms}

For convenience we introduce another piece of notation.
Let the function $\extractshort$ obtain from an $\cpdsord$-expansion
$\satranfull{\sastate}
            {\cha}
            {\sastateset_\branch}
            {\sastateset_1, \ldots, \sastateset_\cpdsord}$
its (unique) corresponding set of transitions.
For example,
$\apply{\extractshort}
       {\satranfull{\sastate_3}
                   {\cha}
                   {\sastateset_\branch}
                   {\sastateset_1, \sastateset_2, \sastateset_3}}$
will return a set containing
$\sastate_3 \satran{\sastate_2} \sastateset_3$
and
$\sastate_2 \satran{\sastate_1} \sastateset_2$
and
$\sastate_1 \satrancol{\cha}{\sastateset_\branch} \sastateset_1$
for some $\sastate_2$ and $\sastate_1$.

Algorithm \ref{alg:main} gives the main loop and introduces the global sets of transitions
$\transdone$ and $\transtodo$, and two arrays $\sources[\opord]$ and
$\targs[\opord]$ containing sources and targets for each order.  The algorithm
processes $\pop{\cpdsord}$ and $\collapse{\cpdsord}$ rules like the naive
algorithm and creates trip-wires for the alternating transitions.
Algorithm~\ref{alg:update-cpdsrules} gives the main steps processing a new
transition.  In most cases a new transition is created, however, for
\emph{push} rules we create a trip-wire.  We describe some of the algorithms
informally below.

In $\createtrip$ we create a trip-wire with a new target
$\tuple{\sastateset_\opord, \sastateset_\opord, \emptyset, \emptyset}$.  This is
added using an $\addtarget$ procedure which also checks $\transdone$ to create
further targets.  E.g., a new target $\tuple{\sastateset_\opord,
\countdown_\opord, \sastateset_{\mathit{lbl}}, \sastateset'_\opord}$ combines
with an existing $\sastate_\opord \satran{\sastate_{\opord-1}}
\sastateset''_\opord$ to create a new target $\tuple{\sastateset,
\countdown_\opord \setminus \set{\sastate_\opord}, \sastateset_{\mathit{lbl}}
\cup \set{\sastate_{\opord-1}}, \sastateset'_\opord \cup \sastateset''_\opord}$.
(This step corrects a bug of Schwoon\etal)  Similarly
$\updatetrip$ updates existing targets by new transitions.  In all cases, when a
source and matching complete target are created, we perform the propagations
as above.

\begin{prop}
\label{prop:fast-alg-correct}
    Given a CPDS $\cpds$ and stack automaton $\saauta_0$, let $\saauta$ be the
    result of Algorithm~\ref{alg:main}.
    We have
    $\stackw \in \slang{\control}{\saauta}$
    iff
    $\config{\control}{\stackw} \in \prestar{\cpds}{\saauta_0}$.
    \qed
\end{prop}

\subsection{Correctness}

We prove Proposition~\ref{prop:fast-alg-correct}.  I.e., that the fast algorithm
is correct.  The proof is in two parts in the following
sub-sections; in particular in Lemma~\ref{lem:fast-alg-only-if} and
Lemma~\ref{lem:fast-alg-if}.

In the sequel, we fix the following notation.  Let
$\sequence{\saauta_\idxi}{\idxi \geq 0}$ be the sequence of automata constructed
by the naive fixed point algorithm.  Then, let
$\sequence{\transdone^\idxj}{\idxj \geq 0}$ be the sequence of sets of
transitions such that $\transdone^\idxj$ is $\transdone$ after $\idxj$
iterations of the main loop of Algorithm~\ref{alg:main}.  Similarly, define
$\sources^\idxj[\opord]$ and $\targs^\idxj[\opord]$.

\subsubsection{Soundness}

We prove that the algorithm is sound.  First, we show two preliminary lemmas
about the data-structures maintained by the algorithm.

\begin{lem}
    \label{lem:targ-imp-trans}
    For all $\idxj \geq 0$ and $\opord \in \set{2, \ldots, \cpdsord}$, if
    $\tuple{\sastateset_\opord, \sastateset_\opord \setminus
    \sastateset^\targmarker_\opord, \sastateset^\targmarker_{\opord-1},
    \sastateset^{\targmarker'}_\opord} \in \targs^\idxj[\opord]$
    with
    $\sastateset^\targmarker_\opord \subseteq \sastateset_\opord$,
    then we have some
    $\satset \subseteq \transdone^\idxj$ that witnesses
    $\sastateset^\targmarker_\opord \satran{\sastateset^\targmarker_{\opord-1}}
    \sastateset^{\targmarker'}_\opord$.
\end{lem}
\proof
    We proceed by induction over $\idxj$ and the order in which targets are
    created.  In the base case we only have $\tuple{\emptyset, \emptyset,
    \emptyset, \emptyset} \in \targs^0[\opord]$.  Setting $\satset = \emptyset$
    witnesses $\emptyset \satran{\emptyset} \emptyset$.

    In the inductive case, consider the location of the call to $\addtarget$:
    $\createtrip$ or $\proctarg$.  When we are in
    $\createtrip$, we have a target of the form $\tuple{\sastateset_\opord,
    \sastateset_\opord, \emptyset, \emptyset}$, hence
    $\sastateset^\targmarker_\opord = \emptyset$ and we trivially have $\satset =
    \emptyset \subseteq \transdone^\idxj$ witnessing $\emptyset \satran{\emptyset}
    \emptyset$.

    Otherwise the call is from $\proctarg$ against a transition
    $\sat = \brac{\sastate_\opord \satran{\sastate_{\opord-1}} \sastateset_\opord''}$
    and a target
    $\mathit{targ} = \tuple{\sastateset_\opord, \sastateset_\opord \setminus
    \sastateset^\targmarker_\opord, \sastateset^\targmarker_{\opord-1},
    \sastateset^{\targmarker'}_\opord}$ already in $\targs^{\idxj-1}$.  Hence, by induction,
    we know that there is some $\satset \subseteq \transdone^\idxj[\opord]$
    witnessing $\sastateset^\targmarker_\opord
    \satran{\sastateset^\targmarker_{\opord-1}} \sastateset^{\targmarker'}_\opord$.  The
    transition $\sat$ is either already in $\transdone^\idxj$ or will be moved
    there at the end of the $\idxj$th iteration.  Combining $\sat$ with
    $\satset$ we have $\satset \cup \set{\sat} \subseteq \transdone^\idxj$
    witnessing $\sastateset^\targmarker_\opord \cup \set{\sastate_\opord}
    \satran{\sastateset^\targmarker_{\opord-1} \cup
    \set{\sastate_{\opord-1}}}
    \sastateset^{\targmarker'}_\opord \cup \sastateset''_\opord$.  Since the new
    target added is $\tuple{\sastateset_\opord, \sastateset_\opord \setminus
    \brac{\sastateset^\targmarker_\opord \cup \set{\sastate_\opord}},
    \sastateset^\targmarker_\branch \cup
    \set{\sastate_{\opord-1}},
    \sastateset^{\targmarker'}_\opord \cup \sastateset_\opord''}$ we are done.
\qed

\begin{lem}
    \label{lem:targ-imp-trans-o1}
    For all $\idxj \geq 0$, if $\tuple{\sastateset_1, \sastateset_1 \setminus
    \sastateset^\targmarker_1, \cha, \sastateset^\targmarker_\branch,
    \sastateset^{\targmarker'}_1} \in \targs^\idxj[1]$
    with
    $\sastateset^\targmarker_1 \subseteq \sastateset_1$,
    then we have some $\satset
    \subseteq \transdone^\idxj$ that witnesses $\sastateset^\targmarker_1
    \satrancol{\cha}{\sastateset^\targmarker_\branch} \sastateset^{\targmarker'}_1$.
\end{lem}
\proof
    The proof is essentially the same as the order-$\opord$ case above.  We
    proceed by induction over $\idxj$ and the order in which targets are
    created.  In the base case we only have $\tuple{\emptyset, \emptyset, \cha,
    \emptyset, \emptyset} \in \targs^0[1]$.  The set $\satset = \emptyset$
    witnesses $\emptyset \satrancol{\cha}{\emptyset} \emptyset$.

    In the inductive case, consider the location of the call to $\addtarget$:
    $\createtrip$ or $\proctarg$.  When in
    $\createtrip$, we have a target of the form $\tuple{\sastateset_1,
    \sastateset_1, \cha, \emptyset, \emptyset}$, hence $\sastateset^\targmarker_1 =
    \emptyset$ and we trivially have $\satset = \emptyset \subseteq
    \transdone^\idxj$ witnessing $\emptyset \satrancol{\cha}{\emptyset}
    \emptyset$.

    Otherwise the call is from $\proctarg$ called with a transition
    $\sat = \brac{\sastate_1 \satrancol{\cha}{\sastateset_\branch} \sastateset_1''}$
    and a target
    $\mathit{targ} = \tuple{\sastateset_1, \sastateset_1 \setminus
    \sastateset^\targmarker_1, \cha, \sastateset^\targmarker_\branch,
    \sastateset^{\targmarker'}_1}$ already in $\targs^{\idxj-1}$.  Hence, by induction, we
    know that there is some $\satset \subseteq \transdone^\idxj[1]$ witnessing
    $\sastateset^\targmarker_1 \satrancol{\cha}{\sastateset^\targmarker_\branch}
    \sastateset^{\targmarker'}_1$.  The transition $\sat$ is either already in
    $\transdone^\idxj$ or will be moved there at the end of the $\idxj$th
    iteration.  Combining $\sat$ with $\satset$ we have $\satset \cup \set{\sat}
    \subseteq \transdone^\idxj$ witnessing $\sastateset^\targmarker_1 \cup
    \set{\sastate_1} \satrancol{\cha}{\sastateset^\targmarker_\branch \cup
    \sastateset_\branch} \sastateset^{\targmarker'}_1 \cup \sastateset''_1$.  Since
    the new target is $\tuple{\sastateset_1, \sastateset_1 \setminus
    \brac{\sastateset^\targmarker_1 \cup \set{\sastate_1}}, \cha,
    \sastateset_\branch \cup \sastateset^\targmarker_\branch,
    \sastateset^{\targmarker'}_1 \cup \sastateset_1''}$ we are done.
\qed

We are now ready to prove the algorithm is sound.

\begin{lem}
    \label{lem:fast-alg-if}
    Given a CPDS $\cpds$ and stack automaton $\saauta_0$, let $\saauta$ be the
    result of Algorithm~\ref{alg:main}.
    We have
    $\stackw \in \slang{\control}{\saauta}$
    implies
    $\config{\control}{\stackw} \in \prestar{\cpds}{\saauta_0}$.
\end{lem}
\proof
    We proceed by induction over $\idxj$ and show every transition appearing in
    $\transdone^\idxj$ appears in $\saauta_\idxi$ for some $\idxi$.  This
    gives the lemma.

    When $\idxj = 0$ the property is immediate, since the only transitions added
    are already in $\saauta_0$, or added to $\saauta_1$ during the first
    processing of the $\pop{\cpdsord}$ and $\collapse{\cpdsord}$ rules.

    In the inductive step, we consider some $\sat$ first appearing in
    $\transtodo^\idxj$ (and thus, eventually in $\transdone^{\idxj'}$ for some
    $\idxj'$).  There are several cases depending on how $\sat$ was added to
    $\transtodo$ (i.e. from where $\addworklist$ was called).  We consider the
    simple cases first.  In all the following cases, $\sat$ was added during
    $\updatebyrules$ called with a transition $\sat'$ appearing in
    $\transtodo^{\idxj-1}$.
    \begin{itemize}
        \item Suppose
              $\sat' = \brac{
                  \sastate_\opord
                  \satran{\sastate_{\opord-1}}
                  \sastateset_\opord}$
              and $\sastate_\opord$ has $\cpdsord$-expansion
              $\satranfullk{\sastate_{\control'}}
                           {\sastate_\opord}
                           {\sastateset_{\opord+1},
                            \ldots,
                            \sastateset_\cpdsord}$.
              Moreover, suppose
              $\sat \in \apply{\extractshort}{\satfull}$
              where
              \[
                    \satfull = \brac{
                      \satranfull{\sastate_\control}
                                 {\cha}
                                 {\sastateset_\branch}
                                 {\emptyset, \dots, \emptyset,
                                  \set{\sastate_{\opord-1}},
                                  \sastateset_{\opord},
                                  \ldots,
                                  \sastateset_\cpdsord}
                    }
              \]
              was added during the processing of
              $\sastate_{\opord-1}$
              against a $\pop{\opord-1}$ rule.
              By induction $\sat'$ appears in $\saauta_\idxi$ for some $\idxi$,
              and hence the transitions in
              $\apply{\extractshort}{\satfull}$
              (and hence $\sat$) are present in $\saauta_{\idxi+1}$.

        \item Suppose
              $\sat' = \brac{
                  \sastate_\opord
                  \satran{\sastate_{\opord-1}}
                  \sastateset_\opord}$
              where $\sastate_\opord$ has $\cpdsord$-expansion
              $\satranfullk{\sastate_{\control'}}
                           {\sastate_\opord}
                           {\sastateset_{\opord+1},
                            \ldots,
                            \sastateset_\cpdsord}$.
              Moreover,
              $\sat \in \apply{\extractshort}{\satfull}$
              where
              \[
                    \satfull = \brac{
                      \satranfull{\sastate_\control}
                                 {\cha}
                                 {\set{\sastate_{\opord-1}}}
                                 {\emptyset, \ldots, \emptyset,
                                  \sastateset_{\opord},
                                  \ldots,
                                  \sastateset_\cpdsord}
                    }
              \]
              was added during the processing of
              $\sastate_{\opord-1}$
              against a
              $\collapse{\opord-1}$ rule.
              By induction $\sat'$ appears in
              $\saauta_\idxi$ for some $\idxi$ and hence the transitions in
              $\apply{\extractshort}{\satfull}$
              (and hence $\sat$) are present in $\saauta_{\idxi+1}$.

        \item Suppose
              $\sat' = \brac{
                  \sastate_1
                  \satrancol{\chb}{\sastateset_\branch}
                  \sastateset_1
              }$
              has $\cpdsord$-expansion
              $\satranfull{\sastate_{\control'}}
                          {\chb}
                          {\sastateset_\branch}
                          {\sastateset_1, \ldots, \sastateset_\cpdsord}$
              and
              $\sat \in \apply{\extractshort}{\satfull}$
              where
              $
                    \satfull = \brac{
                      \satranfull{\sastate_\control}
                                 {\cha}
                                 {\sastateset_\branch}
                                 {\sastateset_1, \ldots, \sastateset_\cpdsord}
                    }
              $
              was added during the processing of $\sat'$ against a
              $\rew{\chb}$ rule.  By induction $\sat'$ appears in
              $\saauta_\idxi$ for some $\idxi$, and hence the transitions in
              $\apply{\extractshort}{\satfull}$
              (and hence $\sat$) are present in $\saauta_{\idxi+1}$.
    \end{itemize}

    In the final case, $\addworklist$ is called during $\procsrccomptarg$.  There
    are two cases depending on the provenance of the source.  In the first case,
    the source was added by a call to $\createtrip$ from $\updatebyrules$ while
    processing a $\cpush{\chb}{\opord}$ rule against
    $\sat' = \brac{
        \sastate_1
        \satrancol{\chb}{\sastateset_\branch}
        \sastateset_1
    }$
    with $\cpdsord$-expansion
    $\satranfull{\sastate_{\control'}}
                {\chb}
                {\sastateset_\branch}
                {\sastateset_1, \ldots, \sastateset_\cpdsord}$.
    Therefore,
    $\sat \in \apply{\extractshort}{\satfull}$
    where
    \[
        \satfull = \brac{
            \satranfull{\sastate_\control}{\cha}{\sastateset'_\branch}{\sastateset'_1,
            \sastateset_2, \ldots, \sastateset_{\opord-1}, \sastateset_{\opord}
            \cup \sastateset_\branch, \sastateset_{\opord+1},
            \sastateset_\cpdsord}
        }
    \]
    was added from a source $\tuple{\srcorig_1, \bot, \cha, \sastateset_1} \in
    \sources^{\idxj}[1]$ with
    \[
        \srcorig_1 =
        \tuple{\control,
               \sastateset_2, \ldots, \sastateset_{\opord-1},
               \sastateset_\opord \cup \sastateset_\branch,
               \sastateset_{\opord+1}, \ldots, \sastateset_\cpdsord} \ .
    \]
    By induction, from $\sat'$ we know that
    $
        \satranfull{\sastate_{\control'}}{\chb}{\sastateset_\branch}{\sastateset_1,
        \ldots, \sastateset_\cpdsord}
    $
    was added to $\saauta_\idxi$ for some $\idxi$.  Now, consider the target
    $\tuple{\sastateset_1, \emptyset, \cha, \sastateset'_\branch,
    \sastateset'_1} \in \targs^\idxj[1]$ that was combined with the source to
    add the new transition.  By Lemma~\ref{lem:targ-imp-trans-o1} we have some
    $\satset \subseteq \transdone^\idxj$
    witnessing
    $\sastateset_1 \satrancol{\cha}{\sastateset'_\branch} \sastateset'_1$
    and hence
    (since all transitions in $\transdone^\idxj$ passed through $\transtodo$)
    by induction we have that the transitions in $\satset$ are in $\saauta_{\idxi'}$ for some $\idxi'$.
    Hence, in
    $\saauta_{\apply{\maxfun}{\idxi, \idxi'}+1}$
    we have $\sat$ as required.

    In the second case we have a source $\tuple{\srcorig_1, \bot, \cha,
    \sastateset^s_1} \in \sources^\idxj[1]$ and a complete target of the form
    $\tuple{\sastateset^s_1, \emptyset, \cha, \sastateset^t_\branch,
    \sastateset^{t'}_1} \in \targs^\idxj[1]$ and the source derived from a call
    to $\createtrip$ in $\procsrccomptarg$.  Note, by
    Lemma~\ref{lem:targ-imp-trans-o1} we have some subset of
    $\transdone^\idxj$
    witnessing
    $\sastateset^s_1 \satrancol{\cha}{\sastateset^t_\branch} \sastateset^{t'}_1$.
    The call to $\createtrip$ implies we have a source
    $\tuple{\srcorig_2, \sastate'_1, \cha, \sastateset^s_2} \in
    \sources^{\idxj}[2]$ and complete target of the form
    $\tuple{\sastateset^s_2, \emptyset, \sastateset^t_1, \sastateset^{t'}_2} \in
    \targs^\idxj[2]$, with $\sastateset^s_1 = \sastateset^t_1 \cup S_1$ where
    $S_1 = \setcomp{\sastate'_1}{\sastate'_1 \neq \bot}$ and
    $\srcorig_1$ is $\srcorig_2$ with the additional order-$2$ component
    $\sastateset^{t'}_2$.
    That is if
    $\srcorig_2 = \tuple{\control, \sastateset_3, \ldots, \sastateset_\cpdsord}$
    then
    $\srcorig_1 =  \tuple{\control,
                          \sastateset^{t'}_2,
                          \sastateset_3, \ldots, \sastateset_\cpdsord}$.
    The proof will now iterate
    $\opord = 2, 3, \ldots$
    until a source is discovered that was added during a call to
    $\createtrip$ from $\updatebyrules$ while processing some $\push{\opord'}$ rule or alternating rule.
    Note that sources not added by $\cpush{\chb}{\opord}$ rules can only be
    added in this way and, for all $\opord < \opord'$, the second component of
    the source ($\sastate'_{\opord-1}$) will be $\bot$.

    Hence, inductively, we have a source $\mathit{src} = \tuple{\srcorig_\opord,
    \sastate'_{\opord-1}, \cha, \sastateset^s_\opord} \in
    \sources^{\idxj}[\opord]$ and complete target $\tuple{\sastateset^s_\opord,
    \emptyset, \sastateset^t_{\opord-1}, \sastateset^{t'}_\opord} \in
    \targs^\idxj[\opord]$ with $\sastateset^s_{\opord-1} =
    \sastateset^t_{\opord-1} \cup S_{\opord-1}$ where $S_{\opord-1} =
    \setcomp{\sastate'_{\opord-1}}{\sastate'_{\opord-1} \neq \bot}$ and
    $\srcorig_{\opord-1}$ is $\srcorig_{\opord}$ with the additional order-$\opord$ component
    $\sastateset^{t'}_\opord$.

    Furthermore, by Lemma~\ref{lem:targ-imp-trans} we have some subset of
    $\transdone^\idxj$
    witnessing
    $\sastateset^s_\opord \satran{\sastateset^t_{\opord-1}} \sastateset^{t'}_\opord$.

    In the first case, suppose $\mathit{src}$ was added due to a call to
    $\createtrip$ in a call to $\procsrccomptarg$.  The call to $\createtrip$ implies we have a
    source of the form $\tuple{\srcorig_{\opord+1}, \sastate'_\opord, \cha,
    \sastateset^s_{\opord+1}} \in \sources^{\idxj}[\opord+1]$ and complete
    target $\tuple{\sastateset^s_{\opord+1}, \emptyset, \sastateset^t_{\opord},
    \sastateset^{t'}_{\opord+1}} \in \targs^\idxj[\opord+1]$, with
    $\sastateset^s_\opord = \sastateset^t_\opord \cup S_\opord$ where $S_\opord
    = \setcomp{\sastate'_\opord}{\sastate'_\opord \neq \bot}$ and
    $\srcorig_\opord$ is $\srcorig_{\opord+1}$ with the additional order-${\opord+1}$ component
    $\sastateset^{t'}_{\opord+1}$.

    For the final cases, first suppose that $\mathit{src}$ was added due to a call to
    $\createtrip$ in $\updatebyrules$ from a $\push{\opord}$ rule.  Then we were
    processing a new transition of the form
    $\sastate''_\opord \satran{\sastate''_{\opord-1}} \sastateset_\opord$
    where $\sastate''_\opord$ has $\cpdsord$-expansion
    $\satranfullk{\sastate_{\control'}}
                 {\sastate''_\opord}
                 {\sastateset_{\opord+1}, \ldots, \sastateset_\cpdsord}$.
    Moreover, we have
    $\srcorig_\opord = \tuple{\control,
                              \sastateset_{\opord+1}, \ldots, \sastateset_\cpdsord}$
    and $\sastate'_{\opord-1}$ has $\cpdsord$-expansion
    $\satranfullk{\sastate_{\control'}}
                 {\sastate'_{\opord-1}}
                 {\sastateset_\opord, \ldots, \sastateset_\cpdsord}$
    and
    $\sastateset^s_\opord = \sastateset_\opord$.
    From the induction and since $\sastate'_{\opord'} =
    \bot$ for all $\opord' < \opord$, we have some order-$1$ transitions
    $\satset_1 \cup \set{\sat_1} \subseteq \transdone^\idxj$
    witnessing
    $\satranfull{\sastateset^t_{\opord-1} \cup \set{\sastate'_{\opord-1}}}
                {\cha}
                {\sastateset^t_\branch}
                {\sastateset^{t'}_1, \ldots, \sastateset^{t'}_{\opord-1}}$
    where $\satset_1$ witnesses some
    $\satranfull{\sastateset^t_{\opord-1}}
                {\cha}
                {\sastateset'_\branch}
                {\sastateset'_1, \ldots, \sastateset'_{\opord-1}}$
    and $\sat_1$ witnesses some
    $\satranfull{\sastate'_{\opord-1}}
                {\cha}
                {\sastateset_\branch}
                {\sastateset_1, \ldots, \sastateset_{\opord-1}}$.
    Thus, because
    $\sastate'_{\opord-1}$
    expands to
    $\satranfullk{\sastate_{\control'}}
                 {\sastate'_{\opord-1}}
                 {\sastateset_\opord, \ldots, \sastateset_\cpdsord}$
    and
    $\sastateset^s_\opord = \sastateset_\opord$
    and letting
    $\sastateset'_\opord = \sastateset^{t'}_\opord$,
    we have $\sat_1$ has $\cpdsord$-expansion
    \[
        \satranfull{\sastate_{\control'}}
                   {\cha}
                   {\sastateset_\branch}
                   {\sastateset_1, \ldots, \sastateset_\cpdsord}
    \]
    and $\satset_1$ has strict $(\opord, \sastateset_\opord)$-expansion
    \[
        \satranfull{\sastateset_\opord}{\cha}{\sastateset'_\branch}{\sastateset'_1,
        \ldots, \sastateset'_\opord}
    \]
    and by induction $\sat_1$ and $\satset_1$ are in $\saauta_\idxi$ for some $\idxi$.
    Since we have
    \[
        \srcorig_1 =
            \tuple{\control,
                   \sastateset_2 \cup \sastateset'_2,
                   \ldots
                   \sastateset_{\opord-1} \cup \sastateset'_{\opord-1},
                   \sastateset'_\opord,
                   \sastateset_{\opord+1},
                   \ldots,
                   \sastateset_\cpdsord}
    \]
    we have
    $\sat \in \apply{\extractshort}{\satfull}$
    where
    \[
        \satfull = \brac{
            \satranfull{\sastate_\control}{\cha}{\sastateset_\branch \cup
            \sastateset'_\branch}{\sastateset_1 \cup \sastateset'_1, \ldots,
            \sastateset_{\opord-1} \cup \sastateset'_{\opord-1},
            \sastateset'_\opord, \sastateset_{\opord+1}, \ldots,
            \sastateset_\cpdsord}
        }
    \]
    which was added by the naive saturation algorithm from the
    $\push{\opord}$ rule and the transitions in $\saauta_\idxi$.
    Hence, we satisfy the lemma.

    Otherwise, $\mathit{src}$ was added due to a call to
    $\createtrip$ during initialisation from an alternating rule
    $\cpdsalttran{\control}{\controlset}$.
    Then we have $\opord = \cpdsord$,
    $\srcorig_\opord = \tuple{\control}$,
    $\sastate'_{\opord-1} = \bot$,
    and
    $\sastateset^s_\opord = \setcomp{\sastate_{\control'}}{\control' \in \controlset}$.
    From the induction we have a set of order-$1$ transitions
    $\satset \subseteq \transdone^\idxj$
    with strict $(\opord-1, \sastateset^t_{\opord-1})$-expansion
    $\satranfull{\sastateset^t_{\opord-1}}
                {\cha}
                {\sastateset^t_\branch}
                {\sastateset^{t'}_1, \ldots, \sastateset^{t'}_{\opord-1}}$.
    Thus $\satset$ has the strict $(\opord, \sastateset^s_\opord)$-expansion (recalling $\opord = \cpdsord$)
    $\satranfull{\sastateset^s_\opord}
                {\cha}
                {\sastateset^{t}_\branch}
                {\sastateset^{t'}_1, \ldots, \sastateset^{t'}_\cpdsord}$.
    Moreover, by induction, we have $\satset$ in $\saauta_\idxi$ for some $\idxi$.
    We added $\sat$ as part of
    $\satranfull{\sastate_\control}
                {\cha}
                {\sastateset^{t}_\branch}
                {\sastateset^{t'}_1, \ldots, \sastateset^{t'}_\cpdsord}$
    which was added by the naive saturation algorithm given the alternating rule and transitions in $\saauta_\idxi$.
\qed

\subsubsection{Completeness}

We prove that the algorithm is complete.  For this we need some preliminary
lemmas stating properties of the data-structures maintained by the algorithm.

\begin{lem}
    \label{lem:source-imp-targ}
    For all $\opord \geq 2$ and $\idxj \geq 0$, all $\satset \subseteq
    \transdone^\idxj$ witnessing $\sastateset^\targmarker_\opord
    \satran{\sastateset^\targmarker_{\opord-1}} \sastateset'_\opord$, and all
    $\tuple{\srcorig_\opord, \sastate_{\opord-1}, \cha, \sastateset_\opord} \in
    \sources^\idxj[\opord]$ such that $\sastateset^\targmarker_\opord \subseteq
    \sastateset_\opord$, we have $\tuple{\sastateset_\opord,
    \sastateset_\opord \setminus \sastateset^\targmarker_\opord,
    \sastateset^\targmarker_{\opord-1}, \sastateset'_\opord}$ in
    $\targs^\idxj[\opord]$.
\end{lem}
\proof
    Let $\idxj_1$ be the iteration of Algorithm~\ref{alg:main} where
    $\tuple{\srcorig_\opord, \sastate_{\opord-1}, \cha, \sastateset_\opord}$ was
    first added to $\sources^{\idxj_1}[\opord]$.  We perform an induction over
    $\idxj_1$.  The base case is trivial.  In the inductive
    case, the only position where a source may be added is in the $\createtrip$
    procedure.  After adding the source, we re-establish the induction hypothesis.  There are two cases.

    Let $\mathit{targ} = \tuple{\sastateset_\opord, \sastateset_\opord,
    \emptyset, \emptyset}$.  If $\mathit{targ}$ is already in $\targs^{\idxj_1}$
    then we observe that a target of the form $\tuple{\sastateset_\opord,
    \sastateset_\opord, \ldots}$ is only created in $\createtrip$ (targets are
    also created in $\proctarg$, but these targets are obtained by removing a
    state from the second component of an existing target, hence the two first
    components cannot be equal).  This implies there exists a source
    $\tuple{\_, \_, \_, \sastateset_\opord} \in \sources^{\idxj'}[\opord]$ for
    some $\idxj' < \idxj_1$.  This gives the result by induction since
    $\satset$ and the desired target depend only on the final component of the
    source.

    If $\mathit{targ}$ is not in $\targs^{\idxj_1}[\opord]$, then we add it.
    Next, split $\satset = \satset_1 \cup \satset_2$ such that $\satset_1$
    contains all $\sat \in \satset$ appearing in $\transdone^{\idxj_1-1}$.  The
    balance is contained in $\satset_2$.  The algorithm proceeds to call
    $\proctarg$ on $\mathit{targ}$ and all $\sat \in \transdone^{\idxj_1}$.  In
    particular, this includes all $\sat \in \satset_1$.

    We aim to prove that, after the execution of this loop, we have
    $(\sastateset_\opord, \sastateset_\opord \setminus \sastateset^1_\opord,
    \sastateset^1_{\opord-1}, \sastateset^{1'}_\opord) \in
    \targs^{\idxj_1}[\opord]$ when $\satset_1$ witnesses $\sastateset^1_\opord
    \satran{\sastateset^1_{\opord-1}} \sastateset^{1'}_\opord$.

    Let $\sat_1, \ldots, \sat_\numof$ be a linearisation of $\satset_1$ in the
    order they appear in iterations over $\transdone$ (we assume a fixed order
    here for convenience, though the proof can generalise if the order changes
    between iterations).  Additionally, let $\satset^{\idxz} = \set{\sat_1,
    \ldots, \sat_\idxz}$ witness $\sastateset^{\sat_\idxz}_\opord
    \satran{\sastateset^{\sat_\idxz}_{\opord-1}}
    \sastateset^{\sat'_\idxz}_\opord$.  We show after $\satset^\idxz$ has been
    processed, we have $(\sastateset_\opord, \sastateset_\opord \setminus
    \sastateset^{\sat_\idxz}_\opord, \sastateset^{\sat_\idxz}_{\opord-1},
    \sastateset^{\sat'_\idxz}_\opord) \in \targs^{\idxj_1}[\opord]$.  This gives
    us the property once $\idxz = \numof$.
    That is
    $\satset^{\numof}$
    witnesses
    $\sastateset^{\sat_\numof}_\opord
     \satran{\sastateset^{\sat_\numof}_{\opord-1}}
     \sastateset^{\sat'_\numof}_\opord$
    which is
    $\sastateset^1_\opord
     \satran{\sastateset^1_{\opord-1}} \sastateset^{1'}_\opord$.

    In the base case $\idxz = 0$ and we
    are done.  Otherwise, we know that $\mathit{targ}_\idxz = (\sastateset_\opord,
    \sastateset_\opord \setminus \sastateset^{\sat_\idxz}_\opord,
    \sastateset^{\sat_\idxz}_{\opord-1}, \sastateset^{\sat'_\idxz}_\opord) \in
    \targs^{\idxj_1}[\opord]$ and prove the case for $(\idxz + 1)$.  Consider
    the call to $\addtarget$ that added $\mathit{targ}_\idxz$.  Now take the
    iteration against $\transdone$ that processes $\sat_{\idxz+1}$.  This
    results in the addition of $(\sastateset_\opord, \sastateset_\opord
    \setminus \sastateset^{\sat_{\idxz+1}}_\opord,
    \sastateset^{\sat_{\idxz+1}}_{\opord-1},
    \sastateset^{\sat'_{\idxz+1}}_\opord)$ as required.

    Hence, we have $(\sastateset_\opord, \sastateset_\opord \setminus
    \sastateset^1_\opord, \sastateset^1_{\opord-1}, \sastateset^{1'}_\opord) \in
    \targs^{\idxj_1}[\opord]$.  Now, let $\sat_1, \ldots, \sat_\numof$ be a
    linearisation of $\satset_2$ in the order they are added to $\transdone$.
    Additionally, we write $\sastateset^{\sat_\idxz}_\opord
    \satran{\sastateset^{\sat_\idxz}_{\opord-1}}
    \sastateset^{\sat_\idxz'}_\opord$ for the state-sets and transitions
    witnessed by $\satset_1 \cup \set{\sat_1, \ldots, \sat_\idxz}$.

    We show after $\sat_\idxz$ has been added to $\transdone$ on the $\idxj'$th
    iteration, we have that $(\sastateset_\opord, \sastateset_\opord \setminus
    \sastateset^{\sat_\idxz}_\opord, \sastateset^{\sat_\idxz}_{\opord-1},
    \sastateset^{\sat'_\idxz}_\opord) \in \targs^{\idxj'}[\opord]$ for some
    $\idxj'$.  In the base case $\idxz = 0$ and we are done by the argument
    above.  Otherwise, we know that $\mathit{targ}_\idxz = (\sastateset_\opord,
    \sastateset_\opord \setminus \sastateset^{\sat_\idxz}_\opord,
    \sastateset^{\sat_\idxz}_{\opord-1}, \sastateset^{\sat'_\idxz}_\opord) \in
    \targs^{\idxj'}[\opord]$ and prove the case for $(\idxz + 1)$.  Consider the
    call to $\updatetrip$ with $\sat_{\idxz+1}$.  This results in the addition
    of the target $(\sastateset_\opord, \sastateset_\opord \setminus
    \sastateset^{\sat_{\idxz+1}}_\opord,
    \sastateset^{\sat_{\idxz+1}}_{\opord-1},
    \sastateset^{\sat'_{\idxz+1}}_\opord)$ via the call to $\proctarg$.  When
    $\idxz = \numof$, we have the lemma as required.
\qed

\begin{lem}
    \label{lem:source-imp-targ-o1}
    For all $\idxj \geq 0$, all $\satset \subseteq \transdone^\idxj$ witnessing
    $\sastateset^\targmarker_1 \satrancol{\cha}{\sastateset^\targmarker_\branch}
    \sastateset'_1$, and all $\tuple{\srcorig_1, \bot, \cha, \sastateset_1} \in
    \sources^\idxj[1]$ such that $\sastateset^\targmarker_\opord \subseteq
    \sastateset_1$, we have $\tuple{\sastateset_1, \sastateset_1 \setminus
    \sastateset^\targmarker_1, \cha, \sastateset^\targmarker_\branch, \sastateset'_1}$
    in $\targs^\idxj[1]$.
\end{lem}
\proof
    The proof is essentially the same as the proof when $\opord \geq 2$.  Let
    $\idxj_1$ be the iteration of Algorithm~\ref{alg:main} where
    $\tuple{\srcorig_1, \sastate_{\opord-1}, \cha, \sastateset_1}$ was first
    added to $\sources^{\idxj_1}[1]$.  We perform an induction over $\idxj_1$.
    In the base case the lemma is trivially true.  In the inductive case, the
    only position where a source may be added is in the $\createtrip$ procedure.
    After adding the source, the induction hypothesis needs to be
    re-established.  There are two cases.

    Let $\mathit{targ} = \tuple{\sastateset_1, \sastateset_1, \cha, \emptyset,
    \emptyset}$.  If $\mathit{targ}$ is already in $\targs^{\idxj_1}$ then we
    observe that a target of the form $\tuple{\sastateset_1, \sastateset_1,
    \ldots}$ is only created in $\createtrip$.  This implies the existence of a
    source $\tuple{\_, \_, \_, \sastateset_1} \in \sources^{\idxj'}[1]$ for some
    $\idxj' < \idxj_1$.  This implies the result by induction since neither
    $\satset$ nor the desired target depend any but the final component of the
    source.

    If $\mathit{targ}$ is not in $\targs^{\idxj_1}[1]$, then we add it.  Next,
    split $\satset = \satset_1 \cup \satset_2$ such that $\satset_1$ contains
    all $\sat \in \satset$ appearing in $\transdone^{\idxj_1-1}$.  The balance
    is contained in $\satset_2$.  The algorithm proceeds to call $\proctarg$ on
    $\mathit{targ}$ and all $\sat \in \transdone^{\idxj_1}$.  In particular,
    this includes all $\sat \in \satset_1$.

    We aim to prove that, after the execution of this loop, we have
    $(\sastateset_1, \sastateset_1 \setminus \sastateset^1_1, \cha,
    \sastateset^1_\branch, \sastateset^{1'}_1) \in \targs^{\idxj_1}[1]$ when
    $\satset_1$ witnesses $\sastateset^1_1
    \satrancol{\cha}{\sastateset^1_\branch} \sastateset^{1'}_1$.

    Let $\sat_1, \ldots, \sat_\numof$ be a linearisation of $\satset_1$ in the
    order they appear in iterations over $\transdone$.  Additionally, let
    $\satset^{\idxz} = \set{\sat_1, \ldots, \sat_\idxz}$ witness
    $\sastateset^{\sat_\idxz}_1
    \satrancol{\cha}{\sastateset^{\sat_\idxz}_\branch}
    \sastateset^{\sat'_\idxz}_1$.  We show after $\satset^\idxz$ has been
    processed, we have $(\sastateset_1, \sastateset_1, \setminus
    \sastateset^{\sat_\idxz}_1, \cha, \sastateset^{\sat_\idxz}_\branch,
    \sastateset^{\sat'_\idxz}_1) \in \targs^{\idxj_1}[1]$.  This gives us the
    property once $\idxz = \numof$.  In the base case $\idxz = 0$ and we are
    done.  Otherwise, we know that $\mathit{targ}_\idxz = (\sastateset_1,
    \sastateset_1 \setminus \sastateset^{\sat_\idxz}_1, \cha,
    \sastateset^{\sat_\idxz}_\branch, \sastateset^{\sat'_\idxz}_1) \in
    \targs^{\idxj_1}[1]$ and prove the case for $(\idxz + 1)$.  Consider the
    call to $\addtarget$ that added $\mathit{targ}_\idxz$.  Now take the
    iteration against $\transdone$ that processes $\sat_{\idxz+1}$.  This
    results in the addition of $(\sastateset_1, \sastateset_1 \setminus
    \sastateset^{\sat_{\idxz+1}}_1, \cha, \sastateset^{\sat_{\idxz+1}}_\branch,
    \sastateset^{\sat'_{\idxz+1}}_1)$ as required.

    Hence, we have $(\sastateset_1, \sastateset_1 \setminus \sastateset^1_1,
    \cha, \sastateset^1_\branch, \sastateset^{1'}_1) \in \targs^{\idxj_1}[1]$.
    Now, let $\sat_1, \ldots, \sat_\numof$ be a linearisation of $\satset_2$ in
    the order they are added to $\transdone$.  Additionally, we write
    $\sastateset^{\sat_\idxz}_1
    \satrancol{\cha}{\sastateset^{\sat_\idxz}_\branch}
    \sastateset^{\sat_\idxz'}_1$ for the state-sets and transitions witnessed by
    $\satset_1 \cup \set{\sat_1, \ldots, \sat_\idxz}$.

    We show after $\sat_\idxz$ is added to $\transdone$ on the $\idxj'$th
    iteration, we have that $(\sastateset_1, \sastateset_1 \setminus
    \sastateset^{\sat_\idxz}_1, \cha, \sastateset^{\sat_\idxz}_\branch,
    \sastateset^{\sat'_\idxz}_1) \in \targs^{\idxj'}[1]$ for some $\idxj'$.  In
    the base case $\idxz = 0$ and we are done by the argument above.  Otherwise,
    we know $\mathit{targ}_\idxz = (\sastateset_1, \sastateset_1 \setminus
    \sastateset^{\sat_\idxz}_1, \cha, \sastateset^{\sat_\idxz}_\branch,
    \sastateset^{\sat'_\idxz}_1) \in \targs^{\idxj'}[1]$ and prove the case for
    $(\idxz + 1)$.  Consider the call to $\updatetrip$ with $\sat_{\idxz+1}$.
    This results in the addition of the target $(\sastateset_1, \sastateset_1
    \setminus \sastateset^{\sat_{\idxz+1}}_1, \cha,
    \sastateset^{\sat_{\idxz+1}}_\branch, \sastateset^{\sat'_{\idxz+1}}_1)$ via
    the call to $\proctarg$.  When $\idxz = \numof$, we have the lemma as
    required.
\qed

\begin{lem}
    \label{lem:source-targ-imp-trip}
    For all $\opord > 1$ and $\idxj \geq 0$, if we have $\tuple{\srcorig_\opord,
    \sastate_{\opord-1}, \cha, \sastateset_\opord} \in \sources^\idxj[\opord]$
    with
    $\srcorig_\opord = \tuple{\control,
                              \sastateset_{\opord+1},
                              \ldots,
                              \sastateset_{\cpdsord}}$
    and also $\tuple{\sastateset_\opord, \emptyset, \sastateset_{\opord-1},
    \sastateset'_\opord} \in \targs^\idxj[\opord]$, then it is the case that
    there exists $\idxj' \geq 0$ such that we have
    $\tuple{\srcorig_{\opord-1}, \bot, \cha, \sastateset_{\opord-1} \cup S}
     \in \sources^{\idxj'}[\opord-1]$ where
    $\srcorig_{\opord-1} =
     \tuple{\control,
            \sastateset'_\opord,
            \sastateset_{\opord+1}, \ldots, \sastateset_\cpdsord}$
    and
    $S = \setcomp{\sastate_{\opord-1}}{\sastate_{\opord-1} \neq \bot}$.
\end{lem}
\proof
    Let $\idxj_1$ be the smallest such that $\tuple{\srcorig_\opord,
    \sastate_{\opord-1}, \cha, \sastateset_\opord} \in
    \sources^{\idxj_1}[\opord]$ and $\idxj_2$ be the smallest such that
    $\tuple{\sastateset_\opord, \emptyset, \sastateset_{\opord-1},
    \sastateset'_\opord} \in \targs^{\idxj_2}[\opord]$.

    In the case $\idxj_1 \leq \idxj_2$, we consider the $\idxj_2$th iteration of
    Algorithm~\ref{alg:main} at the moment where the target is added to
    $\targs^{\idxj_2}[\opord]$.  This has to be a result of the call to
    $\addtarget$ during Algorithm~\ref{alg:proc-tran}.  The only other place
    $\addtarget$ may be called is during Algorithm~\ref{alg:create-tripwire};
    however, this implies the target is of the form $\tuple{\sastateset_\opord,
    \sastateset_\opord, \emptyset, \emptyset}$ and hence, for the target to be
    complete, it must be $\tuple{\emptyset, \emptyset, \emptyset, \emptyset}$
    and hence $\idxj_2 = 0$, and since $\idxj_1 > 0$ (since there are initially
    no sources) we have a contradiction.  Hence, the target is added during
    Algorithm~\ref{alg:proc-tran} and the procedure goes on to call
    $\procsrccomptarg$ against each matching source in
    $\sources^{\idxj_2}[\opord]$, including $\tuple{\srcorig_\opord,
    \sastate_{\opord-1}, \cha, \sastateset_\opord}$.  This results in the
    addition of $\tuple{\srcorig_{\opord-1}, \bot, \cha,
    \sastateset_{\opord-1} \cup S}$ to $\sources^{\idxj_2}[\opord-1]$, if it is
    not there already, satisfying the lemma.

    In the case $\idxj_1 > \idxj_2$, we consider the $\idxj_1$th iteration of
    Algorithm~\ref{alg:main} at the moment where the source is added.  This is
    necessarily in the $\createtrip$ procedure.  Since $\tuple{\sastateset_\opord,
    \emptyset, \sastateset_{\opord-1}, \sastateset'_\opord} \in
    \targs^{\idxj_1}[\opord]$ and since this target must have been obtained from
    a target of the form $\tuple{\sastateset_\opord, \sastateset_\opord,
    \emptyset, \emptyset}$, we know that $\tuple{\sastateset_\opord,
    \sastateset_\opord, \emptyset, \emptyset} \in \targs^{\idxj_1}[\opord]$ and
    thus the procedure calls $\procsrccomptarg$ for each complete target
    including $\tuple{\sastateset_\opord, \emptyset, \sastateset_{\opord-1},
    \sastateset'_\opord}$.  Thus we add
    $\tuple{\srcorig_{\opord-1}, \bot, \cha, \sastateset_{\opord-1} \cup S}$
    to
    $\sources^{\idxj_2}[\opord-1]$,
    if it is not there already, satisfying the lemma.
\qed

\begin{lem}
    \label{lem:source-targ-imp-tran}
    For all $\idxj \geq 0$, if $\tuple{\srcorig_1, \bot, \cha, \sastateset_1}
    \in \sources^\idxj[1]$ and $\tuple{\sastateset_1, \emptyset, \cha,
    \sastateset_\branch, \sastateset'_1} \in \targs^\idxj[1]$, if
    $\srcorig_1 = \tuple{\control, \sastateset_2, \ldots, \sastateset_\cpdsord}$,
    then for each $\sat$ in
    \[
        \apply{\extractshort}{\satranfull{\sastate_\control}{\cha}{\sastateset_\branch}{\sastateset'_1,
        \sastateset_2, \ldots, \sastateset_\cpdsord}}
    \]
    there exists some $\idxj' \geq 0$ such that $\sat \in \transdone^{\idxj'}$.
\end{lem}
\proof
    As before, the proof of this order-$1$ case is very similar to the
    order-$\opord$ proof.

    Let $\idxj_1$ be the smallest such that $\tuple{\srcorig_1, \bot, \cha,
    \sastateset_1} \in \sources^{\idxj_1}[1]$ and $\idxj_2$ be the smallest such
    that $\tuple{\sastateset_1, \emptyset, \cha, \sastateset_\branch,
    \sastateset'_1} \in \targs^{\idxj_2}[1]$.

    In the case $\idxj_1 \leq \idxj_2$, we consider the $\idxj_2$th iteration of
    Algorithm~\ref{alg:main} at the moment where the target is added to
    $\targs^{\idxj_2}[1]$.  This has to be a result of the call to $\addtarget$
    during Algorithm~\ref{alg:proc-tran}.  The only other place $\addtarget$ may
    be called is during Algorithm~\ref{alg:create-tripwire}; however, this
    implies the target is of the form $\tuple{\sastateset_1, \sastateset_1,
    \emptyset, \emptyset}$ and hence, for the target to be complete, it must be
    $\tuple{\emptyset, \emptyset, \emptyset, \emptyset}$ and hence $\idxj_2 =
    0$, and since $\idxj_1 > 0$ (since there are initially no sources) we have a
    contradiction.  Hence, the target is added during
    Algorithm~\ref{alg:proc-tran} and the procedure goes on to call
    $\procsrccomptarg$ against each matching source in $\sources^{\idxj_2}[1]$,
    including $\tuple{\srcorig_1, \bot, \cha, \sastateset_1}$.  This results in
    the addition of
    $\satranfull{\sastate_\control}{\cha}{\sastateset_\branch}{\sastateset'_1,
    \sastateset_2, \ldots, \sastateset_\cpdsord}$ satisfying the lemma.

    In the case $\idxj_1 > \idxj_2$, we consider the $\idxj_1$th iteration of
    Algorithm~\ref{alg:main} at the moment where the source is added.  This is
    necessarily in the $\createtrip$ procedure.  Since $\tuple{\sastateset_1,
    \emptyset, \cha, \sastateset_\branch, \sastateset'_1} \in
    \targs^{\idxj_1}[1]$ and since this target must have been obtained from a
    target of the form $\tuple{\sastateset_1, \sastateset_1, \cha, \emptyset,
    \emptyset}$, we know that $\tuple{\sastateset_1, \sastateset_1, \cha, \emptyset,
    \emptyset} \in \targs^{\idxj_1}[1]$ and thus the procedure calls
    $\procsrccomptarg$ against each complete target including
    $\tuple{\sastateset_1, \emptyset, \cha, \sastateset_\branch,
    \sastateset'_1}$.  This results in the addition of
    $\satranfull{\sastate_\control}{\cha}{\sastateset_\branch}{\sastateset'_1,
    \sastateset_2, \ldots, \sastateset_\cpdsord}$ satisfying the lemma.
\qed

We are now ready to prove completeness.

\begin{lem}
    \label{lem:fast-alg-only-if}
    Given a CPDS $\cpds$ and stack automaton $\saauta_0$, let $\saauta$ be the
    result of Algorithm~\ref{alg:main}.  We have
    $\config{\control}{\stackw} \in \prestar{\cpds}{\saauta_0}$
    implies
    $\stackw \in \slang{\control}{\saauta}$.
\end{lem}
\proof
    We know
    (from the correctness of saturation (Theorem~\ref{thm:sat-correct}))
    that the fixed point of
    $\sequence{\saauta_\idxi}{\idxi \geq 0}$ is an automaton recognising
    $\prestar{\cpds}{\saauta_0}$.  We prove, by induction, that for each
    transition $\sat$ appearing in $\saauta_\idxi$ for some $\idxi$, there
    exists some $\idxj$ such that $\sat$ appears in $\transdone^\idxj$.

    In the base case we have all
    transitions in $\saauta_0$ in $\transtodo$ at the beginning of
    Algorithm~\ref{alg:main}.  Since the main loop continues until $\transtodo$
    has been completely transferred to $\transdone$, the result follows.

    Now, let $\sat$ be an order-$\opord$ transition appearing for the first time
    in $\saauta_\idxi$ ($\idxi > 0$).  We case split on the pushdown
    operation that led to the introduction of the transition.  Let
    $\cpdsruler = \cpdsrule{\control}{\cha}{\genop}{\control'}$ be the rule that
    led to the new transition.  We first consider simple cases.
    \begin{itemize}
        \item When $\genop = \pop{\opord}$, then
              when $\opord = \cpdsord$, we added $\sat$ as part of
              $\satranfull{\sastate_{\control}}
                          {\cha}
                          {\emptyset}
                          {\emptyset, \ldots, \emptyset, \set{\sastate_{\control'}}}$.
              In this case we also added $\sat$
              to $\transtodo$ as part of the initialisation
              Algorithm~\ref{alg:main}.
              Otherwise $\cpdsord > \opord$ and there was a state
              $\sastate_\opord$
              with $\cpdsord$-expansion
              $\satranfullk{\sastate_{\control'}}
                           {\sastate_\opord}
                           {\sastateset_{\opord+1}, \dots, \sastateset_\cpdsord}$
              in $\saauta_\idxi$ and we added to
              $\saauta_{\idxi+1}$
              \[
                  \satfull = \brac{
                      \satranfull{\sastate_{\control}}{\cha}{\emptyset}{\emptyset,
                      \ldots, \emptyset, \set{\sastate_\opord},
                      \sastateset_{\opord+1}, \ldots, \sastateset_\cpdsord}
                  }
              \]
              and
              $\sat \in \apply{\extractshort}{\satfull}$.
              By induction we have $\idxj$
              such that
              $\sastate_{\opord+1} \satran{\sastate_\opord} \sastateset_{\opord+1}$
              appears in $\transdone^\idxj$.  Consider
              the $\idxj$th iteration of Algorithm~\ref{alg:main} when
              $\updatebyrules$ is called on $\sat'$.  The $\pop{\opord'}$ loop
              immediately adds $\satfull$ which involves adding $\sat$ to $\transtodo$, giving us some $\idxj' > \idxj$
              such that $\sat$ appears in $\transdone^{\idxj'}$.

        \item When $\genop = \collapse{\opord}$, when $\opord = \cpdsord$, we
              have
              $\sat \in \apply{\extractshort}{\satfull}$
              where the addition of
              $\satfull = \brac{
                  \satranfull{\sastate_{\control}}
                             {\cha}
                             {\set{\sastate_{\control'}}}
                             {\emptyset, \ldots, \emptyset}
               }$
              led to the addition of $\sat$.
              In this case we also added $\sat$
              to $\transtodo$ as part of the initialisation steps of
              Algorithm~\ref{alg:main}.  Otherwise, $\cpdsord > \opord$ and from
              a state $\sastate_\opord$ with $\cpdsord$-expansion
              $\satranfullk{\sastate_{\control'}}
                           {\sastate_\opord}
                           {\sastateset_{\opord+1}, \dots, \sastateset_\cpdsord}$
              we added
              \[
                  \satfull = \brac{
                      \satranfull{\sastate_{\control}}{\cha}{\set{\sastate_\opord}}{\emptyset,
                      \ldots, \emptyset, \sastateset_{\opord+1}, \ldots,
                      \sastateset_\cpdsord}
                  }
              \]
              with
              $\sat \in \apply{\extractshort}{\satfull}$.
              By induction we have $\idxj$ such that
              $\sastate_{\opord+1}
               \satran{\sastate_\opord}
               \sastateset_{\opord+1}$
              appears in $\transdone^\idxj$.  Consider
              the $\idxj$th iteration of Algorithm~\ref{alg:main} when
              $\updatebyrules$ is called on $\sat'$.  The $\collapse{\opord}$ loop
              immediately adds $\satfull$ and hence $\sat$ to $\transtodo$, giving us some $\idxj' > \idxj$
              such that $\sat$ appears in $\transdone^{\idxj'}$.

        \item when $\genop = \rew{\chb}$ then from a transition with $\cpdsord$-expansion
              $\satranfull{\sastate_{\control'}}
                          {\chb}
                          {\sastateset_\branch}
                          {\sastateset_1, \dots, \sastateset_\cpdsord}$
              we added
              $\satfull = \brac{
                    \satranfull{\sastate_{\control}}
                               {\cha}
                               {\sastateset_\branch}
                               {\sastateset_1, \dots, \sastateset_\cpdsord}
               }$
              with
              $\sat \in \apply{\extractshort}{\satfull}$.
              By induction, we know that the order-$1$ transition $\sat'$
              with $\cpdsord$-expansion
              $\satranfull{\sastate_{\control'}}
                          {\chb}
                          {\sastateset_\branch}
                          {\sastateset_1, \ldots, \sastateset_\cpdsord}$
              appears in $\transdone^\idxj$ for some $\idxj$.
              Consider the $\idxj$th iteration of the main loop of
              Algorithm~\ref{alg:main}.  During this iteration $\sat'$ is passed
              to $\updatebyrules$, and the loop handling rules containing
              $\rew{\chb}$ adds $\satfull$ to the worklist.  Since
              $\sat \in \apply{\extractshort}{\satfull}$
              there must be some $\idxj'$ such that
              $\sat$ appears in $\transdone^{\idxj'}$.
    \end{itemize}
    We now consider the push rules, which require more intricate reasoning.
    \begin{itemize}
        \item when $\genop = \push{\opord}$, we had a transition with $\cpdsord$-expansion
              $\satranfull{\sastate_{\control'}}
                          {\cha}
                          {\sastateset_\branch}
                          {\sastateset_1,
                           \ldots,
                           \sastateset_\opord,
                           \ldots,
                           \sastateset_\cpdsord}$
              and $\satset$ with strict $(\opord, \sastateset_\opord)$-expansion
              $\satranfull{\sastateset_\opord}
                          {\cha}
                          {\sastateset'_\branch}
                          {\sastateset'_1, \ldots, \sastateset'_\opord}$
              in $\saauta_\idxi$, and we added
              \[
                  \satfull = \brac{
                      \satranfull{\sastate_{\control}}{\cha}{\sastateset_\branch
                      \cup \sastateset'_\branch}{\sastateset_1 \cup \sastateset'_1,
                      \ldots, \sastateset_{\opord-1} \cup \sastateset'_{\opord-1},
                      \sastateset'_\opord, \sastateset_{\opord+1}, \ldots,
                      \sastateset_\cpdsord}
                  }
              \]
              with
              $\sat \in \apply{\extractshort}{\satfull}$.
              Since
              $\sat_1 = \brac{
                \sastate_\opord \satran{\sastate_{\opord-1}} \sastateset_\opord
              }$
              where $\sastate_\opord$ has $\cpdsord$-expansion
              $\satranfullk{\sastate_{\control'}}
                           {\sastate_\opord}
                           {\sastateset_{\opord+1}, \ldots, \sastateset_\cpdsord}$
              already exists by the assumption of this case, by induction
              there is some $\idxj$ where $\sat_1$ first appears in $\transdone^\idxj$.
              Also by induction, for each $\sat'
              \in \satset$, there is some $\idxj'$ such that $\sat'$ first
              appears in $\transdone^{\idxj'}$.

              Consider the $\idxj$th iteration where $\sat_1$ is added to
              $\transdone$.
              Let
              $\srcorig_\opord =
               \tuple{\control,
                      \sastateset_{\opord+1}, \ldots, \sastateset_\cpdsord}$.
              During the call to $\updatebyrules$ we call
              $\createtrip$ in the loop handling push rules with the arguments
              $\srcorig_\opord$,
              $\sastate_{\opord-1}$,
              $\cha$,
              and
              $\sastateset_\opord$.

              The call ensures $\tuple{\srcorig_\opord, \sastate_{\opord-1},
              \cha, \sastateset_\opord} \in \sources^\idxj[\opord]$.
              Observe there is a unique $\sastateset''_{\opord-1}$ such that the
              $(\opord-1, \sastateset''_{\opord-1})$-expansion
              $\satranfull{\sastateset''_{\opord-1}}
                          {\cha}
                          {\sastateset'_\branch}
                          {\sastateset'_1, \ldots, \sastateset'_{\opord-1}}$
              of $\satset$ is strict.
              Now take $\idxj'$ and
              $\satset_\opord \subseteq \transdone^{\idxj'}$
              such that $\satset_\opord$ witnesses
              $\sastateset_\opord \satran{\sastateset''_{\opord-1}} \sastateset'_\opord$.
              We know such a $\idxj'$ and $\satset_\opord$ exist by induction and because of $\satset$.
              By Lemma~\ref{lem:source-imp-targ} we know that we have
              $\tuple{\sastateset_\opord,
                      \emptyset,
                      \sastateset''_{\opord-1},
                      \sastateset'_\opord} \in \targs^{\idxj'}[\opord]$,
              and then additionally by Lemma~\ref{lem:source-targ-imp-trip} that we have
              $\tuple{\srcorig_{\opord-1},
                      \bot,
                      \cha,
                      \sastateset''_{\opord-1} \cup \set{\sastate_{\opord-1}}}
               \in \sources^{\idxj''}[\opord-1]$ for some $\idxj''$
              where
              $\srcorig_{\opord-1} =
               \tuple{\control,
                      \sastateset'_\opord,
                      \sastateset_{\opord+1}, \ldots, \sastateset_\cpdsord}$.
              Set
              $\sastateset^\sastatelabel_{\opord-1} =
               \sastateset''_{\opord-1} \cup \set{\sastate_{\opord-1}}$.

              We iterate the above argument from $\opord' =
              \opord-1$ down to $\opord' = 1$.  Begin with $\idxj'$ such that
              $\tuple{\srcorig_{\opord'}, \bot, \cha,
              \sastateset^\sastatelabel_{\opord'}} \in
              \sources^{\idxj'}[\opord']$ and there is some
              $\satset_{\opord'} \subseteq \transdone^{\idxj'}$
              that witnesses
              $\sastateset^\sastatelabel_{\opord'}
               \satran{\sastateset^\sastatelabel_{\opord'-1}}
               \sastateset_{\opord'} \cup \sastateset'_{\opord'}$.
              This exists by combining transitions from the expansions in the assumptions of this case.
              By Lemma~\ref{lem:source-imp-targ} we know it to be the case that
              $\tuple{\sastateset^\sastatelabel_{\opord'}, \emptyset,
              \sastateset^\sastatelabel_{\opord'-1}, \sastateset_{\opord'} \cup
              \sastateset'_{\opord'}} \in \targs^{\idxj'}[\opord]$, and then by
              Lemma~\ref{lem:source-targ-imp-trip} we have that
              $\tuple{\srcorig_{\opord'-1},
                      \bot,
                      \cha,
                      \sastateset^\sastatelabel_{\opord'-1}}
               \in \sources^{\idxj''}[\opord-1]$
              for some $\idxj''$ where
              $\srcorig_{\opord'-1} =
               \tuple{\control,
                      \sastateset_{\opord'} \cup \sastateset'_{\opord'},
                      \ldots,
                      \sastateset_{\opord-1} \cup \sastateset'_{\opord-1},
                      \sastateset'_\opord,
                      \sastateset_{\opord+1}, \ldots, \sastateset_\cpdsord}$.

              When $\opord' = 1$, we have some $\idxj'$ with
              $\tuple{\srcorig_1, \bot, \cha, \sastateset^\sastatelabel_1}
               \in \sources^{\idxj'}[1]$
              and some
              $\satset_1 \subseteq \transdone^{\idxj'}$
              witnessing
              $\sastateset^\sastatelabel_1
               \satrancol{\cha}{\sastateset_\branch \cup \sastateset'_\branch}
               \sastateset_1 \cup \sastateset'_1$.
              Note,
              \[
                  \srcorig_1 = \tuple{\control,
                                      \sastateset_2 \cup \sastateset'_2
                                      \ldots,
                                      \sastateset_{\opord-1} \cup \sastateset'_{\opord-1},
                                      \sastateset'_\opord,
                                      \sastateset_{\opord+1},
                                      \ldots,
                                      \sastateset_\cpdsord} \ .
              \]
              By Lemma~\ref{lem:source-imp-targ-o1} we know that
              $\tuple{\sastateset^\sastatelabel_1,
                      \emptyset,
                      \cha,
                      \sastateset_\branch \cup \sastateset'_\branch,
                      \sastateset_1 \cup \sastateset'_1}
               \in \targs^{\idxj'}[\opord]$,
              and then by Lemma~\ref{lem:source-targ-imp-tran} we have $\idxj''$ such that we
          have all $\sat'$ in
              \[
                  \apply{\extractshort}{
                      \satranfull{\sastate_{\control}}
                                 {\cha}
                                 {\sastateset_\branch \cup \sastateset'_\branch}
                                 {\sastateset_1 \cup \sastateset'_1,
                                  \ldots,
                                  \sastateset_{\opord-1} \cup \sastateset'_{\opord-1},
                                  \sastateset'_\opord,
                                  \sastateset_{\opord+1}, \ldots, \sastateset_\cpdsord}
                  }
              \]
              in $\transdone^{\idxj''}$.  This, in particular, includes $\sat$.

        \item when $\genop = \cpush{\chb}{\opord}$ we had an order-$1$ transition $\sat_1$ with $\cpdsord$-expansion
              $\satranfull{\sastate_{\control'}}
                          {\chb}
                          {\sastateset_\branch}
                          {\sastateset_1, \ldots, \sastateset_\cpdsord}$
              and a set of order-$1$ transitions $\satset$ witnessing
              $\sastateset_1
               \satrancol{\cha}{\sastateset'_\branch}
               \sastateset'_1$
              in $\saauta_\idxi$ with
              $\sastateset_\branch \subseteq \sastates_\opord$
              and added
              \[
                  \satfull = \brac{
                      \satranfull{\sastate_{\control}}{\cha}{\sastateset'_\branch}{\sastateset'_1,
                      \sastateset_2, \ldots, \sastateset_{\opord-1},
                      \sastateset_\opord \cup \sastateset_\branch,
                      \sastateset_{\opord+1}, \ldots, \sastateset_\cpdsord}
                  }
              \]
              where
              $\sat \in \apply{\extractshort}{\satfull}$.
              By induction, there is some $\idxj$ where the transition $\sat_1$ is first in
              $\transdone^\idxj$.
              In addition, for each
              $\sat' \in \satset$,
              there is some $\idxj'$ such that $\sat'$ first appears in
              $\transdone^{\idxj'}$.

              Consider the $\idxj$th iteration where $\sat_1$ is added to $\transdone$.
              During the call to $\updatebyrules$ we call $\createtrip$ in the loop handling push rules with the arguments
              \[
                  \srcorig_1 =
                  \tuple{\control,
                         \sastateset_2,
                         \ldots,
                         \sastateset_{\opord-1},
                         \sastateset_\opord \cup \sastateset_\branch,
                         \sastateset_{\opord+1},
                         \ldots,
                         \sastateset_\cpdsord}
              \]
              and $\bot$, $\chb$ and $\sastateset_1$.

              The call ensures
              $\tuple{\srcorig_1, \bot, \chb, \sastateset_1}
               \in \sources^\idxj[1]$.
              Now take $\idxj'$ such that
              $\satset \subseteq \transdone^{\idxj'}$.
              We know that $\satset$ witnesses
              $\sastateset_1
               \satrancol{\cha}{\sastateset_\branch}
               \sastateset'_1$.
              By Lemma~\ref{lem:source-imp-targ-o1} we know that
              $\tuple{\sastateset_1,
                      \emptyset,
                      \cha,
                      \sastateset_\branch,
                      \sastateset'_1} \in \targs^{\idxj'}[1]$, and
              then by Lemma~\ref{lem:source-targ-imp-tran} we have $\idxj''$ such that we have all $\sat'$ in
              \[
                  \apply{\extractshort}{
                      \satranfull{\sastate_{\control}}
                                 {\cha}
                                 {\sastateset'_\branch}
                                 {\sastateset_1 \cup \sastateset'_1,
                                  \sastateset_2,
                                  \ldots,
                                  \sastateset_{\opord-1},
                                  \sastateset_\opord \cup \sastateset_\branch,
                                  \sastateset_{\opord+1},
                                  \ldots,
                                  \sastateset_\cpdsord}
                  }
              \]
              in $\transdone^{\idxj''}$.  This, in particular, includes $\sat$.
    \end{itemize}

    Finally, we consider the alternating rules.  Take a rule
    $\cpdsalttran{\control}{\controlset}$.
    We had $\satset$ with strict $(\cpdsord, \sastateset)$-expansion
    $\satranfull{\sastateset_\cpdsord}
                {\cha}
                {\sastateset'_\branch}
                {\sastateset'_1, \ldots, \sastateset'_\cpdsord}$
    in $\saauta_\idxi$, where
    $\sastateset_\cpdsord = \setcomp{\sastate_{\control'}}{\control' \in \controlset}$,
    and we added
    $
        \satfull = \brac{
            \satranfull{\sastate_{\control}}
                       {\cha}
                       {\sastateset'_\branch}
                       {\sastateset'_1, \ldots, \sastateset'_\cpdsord}
        }
    $
    which has
    $\sat \in \apply{\extractshort}{\satfull}$.
    By induction there is some $\idxj'$ such that
    $\satset \subseteq \transdone^{\idxj'}$.

    During initialisation we call $\createtrip$ with the arguments
    $\tuple{\control}$, $\bot$, $\cha$, and $\sastateset_\cpdsord$.
    The call ensures
    $\tuple{\tuple{\control},
            \bot,
            \cha,
            \sastateset_\cpdsord} \in \sources^\idxj[\opord]$.
    We now iterate from $\opord = \cpdsord$, down to $\opord = 2$.
    At each iteration, there is some $\idxj'$ such that
    $\satset_\opord \subseteq \transdone^{\idxj'}$
    where $\satset_\opord$ witnesses
    $\sastateset_\opord \satran{\sastateset_{\opord-1}} \sastateset'_\opord$
    (this comes from the containment of $\satset$ in some $\transdone^{\idxj''}$).
    Note, this defines $\sastateset_{\opord-1}$ for the next iteration.
    By Lemma~\ref{lem:source-imp-targ} we know that we have
    $\tuple{\sastateset_\opord,
            \emptyset,
            \sastateset_{\opord-1},
            \sastateset'_\opord} \in \targs^{\idxj'}[\opord]$,
    and then additionally by Lemma~\ref{lem:source-targ-imp-trip} that we have
    $\tuple{\srcorig_{\opord-1}, \bot, \cha, \sastateset_{\opord-1}}
     \in \sources^{\idxj''}[\opord-1]$
    for some $\idxj''$
    where
    $\srcorig_{\opord-1} =
     \tuple{\control,
            \sastateset'_\opord, \ldots, \sastateset'_\cpdsord}$.

    For $\opord = 1$ we have $\satset$ witnesses
    $\sastateset_1 \satrancol{\cha}{\sastateset'_\branch} \sastateset'_1$.
    By Lemma~\ref{lem:source-imp-targ} we know that we have
    $\tuple{\sastateset_1,
            \emptyset,
            \cha,
            \sastateset'_\branch,
            \sastateset'_1} \in \targs^{\idxj'}[\opord]$,
    and then by Lemma~\ref{lem:source-targ-imp-tran} we have $\idxj''$ such that we have all $\sat'$ in
    \[
        \apply{\extractshort}{
            \satranfull{\sastate_{\control}}
                       {\cha}
                       {\sastateset'_\branch}
                       {\sastateset'_1,
                        \ldots,
                        \sastateset'_\cpdsord}
        }
    \]
    in $\transdone^{\idxj''}$.  This, in particular, includes $\sat$.
    This completes the proof.
\qed

%% file: experiments.tex
\section{Experimental Results}
\label{sec:experiments} \label{sec:contrib-last}

We compared \cshore with the state-of-the-art verification tools for
higher-order recursion schemes (HORS) available on its release: \trecs~\cite{K09b},
\gtrecst~\cite{gtrecs2} (the successor of \cite{K11b}), and
\travmc~\cite{NRO12}.
In an extension to our original publication~\cite{BCHS13}, we have re-run these experiments to also compare with the verification tools released after \cshore: \preface~\cite{RNO14}, and \horsattwo~\cite{horsat2}.
As a further extension, we have tested the efficacy of the improved fixed point computation in Section~\ref{sec:fastalgorithm} by implementing a naive fixed point computation where, during each iteration, each rule is tested against the current automaton to search for new transitions.

Benchmarks are from the \trecs and \travmc benchmark
suites, plus several larger examples provided by Kobayashi.
The majority of the \travmc benchmarks were
translated into HORS from an extended formalism, HORS with Case statements
(HORSC), using a script by Kobayashi.  For fairness, all tools in our
experiments took a pure HORS as input.  However, the authors of \travmc report
that \travmc performs faster on the original HORSC examples than on their HORS
translations.

In all cases, the benchmarks consist of a HORS (generating a computation tree)
and a property automaton.  In the case of \cshore, the property automaton is a
regular automaton describing branches of the generated tree that are considered
errors.  Thus, following the intuition in Section~\ref{sec:functionalIntro}, we
can construct a reachability query over a (non-alternating) CPDS, where the reachability of a
control state $\qerror$ indicates an erroneous branch (see~\cite{CS12} for more
details).  All other tools check co-reachability properties of HORS and thus
the property automaton describes only valid branches of the computation tree.
In all cases, it was straightforward to translate between the co-reachability
and reachability properties.

The experiments were run on a Dell Latitude e6320 laptop with 4GB of RAM and
four 2.7GHz Intel i7-2620M cores.  We ran \cshore on OpenJDK 8.0 using the argument \hbox{``-Xmx''} to limit RAM usage to
2.5GB.  As advised by the \travmc developers, we ran \travmc and \preface on the Mono JIT
compiler (version 4.6.1) with no command line arguments.  Finally \trecs (version
1.34), \gtrecst (version 3.17), and \horsattwo were compiled with the OCaml version 4.02.3
compilers.  On negative examples, \gtrecst was run with its \texttt{-neg}
argument.  We used the ``ulimit'' command to limit memory usage to 2.5GB and set
a CPU timeout of 600 seconds (per benchmark).  The given runtimes were reported
by the respective tools and are the means of three separate runs on each
example.  Note, \cshore was run until the automaton was completely
saturated.

\begin{sidewaystable}
\scriptsize
\vspace{15cm}
\begin{tabular}{lccccccccccccc}
    \toprule
    Benchmark file                      & Ord & Sz   & T       & TMC    & G      & N      & C      & P      & H     & \tick / \cross & Ctran & Ccpds  & Capprox \\
    \midrule
    \texttt{example3-1} (bug)           & 1   & 8    & 0.000   & 0.111  & ---    & 0.060  & 0.059  & 0.293  & 0.003 &                & 0.027 & 0.032  & 0.016   \\
    \texttt{file}                       & 1   & 8    & 0.000   & 0.032  & ---    & 0.051  & 0.053  & 0.286  & 0.003 &                & 0.026 & 0.027  & 0.022   \\
    \texttt{fileocamlc}                 & 4   & 111  & 0.027   & 0.047  & 0.042  & ---    & 0.222  & 0.295  & 0.010 & \cross         & 0.045 & 0.177  & 0.130   \\
    \texttt{lock2}                      & 4   & 45   & 0.036   & 0.050  & 0.261  & ---    & 0.235  & 0.331  & 0.010 &                & 0.034 & 0.201  & 0.101   \\
    \texttt{order5}                     & 5   & 52   & 0.013   & 0.042  & ---    & 37.152 & 0.250  & 0.315  & 0.010 &                & 0.037 & 0.213  & 0.090   \\
    \texttt{order5-2}                   & 5   & 40   & 0.044   & 0.073  & ---    & ---    & 0.163  & 0.317  & 0.007 &                & 0.034 & 0.129  & 0.070   \\
    \texttt{order5-variant}             & 5   & 55   & 0.043   & 0.042  & 1.094  & ---    & 0.242  & 0.322  & 0.010 &                & 0.038 & 0.204  & 0.077   \\
    \texttt{filepath}                   & 2   & 5956 & 215.401 & ---    & ---    & 0.205  & 0.212  & 0.503  & 0.040 & \tick          & 0.075 & 0.136  & 0.130   \\
    \texttt{filter-nonzero} (bug)       & 5   & 484  & 0.013   & 0.141  & 0.284  & ---    & 1.783  & 0.554  & 0.026 & \cross         & 0.064 & 1.719  & 1.450   \\
    \texttt{filter-nonzero-1}           & 5   & 890  & 0.281   & 96.163 & ---    & ---    & 5.018  & 1.827  & 0.100 &                & 0.093 & 4.925  & 4.244   \\
    \texttt{map-head-filter} (bug)      & 3   & 370  & 0.012   & 0.123  & 0.076  & ---    & 0.298  & 0.393  & 0.013 & \cross         & 0.055 & 0.243  & 0.093   \\
    \texttt{map-head-filter-1}          & 3   & 880  & 0.238   & 0.698  & ---    & 0.242  & 0.229  & 0.366  & 0.016 & \tick          & 0.071 & 0.158  & 0.151   \\
    \texttt{map-plusone}                & 5   & 302  & 0.034   & 0.088  & 0.224  & ---    & 0.827  & 0.398  & 0.013 & \cross         & 0.063 & 0.765  & 0.605   \\
    \texttt{map-plusone-1}              & 5   & 459  & 0.057   & 0.388  & ---    & ---    & 1.443  & 0.478  & 0.037 &                & 0.078 & 1.365  & 1.132   \\
    \texttt{map-plusone-2}              & 5   & 704  & 1.423   & 6.450  & ---    & ---    & 2.750  & 0.588  & 0.081 &                & 0.086 & 2.664  & 2.235   \\
    \texttt{safe-head}                  & 3   & 354  & 0.048   & 0.046  & 0.040  & ---    & 0.246  & 0.364  & 0.012 & \cross         & 0.047 & 0.199  & 0.066   \\
    \texttt{safe-init}                  & 3   & 680  & 0.081   & 0.147  & 0.263  & ---    & 0.486  & 0.416  & 0.016 & \cross         & 0.071 & 0.415  & 0.103   \\
    \texttt{safe-tail}                  & 3   & 468  & 0.061   & 0.051  & 0.052  & ---    & 0.306  & 0.391  & 0.013 & \cross         & 0.058 & 0.248  & 0.093   \\
    \texttt{g41}                        & 4   & 31   & ---     & 0.046  & 0.067  & ---    & ---    & 0.321  & 0.006 & \cross         & 0.027 & ---    & 0.116   \\
    \texttt{cfa-life2}                  & 14  & 7648 & ---     & ---    & ---    & ---    & ---    & 0.857  & 0.173 &                & 0.431 & ---    & ---     \\
    \texttt{cfa-matrix-1}               & 8   & 2944 & 17.358  & ---    & ---    & 16.905 & 17.311 & 0.412  & 0.056 & \tick          & 0.225 & 17.086 & 17.081  \\
    \texttt{cfa-psdes}                  & 7   & 1819 & 17.850  & ---    & ---    & 1.331  & 1.452  & 0.363  & 0.033 & \tick          & 0.143 & 1.309  & 1.301   \\
    \texttt{dna}                        & 2   & 411  & 0.069   & 0.173  & 0.063  & 21.220 & 6.867  & 11.553 & 0.038 & \cross         & 0.120 & 6.746  & 6.303   \\
    \texttt{exp4-5}                     & 4   & 55   & ---     & ---    & 0.306  & ---    & ---    & 0.389  & 0.010 &                & 0.032 & ---    & 2.410   \\
    \texttt{fibstring}                  & 4   & 29   & ---     & 33.340 & 0.066  & ---    & ---    & 0.294  & 0.004 &                & 0.031 & ---    & 0.132   \\
    \texttt{fold\_fun\_list}            & 7   & 1346 & 0.618   & ---    & ---    & 1.262  & 1.284  & 0.327  & 0.020 &                & 0.109 & 1.175  & 1.169   \\
    \texttt{fold\_right}                & 5   & 1310 & 32.123  & ---    & ---    & 1.248  & 1.335  & 0.331  & 0.021 & \tick          & 0.106 & 1.229  & 1.222   \\
    \texttt{jwig-cal\_main}             & 2   & 7627 & 0.127   & 0.053  & ---    & 4.662  & 5.137  & 0.530  & 0.137 &                & 5.087 & 0.050  & 0.044   \\
    \texttt{l}                          & 3   & 35   & ---     & 7.523  & 0.020  & 0.131  & 0.129  & 0.297  & 0.006 &                & 0.030 & 0.100  & 0.092   \\
    \texttt{search-e-church} (bug)      & 6   & 837  & 0.023   & 0.218  & ---    & ---    & 5.708  & 3.623  & 0.038 &                & 0.102 & 5.606  & 1.790   \\
    \texttt{specialize\_cps\_coerce1-c} & 3   & 2731 & ---     & ---    & ---    & 0.463  & 0.503  & 0.433  & 0.206 & \tick          & 0.184 & 0.320  & 0.313   \\
    \texttt{tak} (bug)                  & 8   & 451  & ---     & 2.002  & ---    & ---    & 50.032 & 3.276  & 0.090 &                & 0.084 & 49.948 & 42.078  \\
    \texttt{xhtmlf-div-2} (bug)         & 2   & 3003 & 0.333   & ---    & 13.401 & 3.497  & 3.651  & 1.438  & 1.597 &                & 3.360 & 0.291  & 0.269   \\
    \texttt{xhtmlf-m-church}            & 2   & 3027 & 0.336   & ---    & 5.342  & 3.542  & 3.441  & 0.754  & 1.153 &                & 3.194 & 0.247  & 0.240   \\
    \texttt{zip}                        & 4   & 2952 & 22.606  & ---    & ---    & ---    & 2.567  & 0.728  & 0.060 & \tick          & 0.157 & 2.409  & 1.612   \\
    \bottomrule
    \end{tabular}
    \caption{\label{tbl:results}Comparison of model-checking tools.}
\end{sidewaystable}

Table~\ref{tbl:results} shows trials where at least one tool took over 1s.
This is to save space and because virtual machine ``warm-up'' and HORS to CPDS conversion can skew the results on small benchmarks.
Examples violating their property are marked ``(bug)''.
The order (Ord) and size (Sz) of the schemes were reported by \trecs.
We show reported times in seconds for \trecs (T), \gtrecst (G), \travmc (TMC), \preface (P), \horsattwo (H), and \cshore (C) as well as \cshore implementing a naive fixed point computation for the saturation (N).
A dash ``---'' means analysis failed.
In the next column we mark when \cshore was the fastest (\tick) and slowest (\cross) amongst its previous competitors (not including \preface or \horsattwo).
For \cshore, we then report the times for HORS to CPDS translation (Ctran), CPDS analysis (Ccpds), and building the approximation graph (Capprox).
Capprox is part of Ccpds, and the full time (C) is the sum of Ctran and Ccpds.

Of 35 benchmarks, \cshore outperformed its previous competitors on 7 examples.
In 9 cases, \cshore was the slowest, but in only 2 of those cases did \cshore require more than 1 second.
In general, both \preface and \horsattwo outperform all previous tools.
It is worth noting that \horsattwo, which appears to perform the best, is an adaptation of our saturation algorithm to recursion schemes~\cite{BK13}.

Notably, \cshore does not perform well on \texttt{g41} and \texttt{exp4-5}.  These belong to a class of benchmarks that
stress higher-order model-checkers and indicate that our tool currently does not
always scale well. However, \cshore seems to show a more promising capacity to
scale on larger HORS produced by tools such as \mochi~\cite{KSU11}, which are particularly
pertinent in that they are generated by an actual software verification tool.
We also note that \cshore timed out on the fewest examples of the previous tools despite not always
terminating in the fastest time.

Finally, without the forwards analysis described in
Section~\ref{section:optimisations}, all shown examples except
\texttt{filepath} timed out.
In addition, the naive version of the saturation algorithm performed significantly worse than the improved fixed point computation presented in Section~\ref{sec:fastalgorithm}.

%% file: conclusion.tex
\section{Conclusion}

We have given a full account of the \cshore tool.
This includes the development of a \emph{saturation} algorithm for CPDS that we first introduced in ICALP 2012~\cite{BCHS12}.
This is a backwards reachability algorithm.
To produce a viable implementation we optimised this algorithm using two main approaches.
The first is a preliminary forwards analysis which allows the input CPDS to be pruned and guarded, leading to faster analysis times.
The second is an efficient fixed point computation.
This implementation was first published in ICFP 2013~\cite{BCHS13}.

We have extended these results here by providing a generalisation of the implemented algorithms to alternating CPDS.
Furthermore, we have implemented a naive version of the fixed point iteration required by saturation.
Since this naive implementation is significantly out-performed by our efficient algorithm, we provide justification for the development in Section~\ref{sec:fastalgorithm}.

\cshore remains the only implementation of higher-order model checking using CPDS.
This provides a completely novel approach which was competitive with its contemporary tools.
Since its release, two new tools, \preface and \horsat (and \horsattwo), were developed.
These new tools are currently the fastest model-checkers for HORS.

%% file: acknowledgments.tex
{\footnotesize
\paragraph{Thanks}
Robin Neatherway, Steven Ramsay, and Naoki Kobayashi for help with
benchmarking, {\L}ukasz Kaiser and Royal Holloway for web-hosting, and
Stefan Schwoon.  This work was supported by
Deutsche Forschungsgemeinschaft [232350543], Fond.  Sci.  Math.
Paris, AMIS [ANR 2010 JCJC 0203 01 AMIS], FREC [ANR 2010 BLAN 0202 02 FREC],
VAPF (R\'egion IdF), and the Engineering and Physical Sciences Research
Council [EP/K009907/1].
}